\documentclass{article}
\newcommand{\no}{\noindent}
\newcommand{\e}{\mathrm{e}}
\newcommand{\ep}{\epsilon}
\newcommand{\vev}[1]{\left\langle #1 \right\rangle}
\newcommand{\Sf}{S_{\mathrm{free}}}
\newcommand{\Si}{S_{\mathrm{int}}}
\newcommand{\Gi}{\Gamma_{\mathrm{int}}}
\newcommand{\V}{\mathcal{V}}
\newcommand{\ffbox}[1]{\fbox{$\displaystyle #1 $}}
\newcommand{\Op}{\mathcal{O}}
\newcommand{\D}{\mathcal{D}}
\newcommand{\N}{\mathcal{N}}
\newcommand{\phitwo}{\left[\frac{1}{2} \phi^2\right]}
\newcommand{\phifour}{\left[\frac{1}{4!} \phi^4\right]}
\newcommand{\dphitwo}{\left[\frac{1}{2} \left(\partial_\mu \phi\right)^2
\right]}
\newcommand{\fmslash}[1]{\hbox{$#1$\kern-0.5em\raise0.3ex\hbox{/}}}
\newcommand{\Tr}{\mathrm{Tr}\,}
\newcommand{\ld}[1]{\frac{\overrightarrow{\delta}}{\delta \bar{\psi}
(-#1)}}
\newcommand{\rd}[1]{\frac{\overleftarrow{\delta}}{\delta \psi (#1)}}
\newcommand{\K}[1]{K(#1/\Lambda)}
\newcommand{\Rd}[1]{\frac{\overleftarrow{\delta}}{\delta #1}}
\newcommand{\Ld}[1]{\frac{\overrightarrow{\delta}}{\delta #1}}
\newcommand{\s}{\mathcal{S}}
\newcommand{\si}{\mathcal{S}_{\mathrm{int}}}
\newcommand{\ac}[1]{\left[A_{#1} c\right]}
\newcommand{\cc}{\left[cc\right]}
\usepackage{epsfig}
\begin{document}

\title{The Exact Renormalization Group\footnote{Lectures given at the
    Feza G\"ursey Institute, Istanbul, Turkey in 10 -- 21 September,
  2007.}\\
  -- renormalization theory revisited --}
\author{Hidenori Sonoda\footnote{E-mail: \texttt{hsonoda@kobe-u.ac.jp}}\\
  \small Physics Department, Kobe University, Japan} \date{September
  2007}
\maketitle

\begin{abstract}
    We overview the entire renormalization theory, both perturbative
    and non-perturbative, by the method of the exact renormalization
    group (ERG).  We emphasize particularly on the perturbative
    application of the ERG to the $\phi^4$ theory and QED in the $4$
    dimensional euclidean space.
\end{abstract}

\tableofcontents

\newpage
\section{Introduction}

The subject of the following series of lectures is the exact
renormalization group (ERG), also known as the Wilsonian
renormalization group.  Ken Wilson invented it in the early 70's to
understand the physics behind renormalization of quantum field theory.
Quantum field theory had seen phenomenal success in its applications
to QED via renormalization.  But many, if not most, people felt uneasy
about renormalization.  It was quite common to regard renormalization
as a clever mathematical trick to hide what is not understood under
the rug.  Not many had even imagined there was physics behind the
procedure of renormalization.

To understand the physics of renormalization, Wilson introduced RG
flows in the space of theories.  Using such important notions as
criticality, fixed points, and relevant and marginal parameters, the
continuum limit is defined as the RG trajectories flowing out of a UV
fixed point.  The continuum limit can be constructed by using an
almost critical theory by using the scale dimensions of relevant
parameters.  All this has been explained in the classic review article
of Wilson and Kogut \cite{wk}, especially in sect.~12.

In the following lectures, our emphasis is on the perturbative
applications of ERG.  Only in the last part, we discuss the
non-perturbative aspects of ERG.  Even there the emphasis is on the
relation of Wilson's original ERG differential equation to
Polchinski's which can be applied much more easily to perturbation
theory.

The lectures are organized into four parts.  In the first part we
review the application of ERG to the most relevant aspects of
perturbative renormalization theory using the $\phi^4$ theory for
concreteness: We derive Polchinski's version of the ERG differential
equation that describes the cutoff dependence of the Wilson
action.\cite{Pol} We then show how to incorporate perturbative
renormalizability as asymptotic behaviors of the action as the cutoff
is raised toward infinity.  We show how to determine the cutoff action
by solving the ERG differential equation under the asymptotic
conditions.  Using composite operators as an essential tool, we use
ERG to derive the RG equations of the parameters of the theory, and to
show universality.

In the second and third parts, we show how a momentum cutoff can be
consistent with gauge symmetry.  In the second part we take a
bottom-to-up approach to construct a cutoff action of QED: we first
construct an action that reproduces the Ward identities of the
correlation functions, and then introduce Faddeev-Popov ghosts to
introduce a BRST invariant action.  We finally introduce sources that
generate BRST transformations to build a fully BRST invariant
formalism with nilpotent BRST transformations.  The discussions of YM
theories in the third part are relatively brief.  We outline a
construction of a fully BRST invariant cutoff action with sources.
The sources are necessary to show the possibility of satisfying BRST
invariance.  Once the possibility is known, practical calculations can
be done without the sources.  Part of the new results contained in the
second and third parts of the lectures have been obtained in
collaboration with Y.~Igarashi and K.~Itoh.

Finally, in the last part we discuss the non-perturbative aspects of
ERG.  In the first subsection we give a non-perturbative derivation of
a general class of ERG differential equations.  We then relate
Wilson's original ERG differential equation to Polchinski's explicitly
via an integral formula.  The two sets of ERG differential equations
are completely equivalent.  We then show how to modify Polchinski's
ERG differential equation to accommodate anomalous dimensions.  We end
the last part by going back to perturbation theory, computing the
critical exponents of the Wilson-Fisher fixed point using ERG.

Best efforts have been made to make the lectures as self-contained as
possible.  The reader is warned, though, that my style may be somewhat
idiosyncratic; the ERG is a rich field, and it can accommodate widely
varying viewpoints.  The aim of the following lectures is to give a
coherent overview of the applications of ERG to renormalization
theory, but not to give a balanced review of the field of ERG.  For
the latter and also for further references, consultation with the
existing reviews
\cite{becchi,review1,review2,review3,review4,review5,review6,review7,review8,
  review9} is strongly recommended. The other references I cite are
kept minimal except for the inclusion of refs.\cite{iis,hik} which
happen to be relevant to my current research interest.

\section{The Exact Renormalization Group (ERG)}

\subsection{Notation}

We consider a real scalar theory in $D$-dimensional euclidean space.
The scalar field is given by
\[
\phi (r) = \int_p \phi (p) \,\e^{i p r}\qquad \left(\int_p \equiv \int
\frac{d^D p}{(2\pi)^D}\right)
\]
where
\[
p r \equiv \sum_{\mu=1}^D p_\mu r_\mu
\]
Given an action $S [\phi]$, we define the correlation functions by
\[
\vev{\phi (p_1) \cdots \phi (p_n)}_S \cdot (2\pi)^D \delta^{(D)} (p_1
+ \cdots + p_n) \equiv \frac{\int [d\phi] \, \phi (p_1) \cdots \phi
  (p_n)\, \e^{S [\phi]}}{\int [d\phi] \, \e^{S[\phi]}}
\]
Note that the weight of the functional integral is $\e^S$, not
$\e^{-S}$.  For the higher-point functions $n > 2$, we usually
consider only the connected part:
\[
\vev{\phi (p_1) \cdots \phi (p_n)} \stackrel{n > 2}{\Longrightarrow}
\vev{\phi (p_1) \cdots \phi (p_n)}^{\mathrm{connected}}
\]

We consider a real scalar theory whose propagator is given by
\[
\ffbox{\frac{K(p/\Lambda)}{p^2 + m^2}}
\]
where $K(p)$ is a positive decreasing function of $p^2$ with the
property that
\[
K(p) \left\lbrace\begin{array}{l@{\quad}l}
 = 1 & (p^2 < 1)\\
 \rightarrow 0 & (p^2 \to \infty)\end{array}\right.
\]
\begin{figure}[h]
\begin{center}
\epsfig{file=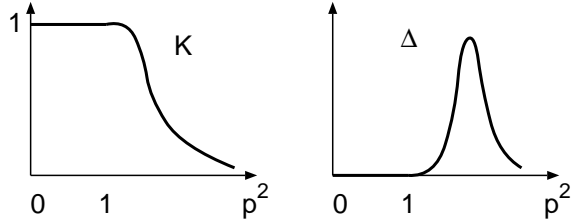, height=3cm}
\caption{The cutoff function $K$ and its derivative $\Delta$}
\end{center}
\end{figure}

The propagation of the momentum modes with $p^2 > \Lambda^2$ is
suppressed sufficiently strongly so that the loop integrals for each
Feynman diagram (to be introduced shortly) become well defined, i.e.,
\textbf{free of UV divergences}.  For later conveniences, we also
define the logarithmic derivative
\[
\Delta (p) \equiv - 2 p^2 \frac{d K(p^2)}{d p^2}
\]
which vanishes for $p^2 < 1$, and is positive for $p^2 > 1$ with
a peak near $p^2 = 1$.

For example, we can take
\[
K(p) \equiv \left\lbrace\begin{array}{c@{\quad}l}
 1 & (p^2 < 1)\\
 1 - \e^{- \frac{1}{\left(p^2 - 1\right)^n}}& (p^2 > 1)
\end{array}\right.
\]
so that
\[
\Delta (p) = \left\lbrace\begin{array}{c@{\quad}l}
 0 & (p^2 < 1)\\
\frac{2 n p^2}{(p^2-1)^{n+1}} \,\e^{- \frac{1}{\left(p^2-1\right)^n}}&
(p^2 > 1)\end{array}\right.
\]
Here, $n$ is a big enough integer.  For $D=4$, the worst divergence is
quadratic, so $n=2$ suffices.

Let us suppose that the action is given by
\[
S[\phi] = \Sf [\phi] + \Si [\phi]
\]
where the free action
\[
\Sf [\phi] \equiv - \frac{1}{2} \int_p \phi (p) \phi (-p) \frac{p^2 +
  m^2}{K(p/\Lambda)}
\]
defines the propagator given above, and the interaction part is given
by
\[
\Si [\phi] \equiv \sum_{n=1}^{\infty} \frac{1}{(2n)!}
\int_{p_1, \cdots, p_{2n}} 
\phi (p_1) \cdots \phi (p_{2n})\, \V_{2n} (p_1, \cdots, p_{2n}) \cdot
(2 \pi)^D \delta^{(D)} (p_1 + \cdots + p_{2n})
\]
We have assumed the $\mathbf{Z}_2$ invariance for
simplicity.\footnote{We also assume that the symmetry is not broken
  spontaneously.  Hence, we will only consider even-point correlation
  functions.}  We assume that the vertex $\V_{2n}$ is local: For our
purposes we only need to assume that $\V_{2n}$ can be expanded in
powers of small momenta.  We denote the $2n$-point vertex graphically
by a blob:
\begin{figure}[h]
\begin{center}
\epsfig{file=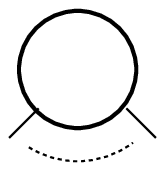}
\caption{A blob represents a $2n$-point vertex $\V_{2n}$.}
\end{center}
\end{figure}
The perturbative expansion of a correlation function is given by the
sum over Feynman diagrams with propagators and vertices.

For example, the diagrams in Fig.~3 contribute to the two-point
function:
\begin{figure}[h]
\begin{center}
\epsfig{file=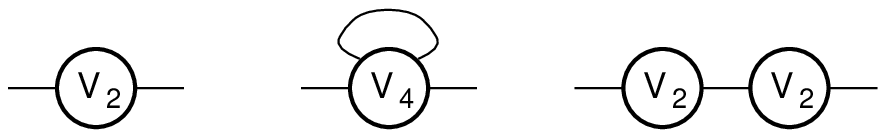}
\caption{These contribute to the two-point function.}
\end{center}
\end{figure}
and those in Fig.~4 contribute to the four-point function.
\begin{figure}[h]
\begin{center}
\epsfig{file=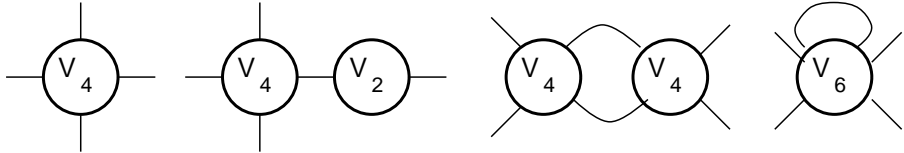}
\caption{These contribute to the four-point function.}
\end{center}
\end{figure}

\subsection{Derivation of the Polchinski ERG differential equation}

We solve the following problem: we decrease the momentum cutoff
$\Lambda$ infinitesimally to $\Lambda \e^{- \Delta t}$.  
\[
\ffbox{\Lambda \longrightarrow \Lambda \e^{- \Delta t}}
\]
We wish to compensate the change by appropriate changes in the
vertices so that the new propagator and vertices give the same
correlation functions.
\begin{center}
\begin{tabular}{ccc}
\ffbox{\Lambda \atop \V_{2n}}&
$\stackrel{\mathrm{equivalent}}{\Leftrightarrow}$&
\ffbox{\Lambda \e^{- \Delta t}\atop \V'_{2n}}
\end{tabular}
\end{center}

To solve this problem, we consider an arbitrary Feynman diagram, and
classify the propagators into the three types:\footnote{There is
  actually one exception to this classification: the free propagator
  contributing to the two-point function.  We will take care of this
  later.}
\begin{enumerate}
\item those connecting two different vertices (type A)
\item those connecting a single vertex (type B)
\item those attaching an external line to a vertex (type C)
\end{enumerate}
\begin{figure}[h]
\begin{center}
\epsfig{file=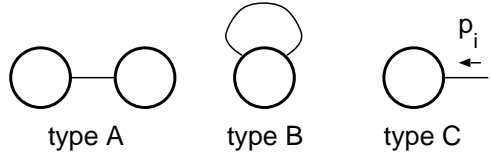}
\caption{The three types of propagators}
\end{center}
\end{figure}

The original propagator is related to the new propagator as follows:
\[
\frac{K(p/\Lambda)}{p^2 + m^2}
= \frac{K(p/(\Lambda \e^{- \Delta t}))}{p^2 + m^2} + \Delta t \,
\frac{\Delta (p/\Lambda)}{p^2 + m^2}
\]
We consider a Feynman graph with the original propagators and
vertices.  Then, we replace each propagator by the sum on the
right-hand side.  At first order in $\Delta t$, we get a
collection of Feynman diagrams in which one of the propagators is
replaced by
\[
\Delta t\, \frac{\Delta (p/\Lambda)}{p^2 + m^2}
\]
If the propagator is either type A or B, we can interpret it as
a correction to the relevant vertex.  (Fig.~6)  Hence, the type A and B
propagators modify the $2n$-point vertex to
\begin{eqnarray*}
&&\V_{2n} (p_1,\cdots,p_{2n})
\Longrightarrow \V'_{2n}(p_1, \cdots, p_{2n}) \equiv
\V_{2n} (p_1, \cdots, p_{2n}) \\
&&\quad + \Delta t \sum_{\mathrm{partitions}} \V_{2k}
  (p_{i_1},\cdots,p_{i_{2k-1}}, q)
\underbrace{\frac{\Delta (q/\Lambda)}{q^2 + m^2}}_{\mathrm{type~A}}
 \V_{2(n+1-k)} (- q, p_{i_{2k}},
\cdots, p_{i_{2n}}) \\
&&\quad + \Delta t \,\frac{1}{2} \int_q \underbrace{\frac{\Delta
    (q/\Lambda)}{q^2 + m^2}}_{\mathrm{type~B}} 
\, \V_{2(n+1)} (q,-q, p_1, \cdots, p_{2n})
\end{eqnarray*}
where the momentum $q$ of the type A propagator is fixed as
\[
q \equiv - \left( p_{i_1} + \cdots + p_{i_{2k-1}}\right)
= p_{i_{2k}} + \cdots + p_{i_{2n}}
\]
but the momentum $q$ of the type B propagator must be integrated over.
\begin{figure}[t]
\begin{center}
\epsfig{file=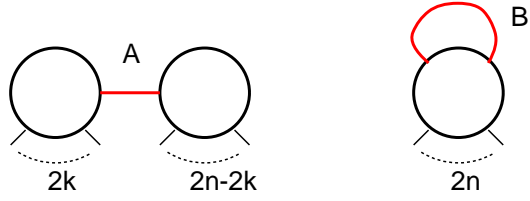}
\caption{Type A and B propagators modify the vertex $\V_{2n}$.}
\end{center}
\end{figure}
Since $\Delta (q/\Lambda)$ is non-vanishing only for $q^2 >
\Lambda^2$, we can consider the change of the vertices as local, i.e.,
in coordinate space the change takes place over a distance of order
$1/\Lambda$.

The change in the type C propagators can be corrected simply by the
multiplicative factor
\[
\frac{K(p_i/\Lambda)}{K(p_i/(\Lambda \e^{-\Delta t}))}
= 1 + \Delta t\, \frac{\Delta (p_i/\Lambda)}{K(p_i/\Lambda)}
\]
for each external line.

Using the modified vertices, we define a new action by
\begin{eqnarray*}
S' [\phi] &=& S [\phi] + \delta S [\phi]\\
 &\equiv& - \frac{1}{2} \int_p \phi (p) \phi (-p) \frac{p^2 +
   m^2}{K(p/(\Lambda \e^{-\Delta t}))} \\
&& + \sum_{n=1}^{\infty} \frac{1}{(2n)!} \int_{p_1,\cdots,p_{2n}}
\V'_{2n} (p_1,\cdots,p_{2n}) \cdot (2\pi)^D \delta^{(D)} (p_1 + \cdots
+ p_{2n})
\end{eqnarray*}
The diagrammatic analysis above implies
\[
\vev{\phi (p_1) \cdots \phi (p_{2n})}_{S}
= \prod_{i=1}^{2n} \frac{K(p_i/\Lambda)}{K(p_i/(\Lambda \e^{-\Delta
    t}))}
\cdot \vev{\phi (p_1) \cdots \phi (p_{2n})}_{S'}
\]
Hence,
\[
\vev{\phi (p_1) \cdots \phi (p_{2n})}_S - \vev{\phi (p_1) \cdots \phi
  (p_{2n})}_{S + \delta S}
 = \Delta t \sum_{i=1}^{2n} \frac{\Delta
  (p_i/\Lambda)}{K(p_i/\Lambda)}
\cdot \vev{\phi (p_1) \cdots \phi (p_{2n})}_S
\]

In fact the above relation is valid only for $2n \ge 4$.  For the
two-point function, the first term in its perturbative expansion is
the free propagator.  It cannot be classified into any of the three
types.
\begin{center}
\parbox{3cm}{$\vev{\phi (p) \phi (-p)}_S =$}
\parbox{1.5cm}{\epsfig{file=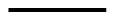}}
\parbox{5cm}{$+ \,\textrm{diagrams with at least one vertex}$}
\end{center}
Hence, it is the difference
\[
\vev{\phi (p) \phi (-p)}_S - \frac{K(p/\Lambda)}{p^2 + m^2}
\]
which contains only the propagators of the three types.  Therefore,
we obtain
\[
\vev{\phi (p) \phi (-p)}_S - \frac{K(p/\Lambda)}{p^2 + m^2}
= \frac{K(p/\Lambda)^2}{K(p/(\Lambda \e^{-\Delta t}))^2} \cdot
\left( \vev{\phi (p) \phi (-p)}_{S'} - \frac{K(p/(\Lambda \e^{-\Delta
        t}))}{p^2 + m^2} \right)
\]
We can rewrite this as
\begin{eqnarray*}
&& \vev{\phi (p) \phi (-p)}_S - \vev{\phi (p) \phi (-p)}_{S + \delta
  S}\\
&& = \Delta t \left[ 2 \frac{\Delta (p/\Lambda)}{K(p/\Lambda)} \cdot
\vev{\phi (p) \phi (-p)}_S 
 - \frac{\Delta (p/\Lambda)}{K(p/\Lambda)}
\, \frac{K(p/\Lambda)}{p^2 + m^2} \right]
\end{eqnarray*}

The expression of the new action $S' = S + \delta S$, which is
equivalent to the original action $S$, is easy to understand
graphically, but quite complicated to write down as a formula.  By using a
functional notation, $\delta \Si$ (the interaction part of $\delta S$)
can be expressed quite simply:
\[
\delta \Si = \Delta t \,
\int_q \frac{\Delta (q/\Lambda)}{q^2 + m^2} \frac{1}{2}
\left\lbrace \frac{\delta \Si}{\delta \phi (q)} \frac{\delta
      \Si}{\delta \phi (-q)} + \frac{\delta^2 \Si}{\delta \phi (q)
      \delta \phi (-q)} \right\rbrace
\]
\no \textbf{HW\#1}: Convince yourself this is the same as the equation
we have obtained using Feynman graphs.

Writing the original $S$ as $S(\Lambda)$ and $S'$ as $S(\Lambda
\e^{-\Delta t})$, we obtain
\[
\delta \Si = \Delta t \cdot \left( - \Lambda \frac{\partial}{\partial
      \Lambda} \right) \Si
\]
Hence, $\Si (\Lambda)$ satisfies the following differential equation:
\[
\ffbox{- \Lambda \frac{\partial \Si (\Lambda)}{\partial \Lambda}
= \int_q \frac{\Delta (q/\Lambda)}{q^2 +
  m^2} \frac{1}{2} \left\lbrace \frac{\delta \Si}{\delta \phi (q)}
    \frac{\delta \Si}{\delta \phi (-q)} + \frac{\delta^2 \Si}{\delta
      \phi (q) \delta \phi (-q)} \right\rbrace}
\]
This was first derived by Polchinksi.\cite{Pol}  The transformation
from $S(\Lambda)$ to $S (\Lambda \e^{-t})$, where $t > 0$, is called
an \textbf{exact renormalization group (ERG)} transformation: it is
exact since we have not introduced any approximation.

Using
\begin{eqnarray*}
\delta \Sf &=& \Sf (\Lambda \e^{- \Delta t}) - \Sf (\Lambda)\\
&=& - \frac{1}{2} \int_p \phi (p) \phi (-p) \left( p^2 + m^2\right)
\left( \frac{1}{K(p/(\Lambda \e^{- \Delta t}))} -
    \frac{1}{K(p/\Lambda)} \right)\\
&=& - \Delta t \,\frac{1}{2} \int_p \phi (p) \phi (-p) \left( p^2 + m^2\right)
\frac{\Delta (p/\Lambda)}{K(p/\Lambda)^2}
\end{eqnarray*}
further, we can rewrite the Polchinski differential equation 
for the entire action as follows:
\[
\ffbox{
\begin{array}{cl}
- \Lambda \frac{\partial}{\partial \Lambda} S (\Lambda)
=& \int_q \frac{\Delta (q/\Lambda)}{q^2 + m^2} \Bigg[
\frac{q^2 + m^2}{K(q/\Lambda)} \phi (q) \frac{\delta
  S(\Lambda)}{\delta \phi (q)}\\
&  \quad + \frac{1}{2} \left\lbrace \frac{\delta S(\Lambda)}{\delta
      \phi (q)} \frac{\delta S(\Lambda)}{\delta \phi (-q)} +
    \frac{\delta^2 S(\Lambda)}{\delta \phi (q) \delta \phi (-q)}
\right\rbrace \Bigg]
\end{array}}
\]
\no \textbf{HW\#2}: Derive this from Polchinski's.  Alternatively,
derive Polchinski's from this.  (You need to ignore an infinite
constant independent of $\phi$.  An additive constant to $S$ does not
affect physics.)

To summarize, the correlation functions satisfy the following ERG
differential equations:
\begin{eqnarray*}
&&- \Lambda \frac{\partial}{\partial \Lambda} \vev{\phi (p) \phi
  (-p)}_{S} = \vev{\phi (p) \phi (-p) \left( - \Lambda
      \frac{\partial S}{\partial \Lambda} \right)}_S^{\mathrm{connected}}\\
&& \qquad\qquad= - 2 \frac{\Delta (p/\Lambda)}{K(p/\Lambda)} \vev{\phi (p) \phi
  (-p)}_S + \frac{\Delta (p/\Lambda)}{p^2 + m^2}\\
&&- \Lambda \frac{\partial}{\partial \Lambda} \vev{\phi (p_1) \cdots
  \phi (p_{2n})} = \vev{\phi (p_1) \cdots \phi (p_{2n}) \left( - \Lambda
      \frac{\partial S}{\partial \Lambda} \right)}_S^{\mathrm{connected}}\\
&& \qquad\qquad = - \sum_{i=1}^{2n} \frac{\Delta (p_i/\Lambda)}{K(p_i/\Lambda)}
\cdot \vev{\phi (p_1) \cdots \phi (p_{2n})}_S
\end{eqnarray*}

\subsection{Physical understanding of ERG}

To gain physical (or intuitive) understanding of ERG, let us forget
about UV divergences or any other difficulties our theory might face
with, for the time being.

The standard propagator can be expressed as the sum:
\[
\frac{1}{p^2 + m^2} = \frac{K(p/\Lambda)}{p^2 + m^2} + \frac{1 -
  K(p/\Lambda)}{p^2 + m^2}
\]
The first term on the right is the cutoff propagator for the low
momentum modes.  The second term is the propagator for the high
momentum modes.

Given a Feynman diagram with the standard propagators and elementary
vertices, we substitute the above sum into each propagator.  The
substitution generates multiple diagrams in which some propagators are
the low-momentum propagators, and the rest are high-momentum
propagators.  Then, we pay attention to those parts of the Feynman
diagram connected only by high-momentum propagators.   The
entire diagram has these connected parts connected further by
low-momentum propagators.  (Fig.~7)
\begin{figure}[b]
\begin{center}
\epsfig{file=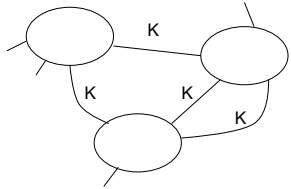, height=2.5cm}
\caption{Each blob contains only high-momentum propagators inside.}
\end{center}
\end{figure}
Suppose a connected part has $2n$ external lines.  Then, the
connected part can be interpreted as a contribution to the vertices
$\V_{2n} (\Lambda)$ of the cutoff theory $S(\Lambda)$.  (Fig.~8)
\begin{figure}[t]
\begin{center}
\epsfig{file=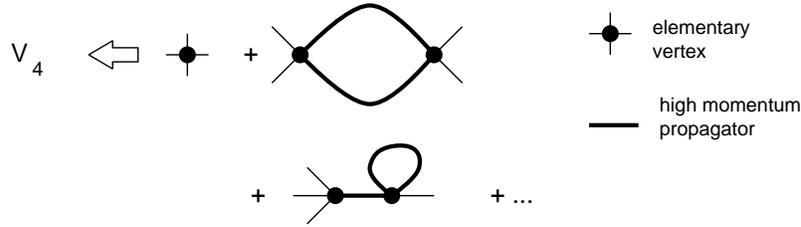, height=3cm}
\caption{$\V_{2n} (\Lambda)$ consists of graphs whose internal lines
  carry only high momenta above $\Lambda$.}
\end{center}
\end{figure}
Thus, we put all the short-distance physics (shorter than the length
$\frac{1}{\Lambda}$) into the vertices.

Now, the problem is the potential UV divergence of the diagrams with
the high-momentum propagators.  For example, the 1st order
contribution to $\V_2$ is given by
\begin{center}
\parbox{1.5cm}{\epsfig{file=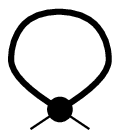}}
\parbox{4cm}{$\displaystyle \stackrel{?}{=} - \frac{\lambda}{2} \int_q
  \frac{1 - K(q/\Lambda)}{q^2 + m^2}$}
\end{center}
which is quadratically divergent since $K(q/\Lambda) \to 0$ as $q^2
\to \infty$.  We will define the diagram, not by the naive loop
integral, but as a finite solution to the ERG differential equation.
It is the role of ERG to make sense out of the naively UV divergent
Feynman diagrams.

\subsection{Perturbative renormalizability}

Polchinski introduced his differential equation in order to simplify
the proof of perturbative renormalizability of the $\phi^4$ theory in
$D=4$.  We will not follow his proof, but merely sketch his ideas
here.  We start from a bare action defined at a large momentum scale
$\Lambda$:
\begin{eqnarray*}
    &&S_\Lambda (\Lambda) = - \frac{1}{2} \int_p \phi (p) \phi (-p) \frac{p^2 +
      m^2}{K(p/\Lambda)} \\
    &&\quad - \frac{1}{2} \int_p \phi (p) \phi
    (-p) \left( \Delta z_0 \cdot p^2 + \Delta m_0^2 \right) 
    - \frac{\lambda_0}{4!} \int_{p_1+p_2+p_3+p_4=0} \phi (p_1) \cdots
    \phi (p_4) 
\end{eqnarray*}
We then reduce the cutoff $\Lambda$ to an arbitrary but fixed finite
momentum scale $\mu$ to obtain an equivalent action $S_\Lambda (\mu)$.
Here we have added a suffix $\Lambda$ to remind ourselves that we
started from the scale $\Lambda$.  We wish to introduce appropriate
$\Lambda$ dependence to the bare parameters
\[
\Delta z_0 (\Lambda),\quad \Delta m_0^2 (\Lambda),\quad \lambda_0 (\Lambda)
\]
such that the continuum limit
\[
\bar{S} (\mu) = \lim_{\Lambda \to \infty} S_\Lambda (\mu)
\]
exists.  (Fig.~9) This is \textbf{perturbative} renormalizability
since we prove this, order by order in a coupling constant $\lambda$.
Polchinski has proven this using his differential equation.
\begin{figure}[t]
\begin{center}
\epsfig{file=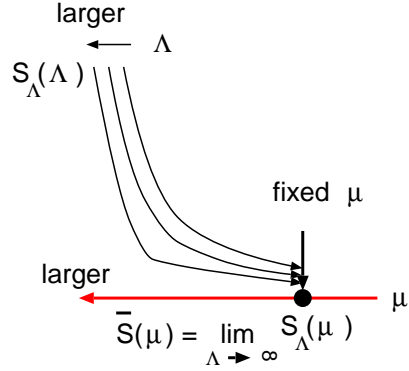}
\caption{Renormalizability amounts to the existence of $\lim_{\Lambda
    \to \infty} S_\Lambda (\mu)$.}
\end{center}
\end{figure}

Why is this renormalizability?  ERG guarantees that the correlation
functions we compute with $\bar{S} (\mu)$ is the same as those with
the bare action $S_\Lambda (\Lambda)$.  In the limit $\Lambda \to
\infty$, the theory is defined all the way up to the zero distance
scale, and the correlation functions are independent of $\Lambda$.
This is what we usually call renormalizability.

Typically, the bare parameters have the following $\Lambda$
dependence:
\[
\left\lbrace\begin{array}{c@{~=~}l}
 \Delta m_0^2 (\Lambda) & \Lambda^2 \cdot g \left(\frac{m^2}{\mu^2}, \ln
     \frac{\Lambda}{\mu},  \lambda\right) + m^2 \cdot z_m 
\left(\frac{m^2}{\mu^2}, \ln \frac{\Lambda}{\mu}, \lambda\right)\\
 \Delta z_0 (\Lambda) & z \left(\frac{m^2}{\mu^2}, \ln
     \frac{\Lambda}{\mu}, \lambda\right)\\ 
 \lambda_0 (\Lambda) & z_\lambda \left(\frac{m^2}{\mu^2}, \ln
     \frac{\Lambda}{\mu}, \lambda\right)
\end{array}\right.
\]
At each order of $\lambda$, the coefficient functions are finite
degree polynomials of $\ln \frac{\Lambda}{\mu}$.  The so-called
\textbf{mass independent} scheme is distinguished by the absence of
the dependence on $\frac{m^2}{\mu^2}$ in the coefficient functions.
Hence, in a mass independent scheme, we find
\[
\left\lbrace\begin{array}{c@{~=~}l}
 \Delta m_0^2 (\Lambda) & \Lambda^2 \cdot g \left(\ln
     \frac{\Lambda}{\mu},  \lambda\right) + m^2 \cdot z_m 
\left(\ln \frac{\Lambda}{\mu}, \lambda\right)\\
 \Delta z_0 (\Lambda) & z \left(\ln
     \frac{\Lambda}{\mu}, \lambda\right)\\ 
 \lambda_0 (\Lambda) & z_\lambda \left(\ln
     \frac{\Lambda}{\mu}, \lambda\right)
\end{array}\right.
\]

\subsection{Direct determination of the renormalized theory}

We now make a crucial observation.  The action $\bar{S} (\mu)$, which
defines the renormalized theory or the continuum limit, satisfies the
ERG differential equation of Polchinski with respect to $\mu$:
\[
\ffbox{
\begin{array}{cl}
- \mu \frac{\partial \bar{S} (\mu)}{\partial \mu}
=& \int_q \frac{\Delta (q/\mu)}{q^2 + m^2} 
\Bigg[ \frac{q^2 + m^2}{K(q/\mu)} \phi (q) \frac{\delta
      \bar{S}}{\delta \phi (q)} \\
& + \frac{1}{2} \left\lbrace
        \frac{\delta \bar{S}}{\delta \phi (q)} \frac{\delta
          \bar{S}}{\delta \phi (-q)} + \frac{\delta^2 \bar{S}}{\delta
          \phi (q) \delta \phi (-q)} \right\rbrace \Bigg]
\end{array}}
\]
Hence, we should be able to determine $\bar{S} (\mu)$ directly without
starting from $S_\Lambda (\Lambda)$ by solving the differential
equation.  But as is the case with any differential equation, we need
what amounts to an initial condition.  We cannot use $\bar{S}
(\Lambda) = S_\Lambda (\Lambda)$, because $S_\Lambda (\Lambda)$ will
not lie on the ERG trajectory of $\bar{S} (\Lambda)$.  Instead we
adopt the following:
\begin{enumerate}
\item $\V_{2n} (\Lambda; p_1, \cdots, p_{2n})$ vanishes as $\Lambda
    \to \infty$ for $2n \ge 6$
\item $\V_2$ satisfies the asymptotic condition
\begin{eqnarray*}
\V_2 (\Lambda; p, -p) &\stackrel{\Lambda \to \infty}{\longrightarrow}&
\Lambda^2 a_2 (m^2/\mu^2, \ln \Lambda/\mu; \lambda) \\
&& + m^2 b_2
(m^2/\mu^2, \ln \Lambda/\mu; \lambda) + p^2 c_2
(m^2/\mu^2, \ln \Lambda/\mu; \lambda)
\end{eqnarray*}
and the conditions at $\Lambda = \mu$, where $\mu$ is fixed but arbitrary:
\[
\left\lbrace\begin{array}{c}
\V_2 (\mu; 0,0) = 0\\
\frac{\partial}{\partial p^2} \V_2 (\mu, p, -p) \Big|_{p^2=0} = 0
\end{array}\right.
\]
\item $\V_4$ satisfies the asymptotic condition
\[
\V_4 (\Lambda; p_1, \cdots, p_4) \stackrel{\Lambda \to
  \infty}{\longrightarrow} a_4 (m^2/\mu^2, \ln \Lambda/\mu; \lambda)
\]
and the condition at $\Lambda = \mu$:
\[
\V_4 (\mu; 0,\cdots,0) = - \lambda
\]
\end{enumerate}
The asymptotic conditions are crucial for renormalizability.  The two
conditions on $\V_2$ have to do with mass and wave function
renormalization, and the condition on $\V_4$ normalizes the coupling
$\lambda$.

The above scheme is not mass independent.  For mass independence, we
can alternatively adopt the following conditions:
\begin{enumerate}
\item $\V_{2n} (\Lambda; p_1, \cdots, p_{2n})$ vanishes as $\Lambda
    \to \infty$ for $2n \ge 6$
\item $\V_2$ satisfies
the asymptotic condition
\[
\V_2 (\Lambda; p, -p) \stackrel{\Lambda \to \infty}{\longrightarrow}
\Lambda^2 a_2 (\ln \Lambda/\mu; \lambda) + m^2 b_2 (\ln \Lambda/\mu;
\lambda) + p^2 c_2 (\ln \Lambda/\mu; \lambda)
\]
and the conditions at $\Lambda = \mu$:
\[
\begin{array}{c@{~=~0}}
\frac{\partial}{\partial m^2} \V_2 (\mu; 0,0)\Big|_{m^2=0} \\
\frac{\partial}{\partial p^2} \V_2 (\mu; p,-p)\Big|_{m^2=p^2=0}
\end{array}
\]
which implies
\[
b_2 (0; \lambda) = c_2 (0; \lambda) = 0
\]
\item $\V_4$ satisfies the asymptotic condition
\[
\V_4 (\Lambda; p_1, \cdots, p_4) \stackrel{\Lambda \to
  \infty}{\longrightarrow} a_4 (\ln \Lambda/\mu; \lambda)
\]
and the condition at $\Lambda = \mu$:
\[
\V_4 (\mu; 0,\cdots,0)\Big|_{m^2 = 0} = - \lambda
\]
which implies
\[
a_4 (0; \lambda) = 0
\]
\end{enumerate}
Note the absence of $m^2$ dependence in the coefficient functions
$a_2, b_2, c_2$, and $a_4$.

Hence, to show renormalizability of $\phi^4$ theory in $D=4$, we must
show that given $m^2, \lambda$, and an arbitrary scale $\mu$, the
solution $\bar{S} (\Lambda)$ of the Polchinski differential equation
is determined uniquely.  This was done for the mass independent scheme
in \cite{hs03}.\footnote{From now on, we only consider $\bar{S}$, and
  omit the bar.}

\subsection{Simple examples}

Let us compute the vertices for $D=4$ to see how the conditions
introduced above determine the vertices uniquely.

\subsection*{First order in $\lambda$}

The four-point vertex is given by
\[
\V_4^{(0)} (p_1,\cdots,p_4) = - \lambda
\]
This is the starting point of perturbative calculations.

The two-point vertex $\V_2^{(1)} (\Lambda)$ satisfies the differential
equation
\[
- \Lambda \frac{\partial \V_2^{(1)} (\Lambda)}{\partial \Lambda} =
\frac{1}{2} \int_q \frac{\Delta (q/\Lambda)}{q^2 + m^2} \V_4^{(0)}
(q,-q,p,-p)
= - \lambda \frac{1}{2} \int_q \frac{\Delta
  (q/\Lambda)}{q^2 + m^2}
\]
This can be expressed diagrammatically as 
\begin{center}
\epsfig{file=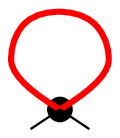}
\end{center}
where we use the notation
\begin{center}
\parbox{1.3cm}{\epsfig{file=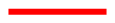}}\parbox{3cm}{$\displaystyle =
  \frac{\Delta (q/\Lambda)}{q^2 + m^2}$}
\end{center}
We can compute the right-hand side further as
\begin{eqnarray*}
&=& - \lambda \frac{1}{2} \int_q \Delta (q/\Lambda) \left(
    \frac{1}{q^2} - \frac{m^2}{q^4} + \frac{m^4}{q^4 (q^2+m^2)}
\right)\\
&=& - \lambda \frac{1}{2} \left( \Lambda^2 \int_q \frac{\Delta
      (q)}{q^2} - m^2 \int_q \frac{\Delta (q)}{q^4} + m^4 \int_q
    \Delta (q/\Lambda) \frac{1}{q^4(q^2+m^2)} \right)
\end{eqnarray*}
Integrating this over $\Lambda$, we obtain
\begin{center}
\parbox{2cm}{$\displaystyle 
\V_2^{(1)} (\Lambda) =$}
\parbox{8cm}{\epsfig{file=v2.eps}\hfill}
\end{center}
\begin{eqnarray*}
&=& - \frac{\lambda}{2} \Bigg(
- \frac{\Lambda^2}{2} \int_q \frac{\Delta (q)}{q^2} 
+ m^2 \ln \frac{\Lambda}{\mu} \int_q \frac{\Delta (q)}{q^4} + m^4
\int_q \frac{1-K(q/\Lambda)}{q^4 (q^2 + m^2)} \\
&& \qquad + \frac{\mu^2}{2} \int_q \frac{\Delta (q)}{q^2}
- m^4 \int_q \frac{1- K(q/\mu)}{q^4 (q^2 + m^2)} \Bigg)
\end{eqnarray*}
which satisfies the condition
\[
\V_2^{(1)} (\Lambda = \mu) = 0
\]
Alternatively, in the mass independent scheme, we obtain
\[
\V_2^{(1)} (\Lambda) = - \frac{\lambda}{2} \left(
- \frac{\Lambda^2}{2} \int_q \frac{\Delta (q)}{q^2} 
+ m^2 \ln \frac{\Lambda}{\mu} \int_q \frac{\Delta (q)}{q^4} + m^4
\int_q \frac{1-K(q/\Lambda)}{q^4 (q^2 + m^2)} \right)
\]
which has the simple asymptotic behavior
\[
\V_2^{(1)} (\Lambda) \to  - \frac{\lambda}{2} \left(
- \frac{\Lambda^2}{2} \int_q \frac{\Delta (q)}{q^2} 
+ m^2 \ln \frac{\Lambda}{\mu} \int_q \frac{\Delta (q)}{q^4} \right)
\]

\subsection*{Second order in $\lambda$}

The ERG differential equation for the six-point vertex is given
diagrammatically as
\begin{center}
\parbox{4.5cm}{$\displaystyle - \Lambda \frac{\partial}{\partial
    \Lambda} \V_6^{(0)} (\Lambda; p_1, \cdots, p_6) =$}
\parbox{3cm}{\epsfig{file=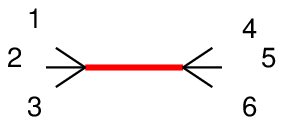}}
\parbox{3cm}{$+ \,\textrm{2 permutations}$}
\end{center}
Solving this, we obtain
\begin{center}
\parbox{3.5cm}{$\displaystyle
\V_6^{(0)} (\Lambda; p_1, \cdots, p_6) =$}
\parbox{3cm}{\epsfig{file=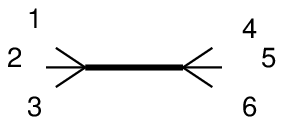}}
\parbox{3cm}{$+ \,\textrm{2 permutations}$}
\end{center}
\[
= \lambda^2 \left( \frac{1 -
      K((p_1+p_2+p_3)/\Lambda)}{(p_1+p_2+p_3)^2 + m^2} + \textrm{2
      permutations} \right)
\]
This vanishes as $\Lambda \to \infty$.

At one-loop the four-point vertex satisfies the ERG differential
equation:
\begin{center}
\parbox{5.5cm}{$\displaystyle
- \Lambda \frac{\partial}{\partial \Lambda}
 \V_4^{(1)} (\Lambda; p_1,\cdots, p_4) = \sum_{i=1}^4 \Bigg($}
\parbox{2cm}{\epsfig{file=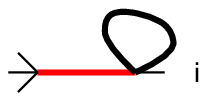}}
\parbox{0.5cm}{$+$}
\parbox{2cm}{\epsfig{file=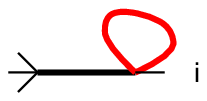}}
\parbox{0.5cm}{$\Bigg)$}\\
\parbox{3cm}{\hfill$+$}
\parbox{2cm}{\epsfig{file=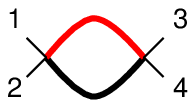}}
\parbox{2cm}{$+ \textrm{2 permutations}$}
\end{center}
The first part on the right-hand side can be integrated to
\[
(-\lambda) \sum_{i=1}^4 \frac{1 - K(p_i/\Lambda)}{p_i^2 + m^2} \cdot
\V_2^{(1)} (\Lambda)
\]
The first graph on the second part gives
\begin{eqnarray*}
    &&\lambda^2 \int_q \frac{1 - K((q+p_1+p_2)/\Lambda)}{(q+p_1+p_2)^2 + m^2}
    \frac{\Delta (q/\Lambda)}{q^2 + m^2}
    = \lambda^2 \int_q \Bigg[ \frac{\left(1 - K(q/\Lambda)\right) \Delta
          (q/\Lambda)}{q^4} \\
&&\qquad + \frac{1 -
          K((q+p_1+p_2)/\Lambda)}{(q+p_1+p_2)^2 + m^2} 
        \frac{\Delta (q/\Lambda)}{q^2 + m^2} - \frac{\left(1 -
              K(q/\Lambda)\right) \Delta 
          (q/\Lambda)}{q^4} \Bigg]
\end{eqnarray*}
The first integral on the right gives a constant
\[
\int_q \frac{\Delta (q) \left( 1 - K(q) \right)}{q^4}
\]
The remaining integral can be integrated over $\Lambda$ to give a
finite result:
\[
\frac{1}{2} \int_q \left[ \frac{1 -
          K((q+p_1+p_2)/\Lambda)}{(q+p_1+p_2)^2 + m^2} 
        \frac{1 - K(q/\Lambda)}{q^2 + m^2} - \frac{\left(1 -
              K(q/\Lambda)\right)^2}{q^4} \right]
\]
Hence, we obtain
\begin{center}
\parbox{2.5cm}{\epsfig{file=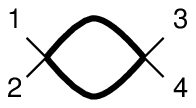}}
\parbox{5cm}{$\displaystyle = - \lambda^2 \ln \frac{\Lambda}{\mu} 
\int_q \frac{\Delta (q) \left( 1 - K(q) \right)}{q^4}$}\\
\parbox{8cm}{$\displaystyle
+ \frac{\lambda^2}{2} \int_q \left[ \frac{1 -
          K((q+p_1+p_2)/\Lambda)}{(q+p_1+p_2)^2 + m^2} 
        \frac{1 - K(q/\Lambda)}{q^2 + m^2} - \frac{\left(1 -
              K(q/\Lambda)\right)^2}{q^4} \right]$}
\end{center}
where we have adopted the mass independence.  This has the
asymptotic property
\begin{center}
\parbox{2.5cm}{\epsfig{file=v4.eps}}
\parbox{7cm}{$\displaystyle \stackrel{\Lambda \to
    \infty}{\longrightarrow}
 - \lambda^2 \ln \frac{\Lambda}{\mu}
\int_q \frac{\Delta (q) \left( 1 - K(q) \right)}{q^4}$}
\end{center}
\no \textbf{HW\#3}: Evaluate the one-loop four-point vertex using the
condition 
\[
\V_4 (\mu; 0,\cdots,0) = - \lambda
\]
instead of the mass independence.  You will see that the asymptotic
behavior acquires $m^2$ dependence.

\subsection*{Integrals}

In general the integrals involving the cutoff function $K$ depend on
the choice of $K$.  Certain integrals are universal, however.  Let us
first consider
\[
I(n) \equiv \int_q \Delta (q) \frac{K(q)^n}{q^4}
\]
in $D=4$.  This is integrable, since the integrand vanishes
exponentially as $q^2 \to \infty$, and also vanishes for $0 < q^2 <
1$.   Since
\[
\Delta (q) = - 2 q^2 \frac{d K(q)}{d q^2}
\]
we obtain
\[
\Delta (q) K(q)^n = \frac{1}{n+1} ( - 2 q^2) \frac{d K
      (q)^{n+1}}{d q^2}
\]
Hence,
\begin{eqnarray*}
I_n &=& \frac{2 \pi^2}{(2 \pi)^4} \frac{1}{n+1} \int_0^\infty
\frac{q^2 dq^2}{2} 
\frac{- 2 q^2}{q^4} \frac{d K (q)^{n+1}}{d q^2}
= \frac{1}{(4 \pi)^2} \frac{2}{n+1} \underbrace{\left[ - K(q)^{n+1}
  \right]_0^\infty}_{= 1}\\
&=& \ffbox{\frac{1}{(4 \pi)^2} \frac{2}{n+1}}
\end{eqnarray*}

Next we consider
\[
J(n) \equiv \int_q \frac{\Delta (q) \left(1 - K(q)\right)^n}{q^4}
\]
in $D=4$, where $n$ is a non-negative integer.  Using the binomial
expansion, we obtain
\begin{eqnarray*}
J(n) &=& \sum_{i=0}^n (-)^i \pmatrix{n\cr i\cr} I(i)
= \frac{2}{(4 \pi)^2} \sum_{i=0}^n (-)^i \frac{1}{i+1} \pmatrix{n\cr
  i\cr}\\
&=& \frac{2}{(4 \pi)^2} \frac{1}{n+1} 
\underbrace{\sum_{i=0}^n (-)^i \pmatrix{n+1\cr
    i+1\cr}}_{= 1}\\
&=& \ffbox{\frac{1}{(4 \pi)^2} \frac{2}{n+1}}
\end{eqnarray*}

Using this result, we can write down the one-loop vertices in the mass
independent scheme as follows:
\begin{center}
\parbox{1.5cm}{\epsfig{file=v2.eps}}
\parbox{9cm}{$\displaystyle = - \frac{\lambda}{2} \left(
- \frac{\Lambda^2}{2} \int_q \frac{\Delta (q)}{q^2} 
+ m^2 \ln \frac{\Lambda}{\mu} \frac{2}{(4\pi)^2} + m^4
\int_q \frac{1-K(q/\Lambda)}{q^4 (q^2 + m^2)} \right)$\hfill}\\
\parbox{2cm}{\epsfig{file=v4.eps}}
\parbox{9cm}{$\displaystyle = \frac{\lambda^2}{(4 \pi)^2} \ln
  \frac{\Lambda}{\mu}$\hfill} \\
\parbox{8cm}{$\displaystyle 
+ \frac{\lambda^2}{2} \int_q \left[ \frac{1 -
          K((q+p_1+p_2)/\Lambda)}{(q+p_1+p_2)^2 + m^2} 
        \frac{1 - K(q/\Lambda)}{q^2 + m^2} - \frac{\left(1 -
              K(q/\Lambda)\right)^2}{q^4} \right]$}
\end{center}
This implies
\[
\left\lbrace
\begin{array}{c@{~=~}l}
b_2 (\ln \Lambda/\mu, \lambda) & - \frac{\lambda}{(4 \pi)^2} \ln
\frac{\Lambda}{\mu} + \cdots\\
c_2 (\ln \Lambda/\mu, \lambda) & \mathrm{O} \left(\lambda^2\right)\\
a_4 (\ln \Lambda/\mu, \lambda) & \frac{3 \lambda^2}{(4 \pi)^2} \ln
\frac{\Lambda}{\mu} + \cdots
\end{array}\right.
\]

\subsection{The continuum limit in terms of a cutoff theory}

Before diving into more detailed analysis of ERG, let us take a moment
to reflect on what ERG gives us.  In deriving ERG, we have made sure
that the entire physics is kept untouched when we lower the cutoff
$\Lambda$.  No matter what $\Lambda$ we use, we can always compute the
same correlation functions.  Let us give this statement a more
concrete expression.

We recall that the two actions $S(\Lambda)$ and $S(\Lambda')$ on the
same ERG trajectory are related as follows:
\[
\left\lbrace\begin{array}{c@{~=~}l}
\vev{\phi (p) \phi (-p)}_{S(\Lambda)} - \frac{K (p/\Lambda)}{p^2+m^2}
& \frac{K(p/\Lambda)^2}{K(p/\Lambda')^2} \left( \vev{\phi (p) \phi
      (-p)}_{S(\Lambda')} - \frac{K (p/\Lambda')}{p^2+m^2} \right)\\
\vev{\phi (p_1) \cdots \phi (p_{2n})}_{S(\Lambda)} & \prod_{i=1}^{2n}
\frac{K(p_i/\Lambda)}{K(p_i/\Lambda')} \cdot \vev{\phi (p_1) \cdots
  \phi (p_{2n})}_{S (\Lambda')}
\end{array}\right.
\]
This can be rewritten as
\[
\left\lbrace\begin{array}{c@{~=~}l}
\frac{1}{K(p/\Lambda)^2} \vev{\phi (p) \phi (-p)}_{S(\Lambda)} +
\frac{1 - 1/K(p/\Lambda)}{p^2 + m^2} & \frac{1}{K(p/\Lambda')^2}
\vev{\phi (p) \phi (-p)}_{S(\Lambda')} + \frac{1 -
  1/K(p/\Lambda')}{p^2 + m^2}\\
\prod_{i=1}^{2n} \frac{1}{K(p_i/\Lambda)} \cdot \vev{\phi (p_1) \cdots
  \phi (p_{2n})}_{S(\Lambda)} & \prod_{i=1}^{2n}
\frac{1}{K(p_i/\Lambda')} \cdot \vev{\phi (p_1) \cdots 
  \phi (p_{2n})}_{S(\Lambda')}
\end{array}\right.
\]
If the action corresponds to a continuum limit, we can take $\Lambda
\to \infty$.  Hence, the correlation functions in the continuum limit
can be obtained from the action $S(\Lambda)$ with a finite cutoff:
\[
\left\lbrace\begin{array}{c@{~=~}l}
\vev{\phi (p) \phi (-p)}_\infty & \frac{1}{K(p/\Lambda)^2}
\vev{\phi (p) \phi (-p)}_{S(\Lambda)} + 
\frac{1 - 1/K(p/\Lambda)}{p^2 + m^2}\\
\vev{\phi (p_1) \cdots \phi (p_{2n})}_\infty & \prod_{i=1}^{2n}
\frac{1}{K(p_i/\Lambda)} \cdot \vev{\phi (p_1) \cdots 
  \phi (p_{2n})}_{S(\Lambda)} 
\end{array}\right.
\]
Here the left-hand sides are the correlation functions calculated with
$S(\Lambda \to \infty)$.  The right-hand sides are independent of
$\Lambda$.  It is surprising that the continuum physics
can be obtained from an action with a finite cutoff, but we must
accept it.  There is a trade off, though.  Our action $S(\Lambda)$ has
an infinite number of interaction terms, and is highly complicated.

The above observation has a surprising consequence.  If we expect the
presence of a symmetry in the continuum limit, whether it is global or
local, it should be present in the cutoff theory, too.  Hence, for
example, we expect to be able to construct gauge theories using a
finite cutoff.  We will discuss QED in some details, and sketch the
construction of YM theories, both using a finite cutoff.

\subsection{ERG for the 1PI $\Gamma [\Phi]$}

This subsection is a digression from the main flow of the lectures.
You can skip this part if you like.  In the literature\footnote{For
  example, see \cite{Morris}.} the ERG differential equation for $\Si$
is sometimes written for its 1PI part $\Gi$.  Roughly speaking, $\Si$
consists of Feynman diagrams with elementary vertices and
high-momentum propagators.  These Feynman graphs are not necessarily
1PI, as we have seen in the case of $\V_4^{(1)}$ in the previous
subsection.  We wish to introduce 1PI vertices $\Gi$ so that $\Si$ is
obtained as the tree diagrams with high-momentum propagators and
vertices $\Gi$.  The purpose of this subsection is to introduce $\Gi$
and ERG differential equation it satisfies.  We will be sketchy since
we will not discuss $\Gi$ in the remainder of the lectures.

Let us consider
\[
\Gamma [\phi] \equiv - \frac{1}{2} \int_p \phi (p) \phi (-p) \frac{p^2
  + m^2}{1 - K(p/\Lambda)} + \Gi [\phi]
\]
We define its Legendre transform by
\[
W[J] \equiv \Gamma [\phi] + \int_p \phi (-p) J(p)
\]
where $\phi$ is determined in terms of $J$ as
\[
\frac{\delta \Gamma [\phi]}{\delta \phi (-p)} + J (p) = 0
\]
This can be rewritten as
\[
\phi (p) = \frac{1 - K(p/\Lambda)}{p^2 + m^2} \left( J(p) +
    \frac{\delta \Gi [\phi]}{\delta \phi (-p)} \right)
\]

In order for $\Gi$ to give the 1PI part of $\Si$, we must demand that
\[
W [J] = \frac{1}{2} \int_p J(p) J(-p) \frac{1-K(p/\Lambda)}{p^2 + m^2}
+ \Si \left[ \frac{1 - K(p/\Lambda)}{p^2 + m^2} J(p) \right]
\]
\begin{figure}[t]
\begin{center}
\epsfig{file=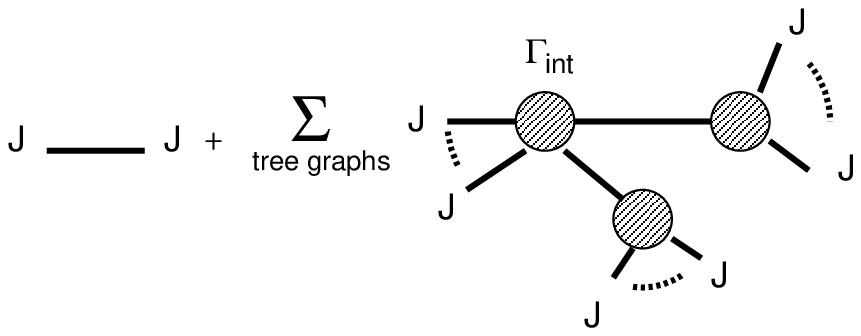}
\caption{$\Gi$ is the 1PI part of $\Si$.}
\end{center}
\end{figure}
(Fig.~10) By substituting
\[
J(p) = \frac{p^2+m^2}{1-K(p/\Lambda)} \phi (p)
\]
we obtain
\[
W \left[ \frac{p^2 + m^2}{1-K(p/\Lambda)} \phi (p) \right] = \frac{1}{2} \int_p
\phi (p) \phi (-p) \frac{p^2 + m^2}{1 - K(p/\Lambda)} + \Si [\phi]
\]
On the other hand, from the definition of the Legendre transform, we obtain
\[
W \left[ \frac{p^2 + m^2}{1-K(p/\Lambda)} \phi (p) \right] = \Gamma [\Phi] +
\int_p \Phi (p) \frac{p^2 + m^2}{1 - K(p/\Lambda)} \phi (-p)
\]
where
\[
\phi (p) = \Phi (p) - \frac{1 - K(p/\Lambda)}{p^2 + m^2} \frac{\delta
  \Gi [\Phi]}{\delta \Phi (-p)}
\]
Hence, we obtain the following relation between $\Gi [\Phi]$ and $\Si
[\phi]$:
\begin{eqnarray*}
&&- \frac{1}{2} \int_p \Phi (p) \Phi (-p) \frac{p^2 + m^2}{1 -
  K(p/\Lambda)} + \Gi [\Phi]\\
&&\qquad = \frac{1}{2} \int_p \phi (p) \phi (-p) \frac{p^2 + m^2}{1 -
  K(p/\Lambda)} + \Si [\phi] - \int_p \phi (p) \frac{p^2 + m^2}{1 -
  K(p/\Lambda)} \Phi (-p)
\end{eqnarray*}
It is a simple algebra to rewrite this as
\[
\Gi [\Phi] = \frac{1}{2} \int_p \left( \phi (p) - \Phi (p) \right)
\left( \phi (-p) - \Phi (-p) \right) \frac{p^2 + m^2}{1 -
  K(p/\Lambda)}
+ \Si [\phi]
\]

Now, we can reverse the direction, and obtain $\Gamma$ by Legendre
transforming $W$:
\[
\Gamma [\Phi] = W [J] - \int_p \Phi (-p) J(p)
\]
where
\[
\Phi (p) = \frac{\delta W[J]}{\delta J(-p)}
\]
By the substitution $J(p) = \frac{p^2+m^2}{1 - K(p/\Lambda)} \, \phi
(p)$, this gives
\[
\Phi (p) = \phi (p) + \frac{1-K(p/\Lambda)}{p^2 + m^2} \frac{\delta
  \Si [\phi]}{\delta \phi (-p)}
\]
Comparing this with the relation between $\phi$ and $\Phi$ obtained
above, we obtain
\[
\frac{\delta \Gi [\Phi]}{\delta \Phi (-p)} = \frac{\delta \Si
  [\phi]}{\delta \phi (-p)}
\]

Now that we have an explicit relation between $\Gi$ and $\Si$, we can
derive the ERG differential equation of $\Gi$ from that of $\Si$.  The
derivation is straightforward, and we only state the result:
\[
- \Lambda \frac{\partial}{\partial \Lambda} \Gi [\Phi]
= \frac{1}{2} \int_q \frac{\Delta (q/\Lambda)}{q^2 + m^2}
\frac{\delta^2 \Si [\phi]}{\delta \phi (-q) \delta \phi (q)}
\]
By differentiating the relation obtained above between the first order
functional derivatives of $\Gi$ and $\Si$, we obtain
\[
\frac{\delta^2 \Si}{\delta \phi (q) \delta \phi (-q)} =
\int_p \frac{\delta \Phi (p)}{\delta \phi (q)} \frac{\delta^2 \Gi
  [\Phi]}{\delta \Phi (p) \delta \Phi (-q)}
\]
where $\frac{\delta \Phi (p)}{\delta \phi (q)}$ is the inverse of
\[
\frac{\delta \phi (q)}{\delta \Phi (p)} = (2 \pi)^4 \delta^{(4)} (p-q)
- \frac{1 - K(q/\Lambda)}{q^2 + m^2} \frac{\delta^2 \Gi [\Phi]}{\delta
  \Phi (p) \delta \Phi (-q)}
\]
Hence, it is given by the geometric series
\begin{eqnarray*}
\frac{\delta \Phi (p)}{\delta \phi (q)}
&=&  (2 \pi)^4 \delta^{(4)} (p-q) + \frac{1 - K(p/\Lambda)}{p^2 + m^2}
\frac{\delta^2 \Gi [\Phi]}{\delta \Phi (-p) \delta \Phi (q)} \\
&&\quad +
\int_r \frac{1 - K(p/\Lambda)}{p^2 + m^2}
\frac{\delta^2 \Gi [\Phi]}{\delta \Phi (-p) \delta \Phi (r)} \frac{1 -
  K(r/\Lambda)}{r^2 + m^2} \frac{\delta^2 \Gi}{\delta \Phi (-r) \delta
  \Phi (q)} + \cdots
\end{eqnarray*}
Therefore, we obtain
\[
\frac{\delta^2 \Si}{\delta \phi (q) \delta \phi (-q)}
= \frac{\delta^2 \Gi}{\delta \Phi (q) \delta \Phi (-q)} +
\int_r \frac{\delta^2 \Gi}{\delta \Phi (q) \delta \Phi (-r)} \frac{1 -
  K(r/\Lambda)}{r^2 + m^2} \frac{\delta^2 \Gi}{\delta \Phi (r) \delta
  \Phi (-q)} + \cdots
\]

Thus, we can express this ERG equation graphically as
\begin{center}
\parbox{1cm}{$\displaystyle - \Lambda \frac{\partial}{\partial
    \Lambda}$}
\parbox{1cm}{\epsfig{file=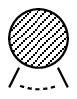}}
\parbox{0.5cm}{$=$}
\parbox{8cm}{\epsfig{file=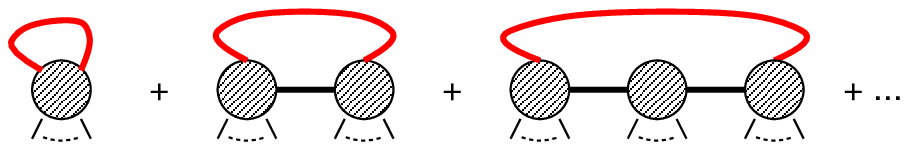}}
\end{center}
This final result is actually easier to obtain diagrammatically.\\
\no \textbf{HW\#4}: From the above diagrammatic ERG equation for
$\Gi$, derive the ERG equation for $\Si$ diagrammatically.

\subsection{Composite operators}

The purpose of this subsection is to introduce composite operators
which play crucial roles in understanding some important properties of
ERG such as its relation to beta functions and anomalous dimensions.
We take $D=4$ for concreteness.

We call
\begin{eqnarray*}
\Op (p) &\equiv& \sum_{n=1}^\infty \frac{1}{(2n)!} \int_{p_1,\cdots,p_{2n}}
\phi (p_1) \cdots \phi (p_{2n}) \Op_{2n} (\Lambda; p_1,\cdots,p_{2n})
\\
&&\qquad \times (2 \pi)^4 \delta^{(4)} (p_1 + \cdots + p_{2n} - p)
\end{eqnarray*}
a \textbf{composite operator} \footnote{Here we assume $\Op$ is even
  in $\phi$.  We can also define $\Op$ odd in $\phi$.  For example,
  $\frac{\delta \Si}{\delta \phi (-p)}$ considered below is odd in
  $\phi$.} if it satisfies the ERG differential equation
\[
- \Lambda \frac{\partial}{\partial \Lambda} \Op (p) = \D \cdot \Op (p)
\]
where $\D$ is the differential operator
\[
\D \equiv \int_q \frac{\Delta (q/\Lambda)}{q^2 + m^2} \left(
    \frac{\delta \Si}{\delta \phi (-q)} \frac{\delta}{\delta \phi
      (-q)} + \frac{1}{2} \frac{\delta^2}{\delta \phi (q) \delta \phi
      (-q)} \right)
\]
We call $\Op$ an operator of dimension $d$, if it satisfies the
asymptotic conditions
\[
\Op_{2n \ge d + 2} (\Lambda; p_1,\cdots,p_{2n}) \stackrel{\Lambda \to
  \infty}{\longrightarrow} 0
\]
Hence, an operator of dimension $2$ has $\Op_{2n \ge 4} \to 0$, and
that of dimension $4$ has $\Op_{2n \ge 6} \to 0$.

It is easy to understand where the above definition of composite
operators comes from.  Let us consider a modified action
\[
S' (\Lambda) \equiv S (\Lambda) + \int_p \ep(-p) \Op (p)
\]
where $\ep$ is an infinitesimal source.  The definition guarantees
that $S'$ satisfies the Polchinski differential equation if $S$
satisfies it.

\no\textbf{HW\#5}: Show that the ERG differential equation for the
composite operator $\Op$ can be rewritten as
\[
- \Lambda \frac{\partial}{\partial \Lambda} \Op 
= \int_q \frac{\Delta (q/\Lambda)}{q^2 + m^2} \left(
\frac{q^2 + m^2}{K(q/\Lambda)} \phi (q) \frac{\delta}{\delta \phi (q)}
+ \frac{\delta S}{\delta \phi (-q)} \frac{\delta}{\delta \phi (q)} +
\frac{1}{2} \frac{\delta^2}{\delta \phi (q) \delta \phi (-q)} \right)
\Op
\]

Let us recall the following $\Lambda$ dependence of the correlation
functions:
\[
\begin{array}{c@{~=~}l}
\vev{\phi (p) \phi (-p)}_{S(\Lambda)} - \frac{K(p/\Lambda)}{p^2 + m^2}
& \frac{K(p/\Lambda)^2}{K(p/\Lambda')^2} \left( \vev{\phi (p) \phi
      (-p)}_{S(\Lambda')} - \frac{K(p/\Lambda')}{p^2 + m^2} \right)\\
\vev{\phi (p_1) \cdots \phi (p_{2n})}_{S(\Lambda)} &
\prod_{i=1}^{2n} \frac{K(p_i/\Lambda)}{K(p_i/\Lambda')} \cdot
\vev{\phi (p_1) \cdots \phi (p_{2n})}_{S(\Lambda')} \quad (n > 1)
\end{array}
\]
Substituting $S'$ into $S$ and taking the part proportional to $\ep
(-p)$, we obtain
\[
\vev{\phi (p_1) \cdots \phi (p_{2n}) \Op (p)}_{S(\Lambda)}
= \prod_{i=1}^{2n} \frac{K(p_i/\Lambda)}{K(p_i/\Lambda')} \cdot
\vev{\phi (p_1) \cdots \phi (p_{2n}) \Op (p)}_{S (\Lambda')}
\]
for any $n \ge 1$.  Equivalently, this gives
\[
\vev{\phi (p_1) \cdots \phi (p_{2n}) \Op (p)}_\infty
= \prod_{i=1}^{2n} \frac{1}{K(p_i/\Lambda)} \cdot
\vev{\phi (p_1) \cdots \phi (p_{2n}) \Op (p)}_{S (\Lambda)}
\]
We can actually reverse the direction, and use either of the above two
relations as the defining property of a composite operator.

The simplest example of a composite operator is $ \Op (p) \equiv
\frac{\delta \Si}{\delta \phi (-p)}$.  To show this, we simply
differentiate the ERG differential equation
\[
- \Lambda \frac{\partial}{\partial \Lambda} \Si = \int_q \frac{\Delta
  (q/\Lambda)}{q^2 + m^2} \frac{1}{2} \left\lbrace \frac{\delta
      \Si}{\delta \phi (q)} \frac{\delta \Si}{\delta \phi (-q)} +
    \frac{\delta^2 \Si}{\delta \phi (q) \delta \phi (-q)}
\right\rbrace
\]
with respect to $\phi (-p)$.  The asymptotic behavior of $\Si$ (in the
mass independent scheme) translates into the following asymptotic
behavior of $\frac{\delta \Si}{\delta \phi (-p)}$:
\[
\begin{array}{c@{~\longrightarrow~}l}
\Op_1 (p) & a_2 (\ln \Lambda/\mu,\lambda) + m^2 b_2  (\ln
\Lambda/\mu,\lambda) + p^2 c_2  (\ln \Lambda/\mu,\lambda)\\
\Op_3 (p_1,p_2,p_3) & a_4  (\ln \Lambda/\mu,\lambda)\\
\Op_{n \ge 5} (p_1,\cdots,p_n) & 0
\end{array}
\]
Hence, $\frac{\delta \Si}{\delta \phi (-p)}$ is dimension $3$.  Since
$\frac{\delta \Si}{\delta \phi (-p)}$ is a composite operator, the
correlation functions
\[
\vev{\frac{\delta \Si}{\delta \phi (-p)} \phi (p_1) \cdots \phi
  (p_{2n-1})}_\infty
= \vev{\frac{\delta \Si}{\delta \phi (-p)} \phi (p_1) \cdots \phi
  (p_{2n-1})}_{S(\Lambda)} \cdot \prod_{i=1}^{2n-1}
\frac{1}{K(p_i/\Lambda)}
\]
are independent of $\Lambda$.

Another example is
\[
\Phi (p) \equiv \phi (p) + \frac{1 - K(p/\Lambda)}{p^2 + m^2}
\frac{\delta \Si}{\delta \phi (-p)}
\]
which corresponds to an elementary field $\phi (p)$.  To show this, we
can again take the $\Lambda$ derivative of $\Phi$:
\[
- \Lambda \frac{\partial \Phi (p)}{\partial \Lambda}
= \frac{\Delta (p/\Lambda)}{p^2+m^2} \frac{\delta \Si}{\delta \phi
  (-p)} + \D \cdot \frac{1 - K(p/\Lambda)}{p^2+m^2} \frac{\delta
  \Si}{\delta \phi (-p)}
\]
Since
\[
\D \cdot \phi (p) = \frac{\Delta (p/\Lambda)}{p^2 + m^2} \frac{\delta
  \Si}{\delta \phi (-p)}
\]
we obtain
\[
- \Lambda \frac{\partial \Phi (p)}{\partial \Lambda}
= \D \cdot \Phi (p)
\]

\no \textbf{HW\#5}: Show that 
\[
K(p/\Lambda) \frac{\delta S}{\delta \phi (-p)} = - (p^2 + m^2) \Phi
(p) + \frac{\delta \Si}{\delta \phi (-p)}
\]
and hence it is a composite operator.  

This is an operator corresponding to the equation of motion.  Since
\begin{eqnarray*}
&&\frac{\delta}{\delta \phi (-p)} \left( \phi (p_1) \cdots \phi (p_n)
    \e^S \right) = \frac{\delta S}{\delta \phi (-p)} \phi (p_1) \cdots
\phi (p_n) \e^S \\
&&\qquad + \sum_{i=1}^n (2 \pi)^4 \delta^{(4)} (p-p_i) \,
\phi (p_1) \cdots \widehat{\phi (p_i)} \cdots \phi (p_n) \e^S
\end{eqnarray*}
is a total derivative, its functional integral vanishes.  Hence,
integrating over $\phi$, we obtain
\[
\vev{\frac{\delta S}{\delta \phi (-p)} \phi (-p)}_{S(\Lambda)} = - 1
\]
and for $n > 1$
\[
\vev{\frac{\delta S}{\delta \phi (-p)} \phi (p_1) \cdots \phi
  (p_n)}_{S(\Lambda)} = 0
\]
if we take the connected part.  Thus, in the continuum limit the
equations of motion are given by
\[
\left\lbrace\begin{array}{c@{~=~}l}
\vev{ K(p/\Lambda) \frac{\delta S}{\delta \phi (-p)} \phi (-p)}_\infty
 & -1\\
\vev{ K(p/\Lambda) \frac{\delta S}{\delta \phi (-p)} \phi (p_1)
 \cdots \phi (p_{2n-1})}_\infty & 0 \qquad (n > 1)
\end{array}\right.
\]

\subsection{Composite operators given in terms of the action}

As for the action itself, the ERG differential equation of a composite
operator does not determine the operator uniquely.  We must supply
appropriate asymptotic conditions.  For later conveniences, we
consider dimension 2 and 4 scalar composite operators.  We adopt the
mass independent scheme for simplicity.

\subsection*{Dimension 2 operator}

There is a unique dimension 2 operator $\Op$ of zero momentum,
specified by the following asymptotic conditions\footnote{This $a_2$
  is not the same as the $a_2$ that gives the asymptotic behavior of
  the vertex $\V_2$.}:
\[
\left\lbrace\begin{array}{c@{~\longrightarrow~}l}
 \Op_2 (\Lambda; p, -p) & a_2 (\ln \Lambda/\mu; \lambda) \\
 \Op_{2n \ge 4} (\Lambda; p_1, \cdots, p_{2n}) & 0
\end{array}\right.
\]
where
\[
a_2 (0; \lambda) = 1
\]
We call this operator $\left[\frac{1}{2} \phi^2\right]$.

\subsection*{Dimension 4 operators}

Dimension 4 operators $\Op$ of zero momentum are specified by the
following asymptotic conditions:
\[
\left\lbrace\begin{array}{c@{~\longrightarrow~}l} \Op_2 (\Lambda;
        p,-p) & \Lambda^2
        a_2 (\ln \Lambda/\mu; \lambda) + m^2 b_2 (\ln \Lambda/\mu;
        \lambda) + p^2 c_2 (\ln \Lambda/\mu; \lambda)\\ 
        \Op_4 (\Lambda; p_1,\cdots,p_4) & a_4 (\ln \Lambda/\mu; \lambda)\\
        \Op_{2n \ge 6} (\Lambda; p_1,\cdots,p_{2n}) & 0
\end{array}\right.
\]
There are two linearly independent operators:
\begin{enumerate}
\item $\left[ \frac{1}{4!} \phi^4 \right]$, satisfying
\[
a_4 (0;\lambda) = 1,\quad b_2 (0; \lambda) = c_2 (0;\lambda) = 0
\]
\item $\left[ \frac{1}{2} (\partial_\mu \phi)^2 \right]$, satisfying
\[
c_2 (0; \lambda) = 1,\quad b_2 (0;\lambda) = a_4 (0; \lambda) = 0
\]
\end{enumerate}
We can regard $m^2 \left[ \frac{1}{2} \phi^2 \right]$ as the third
dimension 4 operator satisfying
\[
b_2 (0; \lambda) = 1,\quad c_2 (0;\lambda) = a_4 (0; \lambda) = 0
\]

\subsection*{Composite operators given in terms of $S$}

The purpose of the remaining part of this subsection is to show that
the three dimension 4 scalar operators
\[
m^2 \left[ \frac{1}{2} \phi^2 \right],\quad
\left[ \frac{1}{2} \left(\partial_\mu \phi\right)^2\right],\quad
\left[ \frac{1}{4!} \phi^4 \right]
\]
are obtained in terms of the action $S$.

As a preparation, we consider the following:
\[
\left\lbrace\begin{array}{c@{~\equiv~}l}
 \Op_1 [f] & \int_p f(p) \phi (p) \frac{\delta S}{\delta \phi
   (p)}\\
 \Op_2 [C] & \int_p C(p^2) \frac{1}{2} \left\lbrace \frac{\delta
       S}{\delta \phi (p)} \frac{\delta S}{\delta \phi (-p)} +
     \frac{\delta^2 S}{\delta \phi (p) \delta \phi (-p)} \right\rbrace
\end{array}\right.
\]
Neither of them is a composite operator by itself, but their
correlation functions can be easily computed.

\begin{enumerate}
\item $\Op_1 [f]$ --- Except for an inessential additive
    constant\footnote{The additive constant does not contribute to the
      connected part.}, $\Op_1 [f]$ is a total derivative
\[
\Op_1 [f] \e^S = \int_p f(p) \frac{\delta}{\delta \phi (p)} \left( \phi (p)
    \e^S \right)
\]
Hence, multiplying this by $\phi (p_1) \cdots \phi (p_{2n})$ and
integrating over $\phi$ by parts, we obtain
\[
\vev{\Op_1 [f] \phi (p_1) \cdots \phi (p_{2n})}_{S} = -
\left(\sum_{i=1}^{2n} f(p_i) \right) \cdot \vev{\phi (p_1) \cdots \phi
  (p_{2n})}_{S} 
\]
\item $\Op_2 [C]$ --- This is also a total derivative:
\[
\Op_2 [C] \e^S = \int_p C (p^2) \frac{1}{2} \frac{\delta^2}{\delta
  \phi (p) \delta \phi (-p)} \,\e^S 
\]
Again, integrating over $\phi$ by parts, we obtain
\[
\left\lbrace\begin{array}{c@{~=~}l}
 \vev{\Op_2 [C] \phi (p) \phi (-p)}_S & C (p^2)\\
 \vev{\Op_2 [C] \phi (p_1) \cdots \phi (p_{2n})}_S & 0\quad (n > 1)
\end{array}\right.
\]
\end{enumerate}

It is interesting to note that the right-hand side of the Polchinski
differential ERG equation is written as
\begin{eqnarray*}
&&\int_q \frac{\Delta (q/\Lambda)}{q^2 + m^2} 
\left[ \frac{q^2 + m^2}{K(q/\Lambda)} \phi (q) \frac{\delta S}{\delta
      \phi (q)} + \frac{1}{2} \left\lbrace 
\frac{\delta S}{\delta \phi (q)} \frac{\delta S}{\delta \phi (-q)} +
\frac{\delta^2 S}{\delta \phi (q) \delta \phi (-q)} \right\rbrace
\right]\\
&&= \Op_1 [f] + \Op_2 [C]
\end{eqnarray*}
where
\[
\left\lbrace\begin{array}{c@{~=~}l}
 f(q) & \frac{\Delta (q/\Lambda)}{K(q/\Lambda)}\\
 C(q^2) & \frac{\Delta (q/\Lambda)}{q^2 + m^2}
\end{array}\right.
\]
Hence, we reproduce
\[
\left\lbrace
\begin{array}{c@{~=~}l}
- \Lambda \frac{\partial}{\partial \Lambda} \vev{\phi (p) \phi (-p)}_S
& \frac{\Delta (p/\Lambda)}{p^2 + m^2} - 2 \frac{\Delta
  (p/\Lambda)}{K(p/\Lambda)} \vev{\phi (p) \phi (-p)}_S\\
- \Lambda \frac{\partial}{\partial \Lambda} \vev{\phi (p_1) \cdots
  \phi (p_{2n})}_S & - \sum_{i=1}^{2n} \frac{\Delta
  (p_i/\Lambda)}{K(p_i/\Lambda)} \cdot \vev{\phi (p_1) \cdots \phi
  (p_{2n})}_S
\end{array}\right.
\]

With the above preparation, we can construct three special operators
$\Op_m, \Op_\lambda$, and $\N$.

\begin{enumerate}
\item $\Op_m$ that generates the change of correlation functions with
    respect to $m^2$ --- by differentiating the correlation functions
    with respect to $m^2$, we obtain
\[
- \partial_{m^2} \vev{\phi (p_1) \cdots \phi (p_{2n})}_S
= \vev{\left( - \partial_{m^2} S\right) \phi (p_1) \cdots \phi
  (p_{2n})}_S\quad (n \ge 1)
\]
Hence, for $n > 1$, we obtain
\[
- \partial_{m^2} \vev{\phi (p_1) \cdots \phi (p_{2n})}_\infty
= \vev{\left( - \partial_{m^2} S\right) \phi (p_1) \cdots \phi
  (p_{2n})}_S \prod_{i=1}^{2n} \frac{1}{K(p_i/\Lambda)}
\]
The case $n=1$ needs extra care.  Since 
\[
\vev{\phi (p) \phi (-p)}_\infty = \frac{1 - 1/K(p/\Lambda)}{p^2 + m^2}
+ \frac{1}{K(p/\Lambda)^2} \vev{\phi (p) \phi (-p)}_S
\]
we obtain
\begin{eqnarray*}
&& - \partial_{m^2} \vev{\phi (p) \phi (-p)}_\infty
= \frac{1 - 1/K(p/\Lambda)}{(p^2 + m^2)^2} +
\frac{1}{K(p/\Lambda)^2} \vev{\left( - \partial_{m^2} S\right) \phi
  (p) \phi (-p)}_S \\
&& = \left\lbrace - \frac{K(p/\Lambda) \left( 1 - K(p/\Lambda) \right)}{(p^2 +
      m^2)^2} + \vev{\left( - \partial_{m^2} S\right) \phi
  (p) \phi (-p)}_S \right\rbrace \frac{1}{K(p/\Lambda)^2}
\end{eqnarray*}
Thus, we obtain
\[
- \partial_{m^2} \vev{\phi (p_1) \cdots \phi (p_{2n})}_\infty
= \vev{\Op_m \phi (p_1) \cdots \phi (p_{2n})}_\infty\quad (n \ge 1)
\]
for the composite operator
\[
\ffbox{\Op_m \equiv \Op_2 [C] - \partial_{m^2} S}
\]
where
\[
C(p^2) \equiv - \frac{K(p/\Lambda) \left( 1 - K(p/\Lambda) \right)}{(p^2 +
  m^2)^2}
\]
\item $\Op_\lambda$ that generates the change with respect to
    $\lambda$ --- by differentiating the correlation functions with
    respect to $\lambda$, we obtain
\[
- \partial_\lambda \vev{\phi (p_1) \cdots \phi (p_{2n})}_S
= \vev{\left( - \partial_\lambda S \right) \phi (p_1) \cdots \phi
  (p_{2n})}_S\quad (n \ge 1)
\]
Hence,
\[
\ffbox{\Op_\lambda \equiv - \partial_\lambda S}
\]
is a composite operator.
\item $\N$ that generates the change of normalization of $\phi$ --- we
    wish to construct a composite operator $\N$ that counts the number
    of fields:
\[
\vev{\N \phi (p_1) \cdots \phi (p_{2n})}_\infty = 2n \vev{\phi (p_1)
  \cdots \phi (p_{2n})}_\infty\quad (n \ge 1)
\]
The starting point is $\Op_1 [f]$, where $f = -1$, which satisfies
\[
\vev{\Op_1 [-1] \phi (p_1) \cdots \phi (p_{2n})}_S = 2n \vev{\phi
  (p_1) \cdots \phi (p_{2n})}_S\quad (n \ge 1)
\]
Again, the case $n=1$ needs extra care.  Since
\[
2 \vev{\phi (p) \phi (-p)}_\infty = 2 \left(
- \frac{K(p/\Lambda) \left( 1 - K(p/\Lambda) \right)}{p^2 + m^2} +
\vev{\phi (p) \phi (-p)}_S \right) \frac{1}{K(p/\Lambda)^2}
\]
we obtain
\[
\ffbox{\N = \Op_2 [C] + \Op_1 [-1]}
\]
where
\[
C(p^2) = - 2 \frac{K(p/\Lambda) \left(1 - K(p/\Lambda) \right)}{p^2 +
  m^2}
\]
\end{enumerate}

To summarize, we have constructed three dimension 4 composite
operators $\Op_m, \Op_\lambda, \N$ with the following properties:
\[
\ffbox{
\begin{array}{c@{~=~}l}
\vev{\Op_m \phi (p_1) \cdots \phi (p_{2n})}_\infty & - \partial_{m^2}
\vev{\phi (p_1) \cdots \phi (p_{2n})}_\infty\\
\vev{\Op_\lambda \phi (p_1) \cdots \phi (p_{2n})}_\infty & - \partial_{\lambda}
\vev{\phi (p_1) \cdots \phi (p_{2n})}_\infty\\
\vev{\N \phi (p_1) \cdots \phi (p_{2n})}_\infty & 2n
\vev{\phi (p_1) \cdots \phi (p_{2n})}_\infty
\end{array}
}
\]
where these operators are given in terms of the action as
\[
\left\lbrace\begin{array}{c@{~\equiv~}l}
\Op_m & - \partial_{m^2} S - \int_q \frac{K(q/\Lambda)
  (1-K(q/\Lambda))}{(q^2 + m^2)^2} \frac{1}{2} \left\lbrace
\frac{\delta S}{\delta \phi (q)} \frac{\delta S}{\delta \phi (-q)}
+ \frac{\delta^2 S}{\delta \phi (q) \delta \phi (-q)}
\right\rbrace \\
\Op_\lambda & - \partial_\lambda S\\
\N & - \int_q \phi (q) \frac{\delta S}{\delta \phi (q)}
- \int_q \frac{K(q/\Lambda)(1 - K(q/\Lambda))}{q^2 + m^2} \left\lbrace 
\frac{\delta S}{\delta \phi (q)} \frac{\delta S}{\delta \phi (-q)}
+ \frac{\delta^2 S}{\delta \phi (q) \delta \phi (-q)} \right\rbrace
\end{array}\right.
\]

These operators are obtained as linear combinations of
$\left[\frac{\phi^2}{2}\right]$, $\left[\frac{\phi^4}{4!}\right]$, and
$\left[ \frac{1}{2} (\partial_\mu \phi)^2 \right]$.  To see this, we
must examine the asymptotic behaviors as $\Lambda \to \infty$.  We
first introduce the following notation\footnote{These are expansions
  in powers of $\frac{1}{\Lambda^2}$}:
\[
\begin{array}{c@{~\longrightarrow~}l}
 \V_4 (\Lambda; q,-q,0,0) & a_4 (\ln \Lambda/\mu; q/\Lambda) +
 \frac{m^2}{\Lambda^2} b_4 (\ln \Lambda/\mu; q/\Lambda) + \cdots\\
 \V_6 (\Lambda; q,-q, 0,\cdots,0) & \frac{1}{\Lambda^2} a_6 (\ln
 \Lambda/\mu; q/\Lambda) + \cdots
\end{array}
\]
Then, using the mass independent scheme, it is straightforward to
derive the following results\footnote{It is ``straightforward'' but
  some work is required.}:
\[
\ffbox{
\begin{array}{c@{~=~}l}
\Op_m & x_m \left[\frac{1}{2} \phi^2\right]\\
\Op_\lambda & \left[ \frac{1}{4!} \phi^4 \right]\\
\N & x_{\N} m^2 \left[\frac{1}{2} \phi^2\right] + y_{\N} \left[
    \frac{1}{4!} \phi^4 \right] + z_{\N} \left[ \frac{1}{2}
    (\partial_\mu \phi)^2 \right]
\end{array}
}
\]
where 
\[
\left\lbrace
\begin{array}{c@{~=~}l}
x_m & 1 - \frac{1}{2} \int_q \frac{K(q)(1-K(q))}{q^4} a_4 (0;q)\\
x_{\N} & 2 + \int_q K(q) (1-K(q)) \left( - \frac{b_4 (0;q)}{q^2} +
    \frac{a_4 (0;q)}{q^4} \right)\\
y_{\N} & - 4 \lambda - \int_q \frac{K(q)(1-K(q))}{q^2} a_6 (0;q)\\
z_{\N} & 2 - \int_q \frac{K(q)(1-K(q))}{q^2} c_4 (0;q)
\end{array}\right.
\]
where $c_4$ is defined by
\[
\frac{1}{\Lambda^2} c_4 (\ln \lambda/\mu; q/\Lambda)
\equiv \frac{\partial}{\partial p^2}
\V_4 (\Lambda; p,-p,q,-q) \Big|_{m^2=p^2=0}
\]
where the angular average over $q_\mu$ is taken.

To conclude this subsection, we invert the relation of $\Op_m,
\Op_\lambda, \N$ to the mass independent composite operators to obtain
\[
\ffbox{
\begin{array}{c@{~=~}l}
 \left[\frac{1}{2} \phi^2\right] & \frac{1}{x_m} \Op_m\\
 \left[\frac{1}{4!} \phi^4\right] & \Op_\lambda\\
 \left[\frac{1}{2} (\partial_\mu \phi)^2 \right] & \frac{1}{z_{\N}}
 \left( \N - \frac{x_{\N}}{x_m} \Op_m - y_{\N} \Op_\lambda \right)
\end{array}
}
\]

\subsection{Beta functions and anomalous dimensions}

To be concrete, we adopt the mass independent scheme to define the
action $S(\Lambda; m^2, \lambda; \mu)$.  The action depends on three
parameters: $m^2, \lambda$, and $\mu$.  Here, $\mu$ is an arbitrary
momentum scale that we have used to specify a unique solution to the
ERG differential equation.  We can physically interpret $m^2$ and
$\lambda$ as the mass and interaction parameters of the theory defined
at the scale $\mu$.

Let us suppose we change the scale $\mu$.  In order to keep the
physics (or equivalently correlation functions) intact, we must change
$m^2$ and $\lambda$ accordingly.  This change is the conventional RG
transformation of the running parameters $m^2$ and $\lambda$.

We first observe that
\[
\Psi \equiv - \mu \frac{\partial}{\partial \mu} S (\Lambda; m^2,
\lambda; \mu)
\]
is a composite operator; for $\mu \ne \mu'$ both $S(\Lambda; m^2,
\lambda; \mu)$ and $S (\Lambda; m^2, \lambda; \mu')$ satisfy the same
ERG differential equation.  From its asymptotic behaviors, we find
$\Psi$ a dimension 4 operator, and hence it is given as a linear
combination of $m^2 \Op_m$, $\Op_\lambda$, and $\N$.  To determine the
coefficients, we must examine the asymptotic behaviors of $\Psi$ in
details.

We first recall the asymptotic behavior of $\Si (\Lambda)$:
\[
\left\lbrace\begin{array}{c@{~\longrightarrow~}l}
 \V_2 (\Lambda; p,-p) & \Lambda^2 a_2 (\ln \Lambda/\mu, \lambda) + m^2
 b_2 (\ln \Lambda/\mu, \lambda) + p^2 c_2 (\ln \Lambda/\mu, \lambda)\\
 \V_4 (\Lambda; p_1,\cdots,p_4) & a_4 (\ln \Lambda/\mu, \lambda)
\end{array}
\right.
\]
Hence, the asymptotic behavior of $\Psi (\Lambda)$ is given by
\[
\left\lbrace\begin{array}{c@{~\longrightarrow~}l}
\Psi_2 (\Lambda; p,-p) & \Lambda^2 \dot{a}_2 (\ln \Lambda/\mu,
\lambda) + m^2 \dot{b}_2 (\ln \Lambda/\mu, \lambda) + p^2 \dot{c}_2
(\ln \Lambda/\mu, \lambda)\\
\Psi_4 (\Lambda; p_1,\cdots,p_4) & \dot{a}_4 (\ln \Lambda/\mu, \lambda)
\end{array}\right.
\]
where we have defined
\[
\left\lbrace\begin{array}{c@{~\equiv~}c@{~=~}l}
 \dot{a}_2 (\ln \Lambda/\mu, \lambda) & - \mu \frac{\partial}{\partial
   \mu} a_2 (\ln \Lambda/\mu, \lambda) & \Lambda
 \frac{\partial}{\partial \Lambda}  a_2 (\ln \Lambda/\mu, \lambda)\\
 \dot{b}_2 (\ln \Lambda/\mu, \lambda) & - \mu \frac{\partial}{\partial
   \mu} b_2 (\ln \Lambda/\mu, \lambda) & \Lambda
 \frac{\partial}{\partial \Lambda}  b_2 (\ln \Lambda/\mu, \lambda)\\
 \dot{c}_2 (\ln \Lambda/\mu, \lambda) & - \mu \frac{\partial}{\partial
   \mu} c_2 (\ln \Lambda/\mu, \lambda) & \Lambda
 \frac{\partial}{\partial \Lambda}  c_2 (\ln \Lambda/\mu, \lambda)\\
 \dot{a}_4 (\ln \Lambda/\mu, \lambda) & - \mu \frac{\partial}{\partial
   \mu} a_4 (\ln \Lambda/\mu, \lambda) & \Lambda
 \frac{\partial}{\partial \Lambda}  a_4 (\ln \Lambda/\mu, \lambda)
\end{array}\right.
\]
We denote the values of these at $\Lambda = \mu$ using the same
notation as follows:
\[
\dot{a}_2 (\lambda) \equiv \dot{a}_2 (0, \lambda),\, 
\dot{b}_2 (\lambda) \equiv \dot{b}_2 (0, \lambda), \,\cdots
\]
Then, comparing the asymptotic behaviors of $\Psi$ with those of 
$\phitwo$, $\phifour$, and $\dphitwo$, we obtain
\[
\Psi = \dot{b}_2 m^2 \phitwo + \dot{c}_2 \dphitwo + \dot{a}_4 \phifour
\]
We can rewrite this using $\Op_m, \Op_\lambda$, and $\N$ as
\begin{eqnarray*}
\Psi &=& \frac{\dot{b}_2}{x_m} m^2 \Op_m  + \dot{a}_4 \Op_\lambda
+ \frac{\dot{c}_2}{z_{\N}}  \left( \N - \frac{x_{\N}}{x_m} m^2 \Op_m -
    y_{\N} \Op_\lambda \right)\\
&=& \frac{1}{x_m} \left( \dot{b}_2 - \frac{x_{\N}}{z_{\N}} \dot{c}_2
\right) m^2 \Op_m + \left(\dot{a}_4 - \frac{y_{\N}}{z_{\N}} \dot{c}_2
\right) \Op_\lambda + \frac{\dot{c}_2}{z_{\N}} \N 
\end{eqnarray*}
Finally, by defining the following functions of $\lambda$
\[
\left\lbrace\begin{array}{c@{~\equiv~}l}
\beta_m (\lambda) & \frac{1}{x_m} \left( \dot{b}_2 -
    \frac{x_{\N}}{z_{\N}} \dot{c}_2 \right)\\ 
\beta (\lambda) & \dot{a}_4 - \frac{y_{\N}}{z_{\N}} \dot{c}_2\\
\gamma (\lambda) & \frac{\dot{c}_2}{z_{\N}}
\end{array}\right.
\]
we obtain
\[
\ffbox{\Psi = \beta_m m^2 \Op_m + \beta \Op_\lambda + \gamma \N}
\]
This implies
\begin{eqnarray*}
- \mu \frac{\partial}{\partial \mu} \vev{\phi (p_1) \cdots \phi
  (p_{2n})}_\infty
&=& \vev{\Psi \phi (p_1) \cdots \phi (p_{2n})}_\infty\\
&=& \left(- \beta_m m^2 \partial_{m^2} - \beta \partial_\lambda + 2 n
    \gamma \right) \vev{\phi (p_1) \cdots \phi (p_{2n})}_\infty
\end{eqnarray*}
Hence, we obtain the standard RG equation\footnote{Our $\beta,
  \beta_m, \gamma$ have the opposite sign convention than the one
  usually used.}
\[
\ffbox{
\left( - \mu \frac{\partial}{\partial \mu}
+ \beta_m  m^2 \partial_{m^2} + \beta \partial_\lambda - 2 n
    \gamma \right) \vev{\phi (p_1) \cdots \phi (p_{2n})}_\infty = 0
}
\]
This result was obtained in \cite{hs06}.

\subsection*{One-loop calculations}

At one-loop, we have
\[
\dot{b}_2 = - \frac{\lambda}{(4 \pi)^2},\quad
\dot{c}_2 = 0, \quad
\dot{a}_4 = \frac{3 \lambda^2}{(4 \pi)^2}
\]
and to the lowest order 
\[
x_m = 1,\quad x_{\N} = 2,\quad y_{\N} = - 4 \lambda,\quad z_{\N} = 2
\]
Hence, we obtain the familiar results:
\[
\beta_m = - \frac{\lambda}{(4 \pi)^2},\quad
\beta = \frac{3 \lambda^2}{(4 \pi)^2},\quad
\gamma = 0
\]
\no \textbf{HW\#6}: Check this result.

\subsection{Universality}

\textbf{Universality} is an important concept in renormalization
theory.  To construct a continuum limit, there are always more than
one way.  The independence of the continuum limit on the particular
method of construction is called universality.  For example, if we use
a lattice to construct a continuum theory, the limit should not depend
on what kind of lattice, whether square or cubic, we use.

In the following we examine universality in two restricted contexts.
First, we wish to show that the continuum limit (i.e., the correlation
functions with the suffix $\infty$) does not depend on the asymptotic
conditions we use to select a particular solution of the ERG
differential equation.  Second, we wish to show that the continuum
limit does not depend on the choice of a cutoff function $K$.

\subsection*{Scheme dependence}

Given the ERG differential equation
\[
- \Lambda \frac{\partial}{\partial \Lambda} S (\Lambda) = \int_q
\frac{\Delta (q/\Lambda)}{q^2 + m^2} \frac{1}{2} \left\lbrace
    \frac{\delta S}{\delta \phi (q)} \frac{\delta S}{\delta \phi (-q)}
    + \frac{\delta^2 S}{\delta \phi (q) \delta \phi (-q)}
\right\rbrace
\]
the solution which corresponds to a renormalized theory has the
following asymptotic behaviors for large $\Lambda$:
\[
\left\lbrace\begin{array}{ccl}
\V_2 (\Lambda; p,-p) & \longrightarrow&
\Lambda^2 a_2 (m^2/\mu^2, \ln \Lambda/\mu,
\lambda)\\
& & + m^2 b_2 (m^2/\mu^2, \ln \Lambda/\mu, \lambda) + p^2 c_2
(m^2/\mu^2, \ln \Lambda/\mu, \lambda)\\
\V_4 (\Lambda; p_1,\cdots,p_4) & \longrightarrow &
 a_4 (m^2/\mu^2, \ln \Lambda/\mu, \lambda)\\
\V_{2n\ge 6} (\Lambda; p_1,\cdots,p_{2n}) & \longrightarrow & 0
\end{array}\right.
\]
The $\Lambda$ independent part of the coefficients $b_2, c_2, a_4$
are not determined by the differential equation, and it must be fixed
by convention.  We have particularly favored the mass independent
scheme in which all the coefficient functions $a_2, b_2, c_2, a_4$ are
free of $\frac{m^2}{\mu^2}$, and 
\[
b_2 = c_2 = 0,\quad a_4 = - \lambda
\]
at $\Lambda = \mu$.

Now, what happens if we change the convention for $b_2, c_2, a_4$?
That would modify the action $S$ into $S + \delta S$ which satisfies
the same ERG differential equation.  Now, an infinitesimal change of
$S$ is given by a composite operator
\[
\delta m^2 \Op_m + \frac{\delta z}{2} \N + \delta \lambda \Op_\lambda
\]
Hence, any infinitesimal change of the convention can be compensated
by infinitesimal changes in $m^2, \lambda$, and the normalization of
the field $\phi$.  Thus, we have nothing to lose by adopting a
particular convention such as the mass independent scheme.

\subsection*{Dependence on the choice of $K$}

In our discussions so far, we have always kept a choice of the cutoff
function $K(p)$.  Does physics depend on the choice of $K$?

Let $S_{K+\delta K} (\Lambda; m^2, \lambda; \mu)$ be the solution of
the ERG equation in the mass independent scheme using the cutoff
function
\[
K + \delta K
\]
infinitesimally different from $K$.

\no \textbf{HW\#7}: Show that
\[
S' (\Lambda)
\equiv S_{K+\delta K} (\Lambda; m^2, \lambda; \mu) + \Op_1 [f] + \Op_2 [C]
\]
where
\[
\left\lbrace\begin{array}{c@{~\equiv~}l}
f (q) & \frac{\delta K(q/\Lambda)}{K(q/\Lambda)} \\
C (q^2) & \frac{\delta K(q/\Lambda)}{q^2 + m^2}
\end{array}\right.
\]
satisfies the same ERG equation as $S_K (\Lambda; m^2, \lambda;
\mu)$.  Alternatively, show that
\[
\left\lbrace
\begin{array}{c}
 \vev{\phi (p) \phi (-p)}_{S' (\Lambda)}
 \frac{1}{\K{p}^2} + \frac{1 - 1/\K{p}}{p^2 + m^2}\\
 \vev{\phi (p_1) \cdots \phi (p_{2n})}_{S' (\Lambda)} 
 \prod_{i=1}^{2n} \frac{1}{\K{p_i}}
\end{array}\right.
\]
are $\Lambda$-independent.

This implies that the difference between $S_{K+\delta K} + \Op_1 +
\Op_2$ and $S_K$ is a dimension 4 composite operator, and we should be
able to write
\[
S_{K+\delta K} (\Lambda; m^2, \lambda; \mu) + \Op_1 [f] + \Op_2 [C]
= S_K (\Lambda; m^2, \lambda; \mu) + \delta \lambda \Op_\lambda +
\delta m^2 \Op_m + \frac{\delta z}{2} \N
\]
in terms of some infinitesimal constants $\delta \lambda, \delta m^2$,
and $\delta z$.

Let us recall the definitions of $\Op_m, \Op_\lambda$, and $\N$:
\[
\left\lbrace\begin{array}{c@{~\equiv~}l}
\Op_m & - \partial_{m^2} S - \int_q \frac{K(q/\Lambda)
  (1-K(q/\Lambda))}{(q^2 + m^2)^2} \frac{1}{2} \left\lbrace
\frac{\delta S}{\delta \phi (q)} \frac{\delta S}{\delta \phi (-q)}
+ \frac{\delta^2 S}{\delta \phi (q) \delta \phi (-q)}
\right\rbrace \\
\Op_\lambda & - \partial_\lambda S\\
\N & - \int_q \phi (q) \frac{\delta S}{\delta \phi (q)}
- \int_q \frac{K(q/\Lambda)(1 - K(q/\Lambda))}{q^2 + m^2} \left\lbrace 
\frac{\delta S}{\delta \phi (q)} \frac{\delta S}{\delta \phi (-q)}
+ \frac{\delta^2 S}{\delta \phi (q) \delta \phi (-q)} \right\rbrace
\end{array}\right.
\]
Substituting this into the above, we obtain
\[
S_{K+\delta K} (\Lambda; m^2 + \delta m^2, \lambda + \delta \lambda;
\mu) = S_K (\Lambda; m^2, \lambda; \mu) - \Op_1 [g] - \Op_2 [D]
\]
where
\[
\left\lbrace\begin{array}{c@{~\equiv~}l}
g (q) &  \frac{\delta
  K(q/\Lambda)}{K(q/\Lambda)} + \frac{1}{2} \delta z\\
D (q^2) & \frac{\delta K (q/\Lambda)}{q^2 + m^2} + \frac{K(q/\Lambda)
  (1-K(q/\Lambda)}{q^2 + m^2} \left( \delta z + \frac{\delta m^2}{q^2
      + m^2} \right)
\end{array}\right.
\]
Recalling
\[
\left\lbrace
\begin{array}{c@{~=~}l}
\vev{\Op_1 [g] \phi (p_1) \cdots \phi (p_{2n})}_S & - \sum_{i=1}^{2n}
g(p_i) \vev{\phi (p_1) \cdots \phi (p_{2n})}_S\\
\vev{\Op_2 [D] \phi (p) \phi (-p)}_S & D (p^2)\\
\vev{\Op_2 [D] \phi (p_1) \cdots \phi (p_{2n})}_S & 0\quad (n \ge 2)
\end{array}\right.
\]
we obtain
\begin{eqnarray*}
&&\vev{\phi (p) \phi (-p)}_{S_{K+\delta K} (\Lambda; m^2 + \delta m^2,
  \lambda + \delta \lambda; \mu)} \\
&&\qquad = \vev{\phi (p) \phi (-p)}_{S_K
  (\Lambda; m^2, \lambda; \mu)} \cdot \left( 1 + \delta z + \frac{2
      \delta K(p/\Lambda)}{K(p/\Lambda)} \right)\\
&&\qquad\qquad - \frac{\delta K(p/\Lambda)}{p^2 + m^2} -
\frac{K(p/\Lambda)(1 - K(p/\Lambda))}{p^2 + m^2} \left( \delta z +
    \frac{\delta m^2}{p^2 + m^2} \right)\\
&&\vev{\phi (p_1) \cdots \phi (p_{2n})}_{S_{K+\delta K} (\Lambda; m^2 +
  \delta m^2, \lambda + \delta \lambda; \mu)} \\
&& \qquad = 
\vev{\phi (p_1) \cdots \phi (p_{2n})}_{S_{K} (\Lambda; m^2, \lambda;
  \mu)}
\cdot \left( 1 + n \delta z + \sum_{i=1}^{2n} \frac{\delta
      K(p_i/\Lambda)}{K(p_i/\Lambda)} \right)\quad (n > 1)
\end{eqnarray*}
The extra terms for the two-point function is just what we need to get
the following results for the $\Lambda$ independent correlation
functions:
\[
\vev{\phi (p_1) \cdots \phi (p_{2n})}^{K+\delta K}_{m^2 + \delta m^2,
  \lambda + \delta \lambda; \mu}
= ( 1 + n \delta z ) \vev{\phi (p_1) \cdots \phi (p_{2n})}^K_{m^2,
  \lambda; \mu}\quad (n \ge 1)
\]
where we have written $m^2, \lambda; \mu$ instead of $\infty$ to
denote the $\Lambda$ independent correlation functions.  Thus, an
infinitesimal change of the cutoff function $K$ can be compensated by
infinitesimal changes of $m^2, \lambda$, and the normalization of
$\phi$.

\no \textbf{HW\#8}: Derive the final results.  (straightforward but
as complicated as HW\#7)

\newpage
\section{Application to QED}

As we have discussed in subsect.~2.7, the correlation functions in the
continuum limit $\Lambda \to \infty$ can be computed using the action
$S(\Lambda)$ with a finite $\Lambda$.  Hence, if the continuum limit
has any symmetry, it must be there also in $S(\Lambda)$.  As a
concrete example, we study QED, the theory of photons and electrons,
which has the U(1) gauge symmetry.

\subsection{Perturbative construction}

To construct QED, we need a real vector field $A_\mu$ for the photon
and spinor fields $\psi$ and $\bar{\psi}$ for the electron/positron.
The free part of the action is given by
\[
\Sf = - \frac{1}{2} \int_k A_\mu (k) A_\nu (-k) \frac{k^2
  \delta_{\mu\nu} - \left(1 - \frac{1}{\xi}\right) k_\mu
  k_\nu}{K(k/\Lambda)} - \int_p \bar{\psi} (-p) \frac{\fmslash{p} + i
  m}{K(p/\Lambda)} \psi (p)
\]
so that the free propagators are given by
\[
\left\lbrace\begin{array}{c@{~=~}l}
\vev{A_\mu (k) A_\nu (-k)}_{\Sf} & \frac{K(k/\Lambda)}{k^2} \left(
    \delta_{\mu\nu} - (1-\xi) \frac{k_\mu k_\nu}{k^2} \right)\\
\vev{\psi (p) \bar{\psi} (-p)}_{\Sf} & \frac{K(p/\Lambda)}{\fmslash{p}
  + i m} = \frac{K(p/\Lambda)}{p^2 + m^2} \left( \fmslash{p} - i m \right)
\end{array}\right.
\]
$\xi$ is the gauge fixing parameter, and $m$ is the electron mass.

The ERG differential equation for $\Si$ is obtained by generalizing
that for the scalar theory as follows:
\begin{eqnarray*}
&&- \Lambda \frac{\partial}{\partial \Lambda} \Si (\Lambda)\\
&=& \int_k \frac{\Delta (k/\Lambda)}{k^2} \left( \delta_{\mu\nu} -
    (1-\xi) \frac{k_\mu k_\nu}{k^2} \right) \frac{1}{2} \left\lbrace
    \frac{\delta \Si}{\delta A_\mu (k)} \frac{\delta \Si}{\delta A_\nu
      (-k)} + \frac{\delta^2 \Si}{\delta A_\mu (k) \delta A_\nu (-k)}
\right\rbrace\\
&&\quad - \int_q \Tr \frac{\Delta (q/\Lambda)}{\fmslash{q} + i m}
\left\lbrace \ld{q} \Si \cdot \Si \rd{q}
 + \ld{q} \Si \rd{q} \right\rbrace
\end{eqnarray*}
where the minus sign for the second integral is due to the Fermi
statistics, and the trace is for the spinor indices.

To solve the ERG differential equation, we must supply asymptotic
conditions for renormalizability.  For example, we can take
\[
\left\lbrace
\begin{array}{ccl}
\frac{\delta \Si}{\delta A_\mu (k) \delta A_\nu (-k)}\Big|_0
&\rightarrow & \delta_{\mu\nu} \left\lbrace \Lambda^2 a_2 (\ln
    \Lambda/\mu) + m^2 b_2 
    (\ln \Lambda/\mu) + k^2 c_2 (\ln \Lambda/\mu) \right\rbrace \\
&& + k_\mu k_\nu d_2 (\ln \Lambda/\mu)\\
\ld{p} \Si \rd{p}\Big|_0 &\rightarrow& \fmslash{p}\, a_f (\ln
\Lambda/\mu) + i m \,b_f 
(\ln \Lambda/\mu)\\
\ld{(p+k)}\frac{\delta \Si}{\delta A_\mu (k)} \rd{p}\Big|_0
&\rightarrow
& a_3 (\ln
\Lambda/\mu) \gamma_\mu\\
\frac{\delta^4 \Si}{\delta A_\mu (k_1) \cdots \delta A_{\delta} (k_4)}
\Big|_0 &\rightarrow& a_4 (\ln \Lambda/\mu) \left( \delta_{\alpha\beta}
    \delta_{\gamma\delta} + \delta_{\alpha\gamma} \delta_{\beta\delta}
    + \delta_{\alpha\delta} \delta_{\beta\gamma} \right)
\end{array}\right.
\]
where $\Big|_0$ means evaluation at vanishing fields, and we choose
the coefficient functions to be independent of $m$ for mass
independence.\footnote{We have assumed the $\gamma_5$ invariance of
  the theory.  Namely, under $\psi \to \gamma_5 \psi, \bar{\psi} \to
  - \bar{\psi} \gamma_5$, the theory is invariant if we change $m$ to
  $-m$ at the same time.  This $\gamma_5$ invariance excludes a term
  linear in $\Lambda$ in the two-point fermionic vertex.}

The $\Lambda$ independent parts of the coefficients, except for $a_2$,
are not fixed by the ERG differential equation, and we must fix them
by convention.  We can take
\[
c_2 (0) = 0
\]
since this is a normalization condition on $A_\mu$.  Likewise we can
take 
\[
a_f (0) = 0
\]
as a normalization of $\psi$.  By normalizing $m$, we can
take
\[
b_f (0) = 0
\]

Now, in addition, we could also take $a_3 (0) = e$, where $e$ is the
coupling constant, but we do not.  We would rather introduce $e$
through the Ward identities.  We will show, in the remainder of this
section, that $a_3 (0)$ and the other coefficients $b_2 (0), d_2 (0),
a_4 (0)$ are all determined by the Ward identities.

\subsection{Ward identities}

Our starting point is the Ward identities in the continuum limit.
For the photon two-point function, the Ward identity is
\[
\frac{1}{\xi} k_\mu \vev{A_\mu (-k) A_\nu (k)}_\infty =
\frac{k_\nu}{k^2}
\]
For the higher-point functions, the identities are given by
\begin{eqnarray*}
&&\frac{1}{\xi} k_\mu \vev{A_\mu (-k) A_{\mu_1} (k_1) \cdots A_{\mu_M}
  (k_M) \psi (p_1) \cdots \psi (p_N) \bar{\psi} (-q_1) \cdots
  \bar{\psi} (-q_N)}_\infty\\
&& \quad = \frac{e}{k^2} \sum_{i=1}^N \left\lbrace
\vev{ \cdots \psi (p_i-k) \cdots}_\infty - \vev{ \cdots \bar{\psi}
  (-q_i - k) \cdots }_\infty \right\rbrace
\end{eqnarray*}
where either $M \ge 2$ or $N \ge 1$.  Note that through these Ward
identities, the coupling constant $e$ is introduced into the theory.

The idea is simple.  We want to transcribe the Ward identities of the
$\Lambda$-independent correlation functions for the
$\Lambda$-dependent correlation functions that are given directly by
the cutoff action $S(\Lambda)$.  Once we do this, it will not be hard
to construct the corresponding invariance of the cutoff action.

\subsection*{Photon two-point function}

Let us first examine the Ward identity for the photon two-point
function.  Since
\[
\vev{A_\mu (-k) A_\nu (k)}_\infty = \frac{1 - 1/K(k/\Lambda)}{k^2}
\left ( \delta_{\mu\nu} - (1-\xi) \frac{k_\mu k_\nu}{k^2} \right) +
\vev{A_\mu (-k) A_\nu (k)}_{S} \frac{1}{K(k/\Lambda)^2}
\]
the Ward identity gives
\[
\frac{1}{\xi} k_\mu \vev{A_\mu (-k) A_\nu (k)}_S = K(k/\Lambda)
\frac{k_\nu}{k^2}
\]

Now, the equation of motion
\[
\vev{\frac{\delta S}{\delta A_\mu (k)} A_\nu (k)}_S = - \delta_{\mu\nu}
\]
gives
\[
- \frac{1}{K(k/\Lambda)}
\left( k^2 \delta_{\mu\alpha} - \left(1 - \frac{1}{\xi}\right) k_\mu
    k_\alpha \right) \vev{A_{\alpha} (-k) A_\nu (k)}_S
+ \vev{\frac{\delta \Si}{\delta A_\mu (k)} A_\nu (k)}_S = -
\delta_{\mu\nu}
\]
Hence, multiplying by $k_\mu$, we obtain
\[
- \frac{1}{\xi} \frac{k^2}{K(k/\Lambda)} k_\mu \vev{A_\mu (-k) A_\nu
  (k)}_S + k_\mu \vev{\frac{\delta \Si}{\delta A_\mu (k)} A_\nu (k)}_S
= - k_\nu
\]
Thus, the Ward identity gives
\[
k_\mu \vev{\frac{\delta \Si}{\delta A_\mu (k)} A_\nu (k)}_S
= 0
\]
This is equivalent to the Ward identity we started from.

\subsection*{Higher-point functions}

We next consider the Ward identities for the higher-point functions.
We first rewrite the left-hand side using the equation of motion,
which gives
\[
\vev{\frac{\delta S}{\delta A_\mu (k)} A_{\mu_1} (k_1) \cdots \psi
  (p_1) \cdots \bar{\psi} (-q_1) \cdots }_S = 0
\]
Multiplying this by $k_\mu$ and substituting $S = \Sf + \Si$,
we can rewrite this as
\begin{eqnarray*}
&&\frac{1}{\xi} \frac{k^2}{K(k/\Lambda)} k_\mu \vev{ A_\mu (-k)
  A_{\mu_1} (k_1) \cdots \psi (p_1) \cdots \bar{\psi} (-q_1) \cdots
}_S \\
&& \, = k_\mu \vev{\frac{\delta \Si}{\delta A_\mu (k)} A_{\mu_1} (k_1)
  \cdots \psi (p_1) \cdots \bar{\psi} (-q_1) \cdots}_S 
\end{eqnarray*}
Since 
\[
\ffbox{J_\mu (-k) \equiv \frac{\delta \Si}{\delta A_\mu (k)}}
\]
is a composite operator corresponding to the charge current, we can
rewrite the Ward identities as
\[
\ffbox{
\begin{array}{c}
k_\mu \vev{J_\mu (-k) A_{\mu_1} (k_1) \cdots \psi (p_1) \cdots
  \bar{\psi} (-q_1) \cdots }_\infty \\
 \quad = e \sum_{i=1}^N \left\lbrace
\vev{ \cdots \psi (p_i-k) \cdots}_\infty - \vev{ \cdots \bar{\psi}
  (-q_i - k) \cdots }_\infty \right\rbrace
\end{array}
}
\]
It is easy to see that the Ward identity for the photon two-point
function can be included as a special case $M=1, N=0$ of the above
identities.

\subsection*{Operator equation}

We wish to express the Ward identities as an operator equation
\[
\ffbox{
k_\mu J_\mu (-k) = \Phi (-k)}
\]
where we define the composite operator $\Phi (-k)$ by
the right-hand side of the Ward identities as
\begin{eqnarray*}
&&\vev{\Phi (-k)  A_{\mu_1} (k_1) \cdots \psi (p_1) \cdots
  \bar{\psi} (-q_1) \cdots }_\infty\\
&&\quad \equiv e \sum_{i=1}^N \left\lbrace
\vev{ \cdots \psi (p_i-k) \cdots}_\infty - \vev{ \cdots \bar{\psi}
  (-q_i - k) \cdots }_\infty \right\rbrace
\end{eqnarray*}
for any $M \ge 0$ and $N \ge 0$.  $\Phi (-k)$ is a composite operator
that converts
\[
\left\lbrace\begin{array}{c@{~\longrightarrow~}l}
\psi (p_i) & e \psi (p_i - k)\\
\bar{\psi} (-q_i) & - e \bar{\psi} (-q_i-k)
\end{array}\right.
\]

As a preparation, we go back to the simpler scalar theory and
construct an analogous operator that generates a shift of momentum:
\[
\vev{\Phi (-k) \phi (p_1) \cdots \phi (p_{2n})}_\infty
= \sum_{i=1}^{2n} \vev{ \cdots \phi (p_i - k) \cdots}_\infty
\]
For $n > 1$, this gives
\[
\vev{\Phi (-k) \phi (p_1) \cdots \phi (p_{2n})}_S 
= \sum_{i=1}^{2n} \frac{K(p_i/\Lambda)}{K((p_i-k)/\Lambda)}
\vev{ \cdots \phi (p_i - k) \cdots}_S
\]
Hence, an operator analogous to $\Op_1 [f]$ will do:
\[
\Phi (-k) \stackrel{?}{=} - \int_q
\frac{K(q/\Lambda)}{K((q-k)/\Lambda)} \phi (q-k) 
\frac{\delta S}{\delta \phi (q)}
\]
But this is not quite correct, since for $n=1$, we must obtain
\begin{eqnarray*}
&&\vev{\Phi (-k) \phi (p+k) \phi (-p)}_S \\
&& =
\frac{K((p+k)/\Lambda)}{K(p/\Lambda)} \vev{\phi (p) \phi (-p)}_S
+ \frac{K(p/\Lambda)}{K((p+k)/\Lambda)} \vev{\phi (p+k) \phi
  (-p-k)}_S\\
&& \quad - K((p+k)/\Lambda) \frac{1 - K(p/\Lambda)}{p^2 + m^2} -
K(p/\Lambda) \frac{1 - K((p+k)/\Lambda)}{(p+k)^2 + m^2}
\end{eqnarray*} 
To generate the extra terms, we need to add an operator analogous to $\Op_2
[C]$:
\begin{eqnarray*}
\Phi (-k) &=&  - \int_q
\frac{K(q/\Lambda)}{K((q-k)/\Lambda)} \phi (q-k) 
\frac{\delta S}{\delta \phi (q)}\\
&&\quad - \int_q K((q+k)/\Lambda) \frac{1 - K(q/\Lambda)}{q^2 + m^2}
\frac{\delta^2 S}{\delta \phi (-q) \delta \phi (q+k)}
\end{eqnarray*}
\no \textbf{HW\#9}: Check the final result.

Generalizing this result to QED, we define a type 1 operator
\[
\Op_1 (-k) \equiv e \int_q \left[
- S \rd{q} \frac{K(q/\Lambda)}{K((q-k)/\Lambda)} \psi (q-k)
+ \frac{K(q/\Lambda)}{K((q+k)/\Lambda)} \bar{\psi} (-q-k) \ld{q} S \right]
\]
and a type 2 operator
\[
\Op_2 (-k) \equiv e \int_q \Tr U (-q-k,q) \left\lbrace
\ld{q} S \cdot S \rd{q+k} + \ld{q} S \rd{q+k} \right\rbrace
\]
where
\[
U(-q-k,q) \equiv K((q+k)/\Lambda) \frac{1 - K(q/\Lambda)}{\fmslash{q}
  + i m} - \frac{1 - K((q+k)/\Lambda)}{\fmslash{q} + \fmslash{k} + i
  m} K(q/\Lambda)
\]
We then define
\[
\Phi (-k) \equiv \Op_1 (-k) + \Op_2 (-k)
\]
This gives the desired relation
\begin{eqnarray*}
&&\vev{\Phi (-k) A_{\mu_1} (k_1) \cdots \psi (p_1) \cdots \bar{\psi}
  (-q_1) \cdots}_\infty\\
&& = e \sum_{i=1}^N \left\lbrace \vev{ \cdots \psi (p_i-k) \cdots}_\infty
- \vev{\cdots \bar{\psi} (-q_i-k) \cdots}_\infty \right\rbrace
\end{eqnarray*}
\no \textbf{HW\#10}: Check this.

To summarize, we have found that the Ward identities of QED can be
expressed as a single operator equation as
\[
k_\mu J_\mu (-k) = \Phi (-k)
\]
This form of the Ward identity was first derived in \cite{hs07}.

\subsection{Perturbative construction}

We wish to show that the Ward identity can be satisfied by choosing
appropriate values of the remaining four parameters: $b_2, d_2, a_3$,
and $a_4$.

The asymptotic behavior of $k_\mu J_\mu$ is easily obtained from that
of the action $\Si (\Lambda)$:
\[
\left\lbrace\begin{array}{l}
\frac{k_\mu \delta J_\mu (-k)}{\delta A_\nu (-k)}\Big|_0 = k_\nu
\left( \Lambda^2 a_2 (\ln \Lambda/\mu) + m^2 b_2 (\ln \Lambda/\mu) +
    k^2 (c_2 + d_2) (\ln \Lambda/\mu) \right)\\
\ld{(p+k)} k_\mu J_\mu (-k) \rd{p}\Big|_0 = \fmslash{k} a_3 (\ln
\Lambda/\mu)\\
\frac{k_\mu \delta^3 J_\mu (-k)}{\delta A_\alpha (k_1) \delta
  A_\beta (k_2) \delta A_\gamma (k_3)}\Big|_0 = \left( k_\alpha
    \delta_{\beta\gamma} + k_\beta \delta_{\gamma\alpha} + k_\gamma
    \delta_{\alpha\beta} \right) a_4 (\ln \Lambda/\mu)
\end{array}\right.
\]
On the other hand, the composite operator $\Phi (-k)$ is a dimension 4
scalar operator of momentum $-k$, vanishing at $k=0$.

Hence, its asymptotic behavior must have the same form as
above\footnote{Here we are assuming what amounts to the CP invariance.
  We exclude the asymptotic behavior
\[
\frac{\delta^2 \Phi (-k)}{\delta A_\alpha (k_1) \delta A_\beta
  (k_2)}\Big|_0 \propto \ep_{\alpha\beta\gamma\delta} k_{1, \gamma}
k_{2, \delta}
\]
In chiral U(1) gauge theory this must vanish automatically.  This is
the anomaly cancellation condition.}:
\[
\left\lbrace\begin{array}{l}
\frac{\delta \Phi (-k)}{\delta A_\nu (-k)}\Big|_0 
= k_\nu \left( \Lambda^2 \bar{a}_2 (\ln \Lambda/\mu) + m^2 \bar{b}_2
    (\ln \Lambda/\mu) + 
    k^2 \bar{d}_2 (\ln \Lambda/\mu) \right)\\
\ld{(p+k)} \Phi (-k) \rd{p}\Big|_0 = \fmslash{k} \bar{a}_3 (\ln
\Lambda/\mu)\\
\frac{\Phi (-k)}{\delta A_\alpha (k_1) \delta
  A_\beta (k_2) \delta A_\gamma (k_3)}\Big|_0 = \left( k_\alpha
    \delta_{\beta\gamma} + k_\beta \delta_{\gamma\alpha} + k_\gamma
    \delta_{\alpha\beta} \right) \bar{a}_4 (\ln \Lambda/\mu)
\end{array}\right.
\]
The two composite operators $k_\mu J_\mu$ and $\Phi$ are the same if
and only if they have the same asymptotic behaviors.  Hence, the Ward
identity is equivalent to the following equations:
\[
\left\lbrace
\begin{array}{c@{~=~}l}
 b_2 (0) & \bar{b}_2 (0)\\
 d_2 (0) & \bar{d}_2 (0)\\
 a_3 (0) & \bar{a}_3 (0)\\
 a_4 (0) & \bar{a}_4 (0)
\end{array}\right.
\]
where we have used the convention $c_2 (0) = 0$.

Now, $\Phi$ is constructed in terms of $S(\Lambda)$.  Hence, we obtain 
\[
\left\lbrace\begin{array}{l}
\frac{\delta \Phi (-k)}{\delta A_\nu (-k)}\Big|_0 
= \int_q \Tr U(-q-k,q) \ld{q} \frac{\delta S}{\delta A_\nu
  (-k)} \rd{q+k} \Big|_0\\
\ld{(p+k)} \Phi (-k) \rd{p}\Big|_0 = e \left( 1 - a_f (\ln
    \Lambda/\mu) \right) \fmslash{k}\\
\qquad + \int_q \ld{(p+k)} \Tr \left[
U(-q-k,q) \ld{q} S \rd{q+k} \right] \rd{p}\Big|_0\\
\frac{\delta^2 \Phi (-k)}{\delta A_\alpha (k_1) \delta A_\beta
  (k_2) \delta A_\gamma (k_3)}\Big|_0 \\
\quad = \int_q \Tr U (-q-k,q) \ld{q} \frac{\delta^3 S}{\delta
  A_\alpha (k_1) \delta A_\beta (k_2) \delta A_\gamma (k_3)}
\rd{q+k}\Big|_0
\end{array}\right.
\]
We note $a_f (0) = 0$ by convention.  Everything on the right-hand
side has an extra loop integral.  Hence, if we know $S$ up to $l$-loop
level, we can determine the right-hand side up to $(l+1)$-loop.  Thus,
by imposing the Ward identities, we can determine the coefficients
$b_2(0), d_2 (0), a_3 (0), a_4 (0)$ at $(l+1)$-loop level by knowing
$S$ up to $l$-loop.  Therefore, we can construct the action
iteratively with the help of the Ward identities.

\subsection*{One-loop calculations}

At tree level, we obtain
\[
b_2 (0) = d_2 (0) = a_4 (0) = 0,\quad a_3 (0) = e
\]

We now compute $b_2 (0), d_2 (0)$ at one-loop.  We find
\[
\frac{\delta \Phi}{\delta A_\nu (-k)}\Big|_0 = e^2 \int_q \Tr
  \gamma_\nu U (-q-k,q)
\]
Changing the integration variable as
\[
q \longrightarrow \Lambda q 
\]
we obtain
\begin{eqnarray*}
&& \frac{\delta \Phi}{\delta A_\nu (-k)}\Big|_0 = e^2 \Lambda^2 \int_q
  \Tr \gamma_\nu \cdot \left\lbrace K\left( q +
  \frac{k}{\Lambda} \right) \frac{1 - K (q)}{\fmslash{q} + i
  m/\Lambda} - \frac{1 - K\left( q +
  \frac{k}{\Lambda}\right)}{\fmslash{q} + \fmslash{k}/\Lambda + i
  m/\Lambda } K(q) \right\rbrace\\ &&\quad\to e^2 k_\nu \left[ - 2
  \Lambda^2 \int_q \frac{1}{q^2} \Delta (q) \left( 1 - K(q) \right) +
  \left( m^2 + \frac{k^2}{3} \right) \int_q \frac{\Delta (q)}{(q^2)^2}
  \right]
\end{eqnarray*}
where we have expanded the integral in powers of $m/\Lambda$
and $k_\mu/\Lambda$.  Using
\[
\int_q \frac{ \Delta (q) K(q)^n}{(q^2)^2}  
= \frac{1}{(4 \pi)^2} \frac{2}{n+1}
\]
(see subsect.~2.6), we obtain
\[
b_2 (0) = \frac{2 e^2}{(4 \pi)^2},\quad
d_2 (0) = \frac{2 e^2}{3 (4 \pi)^2}
\]
Incidentally, the dominant photon mass term is given by
\[
\Lambda^2 a_2 = - 2 \Lambda^2 \int_q \frac{\Delta (q) (1 - K(q))}{q^2}
\]
corresponding to a large positive squared mass.  Both $a_2$ and $b_2$
are non-vanishing, and the photon has mass terms in the action.  But
the mass terms are fixed by the gauge symmetry.

Next we consider $a_3 (0)$ at one-loop:
\begin{eqnarray*}
&&\ld{(p+k)} \Phi (-k) \rd{p} \Big|_0 - e (1 - a_f (\ln \Lambda/\mu))
\fmslash{k} \\
&& = - e^2 \int_q \gamma_\mu U (-q-k,q) \gamma_\nu\\
&& \qquad \times \frac{1 - K ((q-p)/\Lambda)}{(q-p)^2} \left(
\delta_{\mu\nu} - (1-\xi) \frac{(q-p)_\mu (q-p)_\nu}{(q-p)^2}
\right)\\
&& \to - e^3 \fmslash{k} \int_q \frac{1}{(q^2)^2} \left\lbrace \xi
K(q) \left(1 - K(q) \right)^2 + \frac{3 - \xi}{4} \left(1 -
K(q)\right) \Delta (q) \right\rbrace
\end{eqnarray*}
where we have rescaled the integration variables, and
expanded the integral in powers of $k_\mu/\Lambda$, $p_\mu/\Lambda$.
Hence,
\[
a_3 (0) - e = - e^3 \left( \xi \int_q \frac{1}{(q^2)^2} K(q) (1 -
K(q))^2 + \frac{3 - \xi}{4 (4\pi)^2} \right)
\]

Finally, we compute $a_4 (0)$:
\begin{eqnarray*}
&&\frac{\delta^3 \Phi (-k)}{\delta
 A_\alpha (k_1) \delta A_\beta (k_2) \delta A_\gamma
 (k_3)}\Bigg|_0
 = e^3 \int_q \Tr U (-q-k,q) \\
&&\quad \times \Bigg\lbrace \gamma_\gamma \frac{1 - K \left( (q-k_3)/\Lambda
 \right)}{\fmslash{q} - \fmslash{k}_3 + i m} \gamma_\beta 
\frac{1 - K\left( (q-k_2-k_3)/\Lambda \right)}{\fmslash{q} -
 \fmslash{k}_2 - \fmslash{k}_3 + i m} \gamma_\alpha
 + (\textrm{5 permutations}) \,\Bigg\rbrace\\
&& \to 2 e^4 \left( k_\alpha \delta_{\beta\gamma} + k_\beta
  \delta_{\gamma \alpha} + k_\gamma \delta_{\alpha\beta} \right)
\int_q \frac{\Delta (q/\Lambda)}{(q^2)^2} \left(1 - K(q/\Lambda)\right)^2
\end{eqnarray*}
Hence, we obtain
\[
a_4 (0) = \frac{4}{3} \frac{e^4}{(4 \pi)^2}
\]

\subsection{BRST invariance}

As a preparation for YM theories, we wish to formulate the Ward
identity as BRST invariance.  For this purpose we introduce ghost and
antighost fields $c, \bar{c}$.  These are free, and we modify the
action by adding the free ghost kinetic term:
\[
\bar{S} (\Lambda) \equiv S(\Lambda) - \int_k \bar{c} (-k) c(k)
\frac{k^2}{K(k/\Lambda)}
\]

Our Ward identity is given by
\[
k_\mu J_\mu (-k) - \Phi (-k) = 0
\]
We multiply this by $c(k)$ and integrate over $k$ to obtain
\[
\int_k \ep c (k) \left( k_\mu J_\mu (-k) - \Phi (-k) \right) = 0
\]
where $\ep$ is an arbitrary Grassmann constant.  

We first observe
\begin{eqnarray*}
\int_k \ep c(k) k_\mu J_\mu (-k) &=& \int_k \ep c(k) k_\mu
\frac{\delta \Si}{\delta A_\mu (k)}\\
&=& \int_k \ep c(k) k_\mu \left( \frac{\delta \bar{S}}{\delta A_\mu
      (k)} + \frac{1}{\xi} \frac{k_\mu k_\nu}{K(k/\Lambda)} A_\nu (k)
\right)\\
&=& \int_k \left( \ep k_\mu c(k) \frac{\delta \bar{S}}{\delta A_\mu
      (k)} - \frac{1}{\xi} \ep k_\mu A_\mu (k)
    \frac{\overrightarrow{\delta} \bar{S}}{\delta \bar{c} (-k)} \right)
\end{eqnarray*}
Hence, this is the infinitesimal change of the action $\bar{S}$ under
the following change of variables:
\[
\left\lbrace\begin{array}{c@{~\equiv~}l}
\delta_\ep A_\mu (k) & k_\mu \ep c(k)\\
\delta_\ep c(k) & 0\\
\delta_\ep \bar{c} (-k) & - \frac{1}{\xi} k_\mu A_\mu (-k) \ep
\end{array}\right.
\]

We next observe that $\Phi$ can be written as 
\begin{eqnarray*}
&&\Phi (-k) 
= e \int_q \Bigg[\\
&& S \rd{q} \left(
- \frac{K(q/\Lambda)}{K((q-k)/\Lambda)} \psi (q-k)
 - K(q/\Lambda) \frac{1 - K((q-k)/\Lambda)}{\fmslash{q} -
  \fmslash{k} + i m} \ld{(q+k)} S \right)\\
&& + \left( \frac{K(q/\Lambda)}{K((q+k)/\Lambda)} \bar{\psi} (-q-k)
+ S \rd{q+k} \frac{1-K((q+k)/\Lambda)}{\fmslash{q} +
  \fmslash{k} + i m} K(q/\Lambda) \right) \ld{q} S \\
&& + \Tr K(q/\Lambda) \frac{1 - K((q-k)/\Lambda)}{\fmslash{q} -
  \fmslash{k} + i m} \ld{(q+k)} S \rd{q} \\
&& - \Tr \frac{1 - K((q+k)/\Lambda)}{\fmslash{q} + \fmslash{k} + im}
K(q/\Lambda) \ld{q} S \rd{q+k} \quad\Bigg]\\
&&= e \int_q K(q/\Lambda) \left[ - S \rd{q} \Psi (q-k) + \bar{\Psi}
(-q-k) \ld{q} S \right]\\
&&\quad + e \int_q K(q/\Lambda) \Tr \left[ \Psi (q-k) \rd{q}  - \ld{q}
    \bar{\Psi} (-q-k) \right]
\end{eqnarray*}
where
\[
\left\lbrace\begin{array}{c@{~\equiv~}l}
 \Psi (p) & \psi (p) + \frac{1 - K(p/\Lambda)}{\fmslash{p} + i m}
 \ld{p} \Si\\
 \bar{\Psi} (-p) & \bar{\psi} (-p) + \Si \rd{p}
 \frac{1-K(p/\Lambda)}{\fmslash{p} + i m}
\end{array}\right.
\]
are the composite operators corresponding to $\psi (p), \bar{\psi}
(-p)$, respectively.

\no\textbf{HW\#11}: Check the derivation.

Hence, we obtain
\begin{eqnarray*}
- \int_k \ep c(k) \Phi (-k) &=& \int_p \left[ S \rd{q} \delta_\ep \psi
    (q) + \delta_\ep \bar{\psi} (-q) \ld{q} S \right]\\
&& - \int_q \Tr \left[ \delta_\ep \psi (q) \rd{q} + \ld{q} \delta_\ep
    \bar{\psi} (-q) \right]
\end{eqnarray*}
where we define
\[
\left\lbrace\begin{array}{c@{~\equiv~}l}
 \delta_\ep \psi (p) & e K(p/\Lambda) \int_k \ep c(k) \Psi (p-k)\\
 \delta_\ep \bar{\psi} (-p) & - e K(p/\Lambda) \int_k \ep c (k) \bar{\Psi}
 (-p-k)
\end{array}\right.
\]

Thus, altogether, the Ward identity can be written as the BRST
invariance 
\[
\delta_\ep \bar{S}   - \int_q \Tr \left[ \delta_\ep \psi (q) \rd{q} +
    \ld{q} \delta_\ep \bar{\psi} (-q) \right] = 0
\]
where $\delta_\ep \bar{S}$ is the infinitesimal change of the action
$\bar{S}$ under the BRST transformation that is given by
\[
\left\lbrace\begin{array}{c@{~=~}l}
\delta_\ep A_\mu (k) & k_\mu \ep c(k)\\
\delta_\ep c(k) & 0\\
\delta_\ep \bar{c} (-k) & - \frac{1}{\xi} \ep k_\mu A_\mu (-k) \\
\delta_\ep \psi (p) & K(p/\Lambda) \cdot e \int_k \ep c(k) \Psi (p-k)\\
 \delta_\ep \bar{\psi} (-p) & K(p/\Lambda) \cdot 
(-e)\int_k \ep c (k) \bar{\Psi} (-p-k)
\end{array}\right.
\]
Note that the second term of the BRST invariance is the jacobian of
the BRST transformation.  Hence, we can interpret the BRST invariance
as the ``quantum'' invariance of $\bar{S}$ under the BRST
transformation.  By adopting the notation
\[
\Delta_\ep \bar{S} \equiv  - \int_q \Tr \left[ \delta_\ep \psi (q) \rd{q} +
    \ld{q} \delta_\ep \bar{\psi} (-q) \right]
\]
the ``quantum'' BRST invariance of the action is written as
\[
\delta_\ep \bar{S} + \Delta_\ep \bar{S} = 0
\]

The above BRST transformation of the fields has two problems:
\begin{enumerate}
\item it is not nilpotent (For example, $\delta_\ep \delta_\ep \psi
    \ne 0$.)
\item asymmetry --- the transformation of $\psi, \bar{\psi}$ is given
    by the cutoff function times a composite operator, but that of
    either $A_\mu$ or $\bar{c}$ is not.
\end{enumerate}
The first problem is not so serious for QED, but nilpotency is more
important for YM theories; it assures the possibility of satisfying
the BRST invariance by adjusting available parameters.  To obtain
nilpotency, we need to introduce external sources that generate BRST
transformation.  We will discuss it in the next subsection.  

The second problem can be solved.  We modify the BRST transformation
of $A_\mu, \bar{c}$ as follows:
\[
\left\lbrace
\begin{array}{c@{~=~}l}
\delta'_\ep A_\mu (k) & K(k/\Lambda) k_\mu \ep c(k)\\
\delta'_\ep \bar{c} (-k) & K(k/\Lambda) \frac{- 1}{\xi} \ep k_\mu
\mathcal{A}_\mu (-k)
\end{array}\right.
\]
where
\[
\mathcal{A}_\mu (k) \equiv A_\mu (-k) + \frac{1 - K(k/\Lambda)}{k^2}
\left( \delta_{\mu\nu} - \frac{k_\mu k_\nu}{k^2} (1-\xi) \right)
\frac{\delta \Si}{\delta A_\nu (-k)}
\]
is the composite operator corresponding to $A_\mu (-k)$.  Since $c,
\bar{c}$ are non-interacting, $c$ is already a composite
operator.\footnote{The composite operator corresponding to $c$ is
$c + \frac{1-K}{k^2} \frac{\overrightarrow{\delta}}{\delta \bar{c}} \Si
  = c$.}

The action is invariant under the new transformation if it is also
invariant under the difference of the two transformations:
\[
\left\lbrace\begin{array}{c@{~\equiv~}c@{~=~}l}
\delta''_\ep A_\mu (k) & (\delta'_\ep - \delta_\ep) A_\mu (-k) &
\left( K(k/\Lambda) - 1 \right) k_\mu \ep c (k)\\
\delta''_\ep \bar{c} (-k) & (\delta'_\ep - \delta_\ep) \bar{c} (-k)
& \ep \frac{-1}{\xi} k_\mu \left(K(k/\Lambda) \mathcal{A}_\mu (-k) -
    A_\mu (-k) \right)
\end{array}\right.
\]
Since
\[
k_\mu \mathcal{A}_\mu (-k) = k_\mu A_\mu (-k) + \xi
\frac{1-K(k/\Lambda)}{k^2} k_\mu \frac{\delta \Si}{\delta A_\mu (k)}
\]
we find
\begin{eqnarray*}
&&\frac{1}{\xi} k_\mu \left( K(k/\Lambda) \mathcal{A}_\mu (-k) - A_\mu
    (-k) \right) \\
&& = K(k/\Lambda) \left( 1 - K(k/\Lambda)  \right) \left(
- \frac{1}{K(k/\Lambda)}
\frac{1}{\xi} k_\mu A_\mu (-k) + \frac{1}{k^2} k_\mu \frac{\delta
  \Si}{\delta A_\mu (k)} \right)\\
&& =  K(k/\Lambda) \left( 1 - K(k/\Lambda)  \right) \frac{1}{k^2}
k_\mu \frac{\delta \bar{S}}{\delta A_\mu (k)}
\end{eqnarray*}
Hence, we find
\[
\left\lbrace\begin{array}{c@{~=~}l}
\delta''_\ep A_\mu (k) & 
\left( K(k/\Lambda) - 1 \right) k_\mu \ep c (k)\\
\delta''_\ep \bar{c} (-k) & \left( K(k/\Lambda) - 1 \right)
K(k/\Lambda) \ep \frac{1}{k^2}
k_\mu \frac{\delta \bar{S}}{\delta A_\mu (k)}
\end{array}\right.
\]
It is easy to check the invariance of $\bar{S}$ under this:
\begin{eqnarray*}
\delta'' \bar{S} &=& \int_k \left( \delta''_\ep A_\mu (k) \frac{\delta
      \bar{S}}{\delta A_\mu (k)} + \delta''_\ep \bar{c} (-k)
    \frac{\overrightarrow{\delta}}{\delta \bar{c} (-k)} \bar{S}
\right)\\
&=& \int_k \left( K(k/\Lambda) - 1\right) \left(
k_\mu \ep c (k) \frac{\delta \bar{S}}{\delta A_\mu (k)} +
K(k/\Lambda) \ep \frac{k_\mu}{k^2} \frac{\delta \bar{S}}{\delta A_\mu
  (k)} (-) \frac{k^2}{K(k/\Lambda)} c (k) \right)\\
&=& 0
\end{eqnarray*}

Thus, the action $\bar{S}$ is quantum invariant under the following
new BRST transformation\footnote{The jacobian of $\delta_\ep A_\mu,
  \delta_\ep \bar{c}$ is $1$.}:
\[
\ffbox{
\begin{array}{c@{~=~}l}
\delta_\ep A_\mu (k) & K(k/\Lambda) \cdot k_\mu \ep c(k)\\
\delta_\ep \bar{c} (-k) & K(k/\Lambda) \cdot \frac{- 1}{\xi} \ep k_\mu
\mathcal{A}_\mu (-k)\\
\delta_\ep \psi (p) & K(p/\Lambda) \cdot e \int_k \ep c(k) \Psi (p-k)\\
 \delta_\ep \bar{\psi} (-p) & K(p/\Lambda) \cdot 
(-e)\int_k \ep c (k) \bar{\Psi} (-p-k)
\end{array}
}
\]

\subsection{BRST with external sources}

The most elegant way of formulating BRST invariance uses external
sources that generate the BRST transformation.  It is Becchi who first
introduced the BRST invariance of the cutoff action in the presence of
external sources.\cite{becchi} Let $A^*_\mu (-k)$ be the fermionic
source that generates the BRST transformation of $A_\mu (k)$.
Likewise, let us introduce the bosonic sources $\bar{c}^*$, $\psi^*$,
and $\bar{\psi}^*$.  The purpose of this subsection is to show that
the BRST invariance of QED can be formulated in the following form:
\begin{eqnarray*}
&& \e^{-\s} \int_p \K{p} \Bigg(
\Ld{A_\mu^* (-p)} \frac{\delta}{\delta A_\mu (p)}
+ \frac{\delta}{\delta \bar{c}^* (p)} \Ld{\bar{c}(-p)} \\
&&\qquad + \frac{\delta}{\delta \psi_a^* (-p)} \Ld{\psi_a (p)}
+ \frac{\delta}{\delta \bar{\psi}_a^* (p)} \Ld{\bar{\psi}_a (-p)} \Bigg)
\e^{\s}\\
    &&= \int_k \K{k} \left[ \Ld{A_\mu^* (-k)} \s \cdot \frac{\delta \s}{\delta
          A_\mu (k)} + \frac{\delta \s}{\delta \bar{c}^* (k)} \Ld{\bar{c}
          (-k)} \s \right]\\
    &&\, - \int_q \K{q} \,\Tr \left[ \frac{\delta \s}{\delta
          \psi^* (-q)} \cdot 
        \s \rd{q} +  \frac{\delta}{\delta \psi^* (-q)} \s \rd{q}
    \right]\\
    && \,+ \int_q \K{q} \,\Tr \left[ \ld{p} \s
        \cdot \frac{\delta \s}{\delta \bar{\psi}^* (q)}
        +  \ld{q} \frac{\delta \s}{\delta
          \bar{\psi}^* (q)} \right] = 0
\end{eqnarray*}
where $\s$ is the action with external sources.  The meaning of this
equation is clear.  The BRST transformation of the fields is given by
functional derivatives with respect to external sources:
\[
\left\lbrace\begin{array}{c@{~=~}l}
        \delta_\ep A_\mu (k) & \ep \K{k} \Ld{A_\mu^* (-k)} \s\\
        \delta_\ep \bar{c} (-k) & \ep \K{k} \frac{\delta}{\delta
          \bar{c}^* (k)} \s\\ 
        \delta_\ep \psi (q) & \ep \K{q} \frac{\delta}{\delta \psi^*
          (-q)}\s\\ 
        \delta_\ep \bar{\psi} (-q) & \ep \K{q} \frac{\delta}{\delta
          \bar{\psi}^* (q)}\s
\end{array}\right.
\]
At the vanishing sources, we must find
\[
\left\lbrace\begin{array}{c@{~=~}l}
 \Ld{A_\mu^* (-k)} \s & k_\mu c(k)\\
\frac{\delta}{\delta \bar{c}^* (k)} \s & - \frac{1}{\xi} k_\mu
\mathcal{A}_\mu (-k)\\
\frac{\delta}{\delta \psi^* (-q)}\s & e \int_k c(k) \Psi (q-k)\\
\frac{\delta}{\delta \bar{\psi}^* (q)}\s& - e \int_k c(k) \bar{\Psi}
(-q-k)
\end{array}\right.
\]

In this subsection we wish to understand the general structure of the
BRST transformation generated by sources; especially we would like to
introduce a BRST transformation $\delta_Q$ that is nilpotent. In the
next subsection, we will construct $\s$ explicitly in terms of
$\bar{S}$ without sources that we have already constructed.

In order to understand the main properties of the system, it is best
to treat the simplest possible system with the same structure.  So,
instead of QED, we consider the theory of a scalar field $\phi$ in the
presence of an external fermionic source $\phi^*$.  Let $\s [\phi,
\phi^*]$ be a cutoff action satisfying the Polchinski ERG differential
equation
\begin{eqnarray*}
&&- \Lambda \frac{\partial}{\partial \Lambda} \s = \int_q \frac{\Delta
  (q/\Lambda)}{q^2 + m^2} \Bigg( \frac{q^2+m^2}{\K{q}} \phi (q)
    \frac{\delta \s}{\delta \phi (-q)} \\
&&\qquad\qquad +
\frac{1}{2} \left\lbrace \frac{\delta \s}{\delta \phi (q)}
    \frac{\delta \s}{\delta \phi (-q)} + \frac{\delta^2 \s}{\delta
      \phi (q) \delta \phi (-q)} \right\rbrace\Bigg)
\end{eqnarray*}
Note that the external source is a purely classical source that does
not participate in the ERG equation.  Differentiating the equation
with respect to $\phi^*$, we find that $\Ld{\phi^* (-p)} \s$ is a
composite operator.  We also recall that $\K{p} \frac{\delta
  \s}{\delta \phi (-p)}$ is a composite operator.

Now, as a preparation, we wish to show
\begin{enumerate}
\item If $\Op$ is a composite operator, so is
\[
A \equiv \K{q} \left( \Op \frac{\delta \s}{\delta \phi (-q)} + \frac{\delta
      \Op}{\delta \phi (-q)} \right) = \e^{-\s} \K{q}
\frac{\delta}{\delta \phi (-q)} \left( \Op \e^S \right)
\]
\item If $\Op$ is a composite operator, so is
\[
B \equiv \Ld{\phi^* (-p)} \Op + \Ld{\phi^* (-p)} \s \cdot \Op
= \e^{-\s} \Ld{\phi^* (-p)} \left( \Op \e^{\s} \right)
\]
\end{enumerate}
(Proof) Let us consider the $\Lambda$ derivative of $A$:
\begin{eqnarray*}
&&- \Lambda \frac{\partial}{\partial \Lambda} \left( \K{q} \left( \Op
        \frac{\delta \s}{\delta \phi (-q)} + \frac{\delta 
      \Op}{\delta \phi (-q)} \right)\right)\\
&& = \D \Op \cdot \K{q} \frac{\delta \s}{\delta \phi (-q)}
+ \Op \cdot \K{q} \D   \frac{\delta \s}{\delta \phi (-q)}\\
&&\quad + \K{q} \frac{\delta}{\delta \phi (-q)} \D \Op
- \Delta (q/\Lambda) \frac{\delta}{\delta \phi (-q)} \Op \\
&& = \D \left(  \Op \K{q} \frac{\delta \s}{\delta \phi (-q)} \right)
- \int_r \frac{\Delta (r/\Lambda)}{r^2 + m^2} \frac{\delta}{\delta
  \phi (r)} \Op \cdot \K{q} \frac{\delta^2 \s}{\delta \phi (-r) \delta
  \phi (-q)}\\
&&\quad + \K{q} \frac{\delta}{\delta \phi (-q)} \D \Op
- \Delta (q/\Lambda) \frac{\delta}{\delta \phi (-q)} \Op \\
&& = \D \left(  \Op \K{q} \frac{\delta \s}{\delta \phi (-q)} \right)
- \int_r \frac{\Delta (r/\Lambda)}{r^2 + m^2} \frac{\delta}{\delta
  \phi (r)} \Op \cdot \K{q} \frac{\delta^2 \si}{\delta \phi (-r) \delta
  \phi (-q)}\\
&&\quad + \K{q} \frac{\delta}{\delta \phi (-q)} \D \Op
\end{eqnarray*}
Since
\[
\frac{\delta}{\delta \phi (-q)} \D \Op = \D \frac{\delta}{\delta \phi
  (-q)} \Op + \int_r \frac{\Delta (r/\Lambda)}{r^2 + m^2}
\frac{\delta^2 \si}{\delta \phi (-r) \delta \phi (-q)}
\frac{\delta}{\delta \phi (r)} \Op
\]
we obtain
\[
- \Lambda \frac{\partial}{\partial \Lambda} A = \D A
\]
Hence, $A$ is a composite operator.

We next consider $B$:
\begin{eqnarray*}
&& - \Lambda \frac{\partial}{\partial \Lambda} 
\left( \Ld{\phi^* (-p)} \Op + \Ld{\phi^* (-p)} \s \cdot \Op \right)\\
&& = \Ld{\phi^* (-p)} \left( \D \Op \right)
+ \D \Ld{\phi^* (-p)} \s \cdot \Op
+ \Ld{\phi^* (-p)} \s \cdot \D \Op\\
&& = \D \Ld{\phi^* (-p)} \Op + \int_q \frac{\Delta (q/\Lambda)}{q^2 +
  m^2} \Ld{\phi^* (-p)} \frac{\delta \si}{\delta \phi (-q)}\cdot
\frac{\delta}{\delta \phi (q)} \Op\\
&&\quad + \D \left( \Ld{\phi^*(-p)} \s \cdot \Op \right)
- \int_q \frac{\Delta (q/\Lambda)}{q^2 + m^2} \Ld{\phi^* (-p)}
\frac{\delta \si}{\delta \phi (-q)} \cdot \frac{\delta}{\delta \phi
  (q)} \Op\\
&& = \D B
\end{eqnarray*}
Hence, $B$ is a composite operator. \textbf{Q.E.D.}

\no \textbf{HW\#12}: Alternatively you can show
\begin{eqnarray*}
&&\vev{A (q) \phi (p_1) \cdots \phi (p_{2n})}_\infty
= - \sum_{i=1}^{2n} (2\pi)^4 \delta^{(4)}(p_i+q) \vev{\cdots
  \widehat{\phi (p_i)} \Op
  \cdots}_\infty\\
&&\vev{B (p) \phi (p_1) \cdots \phi (p_{2n})}_\infty
= \Ld{\phi^*(-p)} \vev{ \Op\,\phi (p_1) \cdots \phi (p_{2n})}_\infty
\end{eqnarray*}
This shows $A$ and $B$ are composite.

\subsection*{Nilpotent $\delta_Q$}

Now, let us apply what we have found.   Given a composite operator $\Op$,
\[
X (p) \equiv \K{p} \e^{- \s} \frac{\delta}{\delta \phi (-p)} \left(
\e^\s \Op \right) =
K(p/\Lambda) \left( \frac{\delta \s}{\delta \phi (-p)} \Op +
  \frac{\delta \Op}{\delta \phi (-p)} \right)
\]
is also a composite operator according to 1.  Applying 2 to $X$, we
obtain
\[
Y(p) \equiv \e^{-\s} \Ld{\phi^* (p)} \left( \e^{\s} X \right) =
\frac{\overrightarrow{\delta}}{\delta \phi^* (p)} X(p)+
\frac{\overrightarrow{\delta}}{\delta \phi^* (p)} \s \cdot X(p)
\]
is also a composite operator.  We find
\begin{eqnarray*}
Y(p) &=& \K{p} \e^{- \s} \Ld{\phi^* (p)} \frac{\delta}{\delta \phi
  (-p)} \left( \e^{\s} \Op \right)\\
&=& K(p/\Lambda) \Bigg[ \left( \frac{\overrightarrow{\delta}}{\delta
    \phi^* (p)} \s \cdot \frac{\delta \s}{\delta \phi (-p)} +
  \frac{\overrightarrow{\delta}}{\delta \phi^* (p)} \frac{\delta
    \s}{\delta \phi (-p)} \right) \Op \\
&& \quad + \frac{\delta \s}{\delta \phi (-p)} \cdot
  \frac{\overrightarrow{\delta}}{\delta \phi^* (p)} \Op +
  \frac{\overrightarrow{\delta}}{\delta \phi^* (p)} \s \cdot
  \frac{\delta \Op}{\delta \phi (-p)} +
  \frac{\overrightarrow{\delta}}{\delta \phi^* (p)} \frac{\delta
    \Op}{\delta \phi (-p)} \Bigg]
\end{eqnarray*}

Since
\[
K(p/\Lambda) \frac{\delta \s}{\delta \phi (-p)} 
\]
is a composite operator, we find that
\[
 K(p/\Lambda) \left( \frac{\overrightarrow{\delta}}{\delta \phi^* (p)}
  \frac{\delta \s}{\delta \phi (-p)} +
  \frac{\overrightarrow{\delta}}{\delta \phi^* (-p)} \s \cdot
  \frac{\delta \s}{\delta \phi (-p)} \right)
\]
is also a composite operator, according to 2.  (Alternatively, apply 1
to the composite operator $\Ld{\phi^* (p)} \s$.) Thus, it is consistent
with ERG to assume that its integral over $p$ vanishes:
\[
\Sigma [\phi,\phi^*] \equiv \int_p K(p/\Lambda) \left(
  \frac{\overrightarrow{\delta}}{\delta \phi^* (p)} \frac{\delta
    S}{\delta \phi (-p)} + \frac{\overrightarrow{\delta}}{\delta
    \phi^* (p)}S \cdot \frac{\delta S}{\delta \phi (-p)} \right) = 0
\]
This is BRST invariance.  If this equation is satisfied at a
particular $\Lambda$, the ERG differential equation guarantees that it
is satisfied at an arbitrary $\Lambda$.  Becchi's BRST invariance has
precisely this form.

Assuming that the above BRST invariance is satisfied, we define
\begin{eqnarray*}
&&\delta_Q \Op \equiv \int_p Y(p)\\
&&= \int_p K(p/\Lambda) \left[ \frac{\delta \s}{\delta \phi
    (-p)} \cdot \frac{\overrightarrow{\delta}}{\delta \phi^* (p)} \Op
  + \frac{\overrightarrow{\delta}}{\delta \phi^* (p)} \s \cdot
  \frac{\delta \Op}{\delta \phi (-p)} +
  \frac{\overrightarrow{\delta}}{\delta \phi^* (p)} \frac{\delta
    \Op}{\delta \phi (-p)} \right]
\end{eqnarray*}
Hence, given a composite operator $\Op$, $\delta_Q \Op$ defines a
composite operator.  We define $\delta_Q$ by the above, even for $\Op$
which is not a composite operator.  It is straightforward to show that
$\delta_Q$ is nilpotent:
\[
\ffbox{\delta_Q \delta_Q = 0}\qquad \textrm{assuming }\Sigma = 0
\]
as long as the BRST invariance holds.

\no \textbf{HW\#13}: Show the nilpotency.  ($\Sigma = 0$ is necessary.)

Suppose $\Sigma \ne 0$ so that $\delta_Q$ is not necessarily
nilpotent.  We still get an algebraic constraint on $\Sigma$:
\[
\ffbox{\delta_Q \Sigma [\phi,\phi^*] = 0}\qquad \textrm{without
  assuming }\Sigma = 0
\]
\no \textbf{HW\#14}: Derive this.

Now, Becchi's BRST invariance can be interpreted as the invariance
(including the functional measure) of $\s$ under the infinitesimal
field transformation:
\[
\delta \phi (p) \equiv \delta_Q \phi (p) = K(p/\Lambda) \cdot
\underbrace{\frac{\overrightarrow{\delta}}{\delta \phi^*(-p)}
  \s}_{\textrm{composite operator}}
\]
The nilpotency $\delta_Q^2 = 0$, guaranteed by the BRST invariance,
implies
\[
\delta_Q (\delta \phi (p)) = 0
\]
\no \textbf{HW\#15}: Calculate this explicitly and show that $\Sigma = 0$
is necessary for the equality.

Let us now consider generalizing the above results for a theory with
many fields $\phi_i$ either bosonic or fermionic.  Let $\phi_i^*$ be
the corresponding source with the opposite statistics.  The BRST
invariance of the action is given by
\[
\Sigma [\phi, \phi^*]
\equiv \e^{-\s} \int_q \sum_i \Ld{\phi_i^* (-q)} \Ld{\phi_i (q)}
\e^{\s} = 0
\]
Since 
\[
Y(p) = \e^{- \s} \K{p} \sum_i \Ld{\phi_i^* (-p)} \Ld{\phi_i (p)}
\left( \e^{\s} \Op \right)
\]
we arrive at the definition of $\delta_Q$ given by
\begin{eqnarray*}
\delta_Q \Op &\equiv& \int_p \K{p} \Bigg(
\Ld{\phi_i^* (-p)} \s \cdot \Ld{\phi_i (p)} \Op
+ \Ld{\phi_i (p)} \s \cdot \Ld{\phi_i^* (-p)} \Op\\
&& \qquad\qquad + \Ld{\phi_i^* (-p)} \Ld{\phi_i (p)} \Op \Bigg)
\end{eqnarray*}
Especially, we obtain
\[
\delta_Q \phi _i (p) = \K{p} \Ld{\phi_i^* (-p)} \s
\]
so that the BRST invariance is written as the quantum invariance of
the action under the BRST transformation:
\[
\Sigma [\phi,\phi^*]
\equiv \int_q \left( \delta_Q \phi_i (q) \cdot \Ld{\phi_i (q)} \s
+ \Ld{\phi_i (q)} \delta_Q \phi_i (q) \right) = 0
\]

Finally, in the case of QED, the BRST invariance has been already
given at the beginning of this subsection.  We have used
\[
\Ld{A_\mu^* (-k)} \frac{\delta \s}{\delta A_\mu (k)} = 0,\quad
\frac{\delta}{\delta \bar{c}^* (k)} \Ld{\bar{c} (-k)} \s = 0
\]
which we will verify in the next subsection.  
$\delta_Q$ is defined as follows:
\begin{eqnarray*}
&&\delta_Q \Op
\equiv \int_q \K{q} \Bigg[\\
&& \,\frac{\delta \s}{\delta A_\mu (-q)} \cdot \Ld{A_\mu^* (q)} \Op
+ \Ld{A_\mu^* (q)} \s \cdot \frac{\delta}{\delta A_\mu (-q)} \Op
+ \Ld{A_\mu^* (q)} \frac{\delta \Op}{\delta A_\mu (-q)}\\
&&\,+ \Ld{\bar{c} (-q)} \s \cdot \frac{\delta}{\delta \bar{c}^* (q)} \Op
+ \frac{\delta \s}{\delta \bar{c}^* (q)}  \Ld{\bar{c}(-q)} \Op
+ \frac{\delta}{\delta \bar{c}^* (k)} \Ld{\bar{c}(-k)} \Op\\
&&\,- (-)^F \Tr \left( \frac{\delta \Op}{\delta \psi^* (-q)}  \s
    \rd{q} + \Op \rd{q} \frac{\delta \s}{\delta \psi^* (-q)} 
 + \frac{\delta \Op}{\delta \psi^* (-q)} \rd{q}\right)\\
&&\,+ \Tr \left( \ld{q} \s \cdot \Op \Rd{\bar{\psi}^* (q)} + 
\ld{q} \Op \cdot \s \Rd{\bar{\psi}^* (q)} + \ld{q} \Op
\Rd{\bar{\psi}^* (q)} \right) \Bigg]
\end{eqnarray*}
where $(-)^F = \pm 1$ depending on whether $\Op$ is bosonic or
fermionic.  $\delta_Q$ is nilpotent if $\s$ is BRST invariant.

\subsection{Explicit construction of $\s$ in terms of $\bar{S}$}

The purpose of this subsection is to construct the BRST invariant
action $\s$ in the presence of external sources explicitly in terms of
the BRST invariant action $\bar{S}$ without sources.  Since we must
find
\[
\left\lbrace\begin{array}{c@{~=~}l}
 \Ld{A_\mu^* (-k)} \s & k_\mu c(k)\\
\frac{\delta}{\delta \bar{c}^* (k)} \s & - \frac{1}{\xi} k_\mu
\mathcal{A}_\mu (-k)\\
\frac{\delta}{\delta \psi^* (-q)}\s & e \int_k c(k) \Psi (q-k)\\
\frac{\delta}{\delta \bar{\psi}^* (q)}\s& - e \int_k c(k) \bar{\Psi}
(-q-k)
\end{array}\right.
\]
at the vanishing sources, we obtain
\begin{eqnarray*}
\s - \bar{S} &\simeq& \int_k \left[
A_\mu^* (-k) k_\mu c (k) + \bar{c}^* (k) \frac{-1}{\xi} k_\mu
\mathcal{A}_\mu (-k) \right]\\
&& \, + e \int_{q,k} \left[ \psi^* (-q) c (k) \Psi (q-k)
-  c(k) \bar{\Psi} (-q-k) \bar{\psi}^* (q) \right]
\end{eqnarray*}
to first order in external sources.  The question is how to add terms
higher order in external sources so that $\s$ satisfies both ERG and
BRST invariance.

As a preparation we consider a simpler problem.  Suppose we have an
action of a scalar field $\phi$ that satisfies the ERG differential
equation:
\[
- \Lambda \frac{\partial S}{\partial \Lambda} = \int_q \frac{\Delta
  (q/\Lambda)}{q^2 + m^2} \left[ \frac{q^2 + m^2}{\K{q}} \phi (q)
    \frac{\delta S}{\delta \phi (q)} 
+ \frac{1}{2} \left\lbrace \frac{\delta S}{\delta \phi (q)}
    \frac{\delta S}{\delta \phi (-q)} + \frac{\delta^2 S}{\delta \phi
      (q) \delta \phi (-q)} \right\rbrace\right]
\]
We wish to introduce an external source $J(-q)$ that couples to the
scalar field $\phi (q)$.  Since $\frac{\delta S}{\delta J(-q)}$ at
$J=0$ is a composite operator with respect to $S$, we find that the
action $S [\phi, J]$ in the presence of the source is given by
\[
S [\phi, J] - \underbrace{S [\phi, 0]}_{= S[\phi]} \simeq \int_q J(-q)
\Phi (q)
\]
to first order in $J$, where $\Phi$ is the composite operator
\[
\Phi (q) \equiv \phi (q) + \frac{1 - \K{q}}{q^2 + m^2} \frac{\delta
  \Si}{\delta \phi (-q)}
\]
in the absence of $J$.  It is not hard to guess the full source
dependence of $S[\phi, J]$:
\begin{eqnarray*}
S[\phi, J] &=& \Sf [\phi] + \int_q J(-q) \phi (q) + \frac{1}{2} \int_q
J(-q) \frac{1 - \K{q}}{q^2 + m^2} J(q)\\
&&\, + \Si \left[\phi (q) \to \phi (q) + J(q) \frac{1 - \K{q}}{q^2 +
      m^2}\right] 
\end{eqnarray*}
To first order in $J$, this reproduces $S[\phi, J]$ given above.

Let us verify that $S[\phi, J]$ satisfies the ERG differential
equation.  We obtain
\begin{eqnarray*}
&& - \Lambda \frac{\partial}{\partial \Lambda} \left( S[\phi,J] - \Sf
    [\phi] \right)\\
&& = \frac{1}{2} \int_q J(-q) \frac{\Delta (q/\Lambda)}{q^2 + m^2} J(q)
 + \int_q J(q) \frac{\Delta (q/\Lambda)}{q^2 + m^2} \frac{\delta
  \Si}{\delta \phi (q)}\\
&&\quad + \int_q \frac{\Delta (q/\Lambda)}{q^2 + m^2} \frac{1}{2}
\left\lbrace
\frac{\delta \Si}{\delta \phi (q)} \frac{\delta \Si}{\delta \phi (-q)}
+ \frac{\delta^2 \Si}{\delta \phi (q) \delta \phi (-q)} \right\rbrace
\end{eqnarray*}
Since
\[
\left\lbrace
\begin{array}{c@{~=~}l}
\frac{\delta}{\delta \phi (q)} \left( S[\phi, J] - \Sf [\phi]\right)
& J(-q) + \frac{\delta \Si}{\delta \phi (q)}\\
\frac{\delta^2}{\delta \phi (q) \delta \phi (-q)} \left( S[\phi, J] -
    \Sf [\phi]\right) & \frac{\delta^2 \Si}{\delta \phi (q) \delta
  \phi (-q)}
\end{array}\right.
\]
we obtain
\[
- \Lambda \frac{\partial \Si [\phi, J]}{\partial \Lambda}
= \int_q \frac{\Delta (q/\Lambda)}{q^2 + m^2} \frac{1}{2}
\left\lbrace \frac{\delta \Si [\phi,J]}{\delta \phi (q)}
\frac{\delta \Si [\phi, J]}{\delta \phi (-q)} +
\frac{\delta^2 \Si [\phi, J]}{\delta \phi (q) \delta \phi (-q)}
\right\rbrace 
\]
where
\[
\Si [\phi, J] \equiv S [\phi, J] - \Sf [\phi]
\]
Thus, $S [\phi, J]$ satisfies the ERG differential equation.

The external source $J (q)$ is a dimension $3$ parameter of the
theory, which happens to be momentum dependent.  Hence, $J$ enters
into the asymptotic behavior of the action only for the one-point
vertex:
\[
\frac{\delta \Si [\phi, J]}{\delta \phi (-q)}\Big|_0 \to J (q)
\]
Thus, the ERG differential equation determines the $J$-dependence of
the action unambiguously.  In terms of Feynman graphs, $\int_q J(-q)
\phi (q)$ gives a one-point vertex
\begin{center}
\epsfig{file=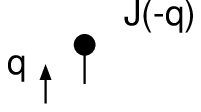}
\end{center}
This vertex can be connected to another vertex only
through a high-momentum propagator $(1-\K{q})/(q^2 + m^2)$.  Multiple
one-point vertices can be connected to an original vertex.  Hence, a
typical vertex in the presence of sources looks like the left figure
below. 
\begin{center}
\epsfig{file=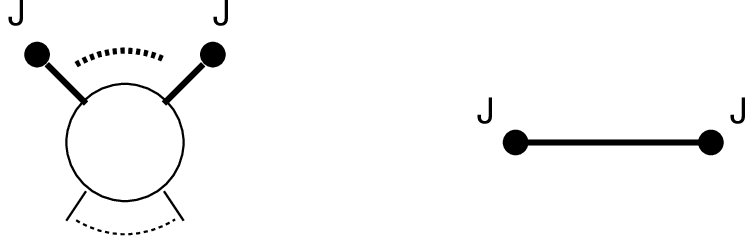}
\end{center}
The only exception is two one-point vertices connected to each other
by a high-momentum propagator.  (the right figure above)

Now, let us apply the above result to QED.  We first tabulate the
fields and their external sources:
\begin{center}
\begin{tabular}{ccc}
\hline
ext. source& coupled field & dimensions of the source\\
\hline
$k_\mu A_\mu^* (-k)$& $c(k)$& $3$\\
$- \frac{1}{\xi} k_\mu \bar{c}^* (k)$& $A_\mu (-k)$& $3$\\
$e \int_k \psi^* (-q-k) c(k)$ & $\psi (q)$& $5/2$\\
$e \int_k c(k) \bar{\psi}^* (q-k)$& $\bar{\psi} (-q)$& $5/2$\\
\hline
\end{tabular}
\end{center}
where we treat the free field $c$ as an external field.  Then, the
full action that satisfies the ERG differential equation in the
presence of the external sources is given by
\begin{eqnarray*}
\s &=& \bar{S}_{\mathrm{free}} + \int_k \left[
k_\mu A_\mu^* (-k) c (k) + \frac{-1}{\xi} k_\mu \bar{c}^* (k) 
A_\mu (-k) \right]\\
&& \, + e \int_{q,k} \left[ \psi^* (-q-k) c (k) \psi (q)
+  \bar{\psi} (-q) c(k) \bar{\psi}^* (q-k) \right]\\
&&\quad - \frac{1}{2 \xi} \int_k \left( 1 - \K{k} \right) \bar{c}^* (-k)
\bar{c}^* (k) \\
&& \quad + e^2 \int_{q,k,l} \psi^* (-q-k) c(k) \frac{1 -
  \K{q}}{\fmslash{q} + i m} c(l) \bar{\psi}^* (q-l)\\
&&\quad + \Si \Big[ A_\mu (k) \to A_\mu (k) - \frac{1 - \K{k}}{k^2} k_\mu
\bar{c}^* (k),\\
&& \qquad \psi (q) \to \psi (q) + \frac{1 - \K{q}}{\fmslash{q}+im} e
\int_k c(k) \bar{\psi}^* (q-k),\\
&& \qquad \bar{\psi} (-q) \to \bar{\psi} (-q) + e \int_k \psi^* (-q-k)
c(k) \frac{1 - \K{q}}{\fmslash{q} + i m}
\Big]
\end{eqnarray*}
We note
\[
\Ld{A_\mu^* (-k)} \s = k_\mu c(k),\quad
\Ld{\bar{c} (-k)} \s = - k^2 c(k)
\]
Hence,
\[
\int_k \K{k} \Ld{A_\mu^* (-k)} \frac{\delta \s}{\delta A_\mu (k)}
= \int_k \K{k} \frac{\delta}{\delta \bar{c}^* (k)} \Ld{\bar{c} (-k)}
\s = 0
\]
The antifield dependence of the action for QED was first derived in
\cite{iis}, but their result corresponds to a different choice of the
cutoff dependence of the external sources.  The above expression has
been obtained independently in \cite{hik}, where its equivalence with
the action of \cite{iis} has also been shown.

\no \textbf{HW\#16}: Verify the BRST invariance $\Sigma = 0$
explicitly using the above action.  (straightforward but tedious)

\newpage
\section{Application to YM theories}

We now consider YM theories.  We first set our notations straight.  We
denote the gauge fields as $A_\mu^a$ and the Faddeev-Popov ghosts by
$c^a, \bar{c}^a$.  We keep the gauge group general.  In terms of the
totally antisymmetric structure constant $f^{abc}$, the covariant
derivatives are given by
\[
\left\lbrace\begin{array}{c@{~\equiv~}l}
 F^a_{\mu\nu} & \partial_\mu A_\nu^a - \partial_\nu A_\mu^a - g
 f^{abc} A_\mu^b A_\nu^c\\
(D_\mu c)^a & \partial_\mu c^a - g f^{abc} A_\mu^b c^c
\end{array}\right.
\]
Hence, for the Fourier transforms, we obtain
\[
\left\lbrace\begin{array}{c@{~=~}l}
 F_{\mu\nu}^a (k) & i k_\mu A_\nu^a (k) - i k_\nu A_\mu^a (k) - g
 f^{abc} \int_l A_\mu^b (k-l) A_\nu^c (l)\\
 (D_\mu c)^a (k) & i k_\mu c^a (k) - g f^{abc} \int_l A_\mu^b (k-l)
 c^c (l)
\end{array}\right.
\]

The classical action
\begin{eqnarray*}
\s_{\mathrm{classical}} &\equiv& \int d^4 x\, \Bigg[ - \frac{1}{4}
F_{\mu\nu}^a F_{\mu\nu}^a - \frac{1}{2 \xi} \left( \partial_\mu
    A_\mu^a \right)^2 - \partial_\mu \bar{c}^a \left( D_\mu c \right)^a \\
&&\, + A_\mu^{a *} \frac{1}{i} \left(D_\mu c\right)^a 
+ \bar{c}^{a *} \frac{1}{\xi}
\frac{1}{i} \partial_\mu A_\mu^a + c^{a *} \frac{g}{2 i} f^{abc} c^b c^c
\Bigg]
\end{eqnarray*}
has the BRST invariance for which the cutoff function is taken as
$1$.  The classical BRST transformation is given by
\[
\left\lbrace\begin{array}{c@{~=~}c@{~=~}l}
 \delta_\ep A_\mu^a  & \ep \Ld{A_\mu^{a *}}
 \s_{\mathrm{classical}} & \ep \frac{1}{i} \left( D_\mu c\right)^a \\
 \delta_\ep \bar{c}^a & \ep \frac{\delta}{\delta \bar{c}^{a *}}
 \s_{\mathrm{classical}} &
 \frac{1}{\xi} \ep \frac{1}{i} \partial \cdot A^a\\
\delta_\ep c^a & \ep \frac{\delta}{\delta c^{a *}}
\s_{\mathrm{classical}}  & \ep \frac{1}{2
  i} g f^{abc} c^b c^c
\end{array}\right.
\]

\subsection{YM theories without sources}

We first summarize the characteristic properties of the YM theories in
the absence of sources.  The correlation functions in the continuum
limit satisfy two important relations:
\begin{enumerate}
\item \textbf{Ward identities}
\begin{eqnarray*}
&&\frac{1}{\xi} k_\mu \vev{A_\mu^a (-k) A_{\nu_1}^{a_1} (k_1) \cdots
  A_{\nu_n}^{a_n} (k_n)}_\infty\\
&& = \sum_{i=1}^n \vev{ A_{\nu_1}^{a_1} (k_1) \cdots
\frac{1}{i} \left( D_{\nu_i} c \right)^{a_i} (k_i) \cdots
  A_{\nu_n}^{a_n} (k_n) \bar{c}^a (-k)}_\infty
\end{eqnarray*}
where $D_\mu c$ is the composite operator for the covariant
derivative. \\
\no (na\"ive proof)  We apply the BRST transformation to
\[
\vev{A_{\nu_1}^{a_1} (k_1) \cdots A_{\nu_n}^{a_n} (k_n) \bar{c}^a
  (-k)}_\infty
\]
Since the action is invariant, we obtain
\begin{eqnarray*}
&&\sum_{i=1}^n \vev{A_{\nu_1}^{a_1} (k_1) \cdots \frac{1}{i} (D_{\nu_i}
  c)^{a_i} (k_i) \cdots \bar{c}^a (-k)}_\infty\\
&&\, + \vev{A_{\nu_1}^{a_1} (k_1) \cdots A_{\nu_n}^{a_n} (k_n)
  \frac{1}{\xi} (-k_\mu) A_\mu^a (-k)}_\infty = 0
\end{eqnarray*}
\item \textbf{The ghost equations of motion} --- The covariant
    derivative $D_\mu c$ must satisfy
\[
\left\lbrace
\begin{array}{l}
k_\mu \vev{ \frac{1}{i} \left( D_\mu c \right)^a (k) \bar{c}^b
  (-k)}_\infty = \delta^{ab}\\
k_\mu \vev{ \frac{1}{i} \left( D_\mu c \right)^a (k)
  A_{\nu_1}^{a_1} (k_1) \cdots A_{\nu_{n-1}}^{a_{n-1}} (k_{n-1}) 
\bar{c}^{a_n} (k_n)}_\infty = 0\quad (n > 1)
\end{array}\right.
\]
\no (na\"ive proof) Since 
\[
\Ld{\bar{c}^a (-k)} S = - k_\mu \frac{1}{i} (D_\mu c)^a (k)
\]
we obtain the above equations of motion.
\end{enumerate}

\subsection{YM theories with sources}

The cutoff action is given in the familiar form:
\[
\s = \Sf + \si
\]
where the free action is given by
\[
\Sf = - \int_k \left[ \frac{1}{2}
A_\mu^a (k) A_\nu^a (-k) \frac{k^2 \delta_{\mu\nu} -
    \left(1 - \frac{1}{\xi}\right) k_\mu k_\nu }{\K{k}} 
+ \bar{c}^a (-k) c^a (k) \frac{k^2}{\K{k}} \right]
\]

Both the Ward identities and ghost equations of motion come from the
invariance of the action $\s$:
\begin{enumerate}
\item \textbf{BRST invariance}
\begin{eqnarray*}
\Sigma &\equiv& \int_k \K{k} \Bigg[ \Ld{A_\mu^{a *} (-k)} \s
\cdot \frac{\delta \s}{\delta A_\mu^a (k)} +
\Ld{A_\mu^{a *} (-k)} \frac{\delta \s}{\delta A_\mu^a
  (k)}\\
&&\quad + \frac{\delta \s}{\delta \bar{c}^{a *} (k)} \Ld{\bar{c}^a
  (-k)} \s + \frac{\delta}{\delta \bar{c}^{a *} (k)}\Ld{\bar{c}^a
  (-k)} \s\\
&&\quad - \frac{\delta \s}{\delta c^{a *} (-k)} \cdot \s \Rd{c^a (k)}
- \frac{\delta \s}{\delta c^{a *} (-k)} \Rd{c^a (k)} \Bigg]
= 0
\end{eqnarray*}
This is the ``quantum'' invariance of the action under the BRST
transformation:
\[
\left\lbrace\begin{array}{c@{~=~}l}
\delta_\ep A_\mu^a (k) & \K{k} \ep \Ld{A_\mu^{a *} (-k)} \s\\
\delta_\ep \bar{c}^a (-k) & \K{k} \ep \frac{\delta \s}{\delta \bar{c}^{a *}
  (k)}\\
\delta_\ep c^a (k) & \K{k} \ep \frac{\delta \s}{\delta c^{a*} (-k)}
\end{array}\right.
\] 
At the vanishing sources, we obtain
\[
\left\lbrace\begin{array}{c@{~=~}l}
\Ld{A_\mu^{a *} (-k)} \s& \frac{1}{i} \left[ (D_\mu c)^a \right]
= k_\mu \mathcal{C}^a (k) + \ac{\mu}^a (k)\\
\frac{\delta \s}{\delta \bar{c}^{a *}(k)} & - \frac{1}{\xi} k_\mu
\mathcal{A}_\mu^a (-k)\\
\frac{\delta \s}{\delta c^{a*} (-k)}& \cc^a (k)
\end{array}\right.
\]
where $\ac{\mu}^a$ is the composite operator corresponding to
\[
i g f^{abc}  A_\mu^b c^c 
\]
and $\cc$ corresponds to
\[
\frac{g}{2 i} f^{abc}  c^b c^c
\]
$\mathcal{C}^a$ and $\mathcal{A}_\mu^a$ are the composite operators
corresponding to the elementary fields $c^a, A_\mu^a$:
\[
\left\lbrace\begin{array}{c@{~\equiv~}l}
\mathcal{C}^a (k) & c^a (k) + \frac{1 - \K{k}}{k^2} \Ld{\bar{c}^a
  (-k)} \si\\
\mathcal{A}_\mu^a (k) & A_\mu^a (k) + \frac{1 - \K{k}}{k^2} \left(
    \delta_{\mu\nu} - (1-\xi) \frac{k_\mu k_\nu}{k^2} \right)
\frac{\delta \si}{\delta A_\nu^a (-k)}
\end{array}\right.
\]
\item \textbf{The ghost equation of motion}
\[
k_\mu \Ld{A_\mu^{a *} (-k)} \s = - \K{k} \Ld{\bar{c}^a (-k)} \s
\]
This is an equality between two composite operators. 
\end{enumerate}

The ghost equation implies that the dependence of $\s$ on $A_\mu^{a*}$
and $\bar{c}^a$ is only through the linear combination
\[
A_\mu^{a *} (-k) - \frac{1}{\K{k}} k_\mu \bar{c}^a (-k)
\]
Using this we can simplify the BRST invariance as follows:
\begin{eqnarray*}
\Sigma &\equiv& \int_k \K{k} \Bigg[ \Ld{A_\mu^{a *} (-k)} \s
\cdot \left( \frac{\delta \s}{\delta A_\mu^a (k)} -
\frac{1}{\K{k}} k_\mu \frac{\delta \s}{\delta \bar{c}^{a*} (k)}
\right)\\
&&\quad + \Ld{A_\mu^{a*} (-k)}  \left( \frac{\delta \s}{\delta A_\mu^a (k)} -
\frac{1}{\K{k}} k_\mu \frac{\delta \s}{\delta \bar{c}^{a*} (k)}
\right)\\
&&\quad - \frac{\delta \s}{\delta c^{a *} (-k)} \cdot \s \Rd{c^a (k)}
- \frac{\delta \s}{\delta c^{a *} (-k)} \Rd{c^a (k)} \Bigg]
= 0
\end{eqnarray*}

As in QED, the source $- \frac{1}{\xi} k_\mu \bar{c}^{a*} (k)$ couples
to the elementary field $A_\mu^a (-k)$.  Hence, the dependence of the
action on $\bar{c}^{a*}$ is obtained as a shift of $A_\mu^a$, and the
action can be written as
\begin{eqnarray*}
\s &=& - \frac{1}{2}\int_k A_\mu^a (k) A_\nu^a (-k) \frac{k^2
  \delta_{\mu\nu} - \left(1 - \frac{1}{\xi}\right) k_\mu k_\nu}{\K{k}}\\
&&\quad + \int_k  \left( A_\mu^{a*} (-k) - \frac{1}{\K{k}}
    k_\mu \bar{c}^a (-k) \right) k_\mu c^a (k)\\
&&\, - \frac{1}{\xi} \int_k k_\mu \bar{c}^{a*} (k) A_\mu^a (-k)
- \frac{1}{2 \xi} \int_k \left( 1 - \K{k} \right) \bar{c}^{a*} (-k)
\bar{c}^{a*} (k) \\
&&\, + \si \Bigg[ A_\mu^a (k) - \frac{1 - \K{k}}{k^2}
    k_\mu \bar{c}^{a*} (k),  A_\mu^{a*} (-k) - \frac{1}{\K{k}}
    k_\mu \bar{c}^a (-k),\\
&&\qquad\qquad c^a (k), c^{a*} (-k) \Bigg]
\end{eqnarray*}
Therefore, we obtain
\begin{eqnarray*}
k_\mu \frac{\delta \s}{\delta \bar{c}^{a*} (k)}\Big|_{\bar{c}*=0} &=&
- \frac{1}{\xi} k_\mu k_\nu A_\nu^a (-k) - \frac{1 - \K{k}}{k^2} k_\mu
k_\nu \frac{\delta \si}{\delta A_\nu^a (k)}\Big|_{\bar{c}*=0}\\
k_\mu \frac{\delta \s}{\delta A_\mu^a (k)}\Big|_{\bar{c}^*=0} &=&
- \frac{1}{\xi} \frac{1}{\K{k}} k^2 k_\nu A_\nu^a (-k) + k_\mu
\frac{\delta \si}{\delta A_\mu^a (k)}\Big|_{\bar{c}^*=0}
\end{eqnarray*}
Hence, we find
\[
k_\mu \frac{\delta \s}{\delta \bar{c}^{a*} (k)}\Big|_{\bar{c}*=0}
= - \frac{1-\K{k}}{k^2} k_\mu k_\nu \frac{\delta \s}{\delta A_\nu^a
  (k)}\Big|_{\bar{c}^*=0}  - \frac{1}{\xi} \frac{1}{\K{k}} k_\mu k_\nu
A_\nu^a (-k)
\]
Thus, we obtain
\begin{eqnarray*}
&&\frac{\delta \s}{\delta A_\mu^a (k)}\Big|_{\bar{c}*=0} -
\frac{1}{\K{k}} k_\mu 
\frac{\delta \s}{\delta \bar{c}^{a*} (k)}\Big|_{\bar{c}*=0}\\
&& = \left( \delta_{\mu\nu} + \frac{1-\K{k}}{\K{k}} \frac{k_\mu
      k_\nu}{k^2} \right) \frac{\delta \s}{\delta A_\nu^a
  (k)}\Big|_{\bar{c}*=0} + \frac{1}{\xi} \frac{1}{\K{k}^2} k_\mu k_\nu
A_\nu^a (-k)\\
&& = \left( \delta_{\mu\nu} + \frac{1-\K{k}}{\K{k}} \frac{k_\mu
      k_\nu}{k^2} \right) \frac{\delta}{\delta A_\nu^a (k)}
\left( \s + \frac{1}{2 \xi} \int_k \frac{k_\mu k_\nu}{\K{k}} A_\mu^a
    (k) A_\nu^a (-k) \right)\Big|_{\bar{c}^*=0}
\end{eqnarray*}

Therefore, taking $\bar{c} = 0$ and $\bar{c}^*=0$, the BRST invariance
gives
\begin{eqnarray*}
&&\Sigma \equiv \int_k \K{k} \Bigg[
\Ld{A_\mu^{a*} (-k)} \s' \cdot
P_{\mu\nu} (k) \frac{\delta \s'}{\delta A_\nu^a (k)}
 + \Ld{A_\mu^{a*} (-k)} P_{\mu\nu} (k) \frac{\delta \s'}{\delta
  A_\nu^a (k)} \\
&& \quad - \frac{\delta \s'}{\delta c^{a*} (-k)} \cdot \s' \Rd{c^a
  (k)} - \frac{\delta \s'}{\delta c^{a*} (-k)} \Rd{c^a
  (k)} \Bigg] = 0
\end{eqnarray*}
where we have defined
\[
P_{\mu\nu} (k) \equiv  \delta_{\mu\nu} + \frac{1-\K{k}}{\K{k}}
\frac{k_\mu k_\nu}{k^2}
\]
and
\begin{eqnarray*}
\s' &\equiv& \s_{\bar{c}^* = 0} + \frac{1}{2 \xi} \int_k \frac{k_\mu
  k_\nu}{\K{k}} A_\mu^a (k) A_\nu^a (-k) \\
&=&  - \frac{1}{2}\int_k A_\mu^a (k) A_\nu^a (-k) \frac{k^2
  \delta_{\mu\nu} - k_\mu k_\nu}{\K{k}}
 + \int_k  A_\mu^{a*} (-k) k_\mu c^a (k)\\
&&\, + \si \Bigg[ A_\mu^a (k),  A_\mu^{a*} (-k),c^a (k), c^{a*} (-k) \Bigg]
\end{eqnarray*}
Once we obtain $\s'$, we can construct $\s$ for the non-vanishing
$\bar{c}, \bar{c}^*$, since we know how the action $\s$ depends on
$\bar{c}$ and $\bar{c}^*$.  We will write $\s'$ as $\s$ from now on.

Our task is now to obtain $\s$ that satisfies the BRST invariance
given above.  From now on, we only have $A_\mu^a, A_\mu^{a*}$ and
$c^a, c^{a*}$ as fields and sources.  Let us define $\delta_Q$ by
\begin{eqnarray*}
\delta_Q &\equiv& \int_k \K{k} \Bigg[
 P_{\mu\nu} (k) \frac{\delta \s}{\delta A_\nu^a (k)} \cdot
 \Ld{A_\mu^{a*} (k)} \\
&&\, + \Ld{A_\mu^{a*} (k)} \s \cdot 
    P_{\mu\nu} (k) \frac{\delta}{\delta
      A_\nu^a (k)}
 + \Ld{A_\mu^{a*} (k)} P_{\mu\nu} (k)
\frac{\delta}{\delta A_\nu^a (k)}\\
&&  - \s \Rd{c^a (k)} \cdot \frac{\delta}{\delta c^{a*} (-k)}
+ \frac{\delta \s}{\delta c^{a*} (-k)} \cdot \Ld{c^a (k)} +
\frac{\delta }{\delta c^{a*} (-k)} \Ld{c^a (k)} \Bigg] = 0
\end{eqnarray*}
This is nilpotent under the assumption $\Sigma = 0$.\\
\no\textbf{HW\#17}: Show this is the case.  (Hint: In general
$\delta_Q$ defined by
\[
\delta_Q \Op \equiv \e^{- \s} \int_q f_{ij} (q) \sum_i \Ld{\phi_i (q)}
\Ld{\phi_j^* (-q)} \left( \e^{\s} \Op \right)
\]
is nilpotent if it is fermionic.)

By the definition of $\delta_Q$ and $\Sigma$, we obtain the algebraic
constraint:
\begin{eqnarray*}
\delta_Q \Sigma &\equiv& \int_k \K{k} \Bigg[
 P_{\mu\nu} (k) \frac{\delta \s}{\delta A_\nu^a (k)} \cdot
 \Ld{A_\mu^{a*} (k)} \Sigma \\
&&\, + \Ld{A_\mu^{a*} (k)} \s \cdot 
    P_{\mu\nu} (k) \frac{\delta}{\delta
      A_\nu^a (k)} \Sigma
 + \Ld{A_\mu^{a*} (k)} P_{\mu\nu} (k) \Sigma
\frac{\delta}{\delta A_\nu^a (k)}\\
&&  - \s \Rd{c^a (k)} \cdot \frac{\delta \Sigma}{\delta c^{a*} (-k)}
- \frac{\delta \s}{\delta c^{a*} (-k)} \cdot \Sigma \Rd{c^a (k)} -
\frac{\delta }{\delta c^{a*} (-k)} \Sigma \Rd{c^a (k)} \Bigg] = 0
\end{eqnarray*}

Now, our task is to construct the loop expansion
\[
\s = \sum_{l=0}^\infty \s_l
\]
Assuming that the BRST invariance holds up to $(l-1)$-loop:
\[
\Sigma_0 = \cdots = \Sigma_{l-1} = 0
\]
we wish to construct $\s_l$ by fine tuning the parameters of the
theory so that
\[
\Sigma_l = 0
\]
is satisfied.  The question is if we have enough parameters to satisfy
this condition.  To answer this, we need the help of the algebraic
constraint $(\delta_Q \Sigma)_l = 0$.

At $l$-loop, the algebraic constraint gives
\begin{eqnarray*}
(\delta_Q \Sigma)_l &\equiv&\int_k \K{k} \Bigg[
P_{\mu\nu} (k) \frac{\delta \s_0}{\delta A_\nu^a (k)}\cdot
\Ld{A_\mu^{a*} (-k)} \Sigma_l \\
&&\, + \Ld{A_\mu^{a*} (-k)} \s_0 \cdot P_{\mu\nu} (k)
\frac{\delta \Sigma_l}{\delta A_\nu^a (k)}\\
&&  - \s_0 \Rd{c^a (k)} \cdot \frac{\delta \Sigma_l}{\delta c^{a*} (-k)}
- \frac{\delta \s_0}{\delta c^{a*} (-k)} \cdot \Sigma_l \Rd{c^a (k)}\Bigg] = 0
\end{eqnarray*}
We note that the second order differentials in the integrand do not
contribute since $\Sigma_{l-1} = 0$.  (If $\Sigma_{l-1} \ne 0$, it
would have contributed to $(\delta_Q \Sigma)_l$,
since the integral over $q$ gives an extra loop.)  Clearly, the
algebraic constraint restricts the possible form of $\Sigma_l$.

As a preparation, let us tabulate the basic properties of the
fields\footnote{We do not need to consider $\bar{c}^a$, since the
  dependence on $\bar{c}^a$ is determined by that on $A_\mu^*$.}:
\begin{center}
\begin{tabular}{cccc}
\hline
field& statistics& dimension& ghost number\\
\hline
$A_\mu^a$& b& $1$& $0$\\
$c^a$& f& $1$& $1$\\
$A_\mu^{a*}$& f& $2$& $-1$\\
$c^{a*}$& b& $2$& $-2$\\
\hline
\end{tabular}
\end{center}
From now on, we examine the case of SU(2) for which the structure
constant is the totally antisymmetric tensor: 
\[
f^{abc} = \ep^{abc}
\]

Now, let us look at the asymptotic behavior of $\s_l$ at a large
cutoff.  Using the coordinate space notation for convenience, the
asymptotic behavior of $\s_l$ is given by
\begin{eqnarray*}
\s_l &\to& \int d^4 x\, \Bigg[ \, a_1 \frac{1}{2} (\partial_\mu
A_\nu^a)^2 + a_2 \frac{1}{2} (\partial_\mu A_\mu^a)^2 + a_3 g
\ep^{abc} \partial_\mu A_\nu^a A_\mu^b A_\nu^c\\
&&\, + a_4 \frac{g^2}{4} A_\mu^a A_\mu^a A_\nu^b A_\nu^b + a_5
\frac{g^2}{4} A_\mu^a A_\nu^a A_\mu^b A_\nu^b \\
&&\, + a_6 A_\mu^{a*} \partial_\mu c^a + a_7  g \ep^{abc} A_\mu^{a*}
A_\mu^b c^c + a_8 \frac{g}{2} \ep^{abc} c^{a*} c^b c^c \,\Bigg]
\end{eqnarray*}
where we have ignored the terms proportional to $\Lambda^2$ since they
are completely determined by the ERG differential equation.  The eight
coefficients $a_1,\cdots,a_8$ are functions of $\ln \Lambda/\mu$ whose
values at $\Lambda = \mu$ are free to choose.  The above $\s_l$ is the
most general bosonic expression with zero ghost number.  Especially,
for $l=0$, we choose the following asymptotic form
\begin{eqnarray*}
\s_0 &\to& \int d^4 x \, \Bigg[\,
\frac{1}{2} (\partial_\mu A_\nu^a)^2 - \frac{1}{2} (\partial_\mu A_\mu^a)^2\\
&&\, + g \ep^{abc} \partial_\mu A_\nu^a A_\mu^b A_\nu^c
+ \frac{g^2}{4} \left( A_\mu^a A_\mu^a A_\nu^b A_\nu^b - A_\mu^a
    A_\nu^a A_\mu^b A_\nu^b \right) \\
&&\, + A_\mu^{a*} \frac{1}{i} \left( \partial_\mu c^a - g \ep^{abc}
    A_\mu^b c^c \right) + c^{a*} \frac{g}{2i} \ep^{abc} c^b c^c
\,\Bigg]
\end{eqnarray*}
as the starting point of the perturbative construction.

We next consider the asymptotic behavior of $\Sigma_l$.  This is a
dimension $5$ operator with ghost number $1$.  There
are eleven terms:
\begin{eqnarray*}
\Sigma_l &\to& \int d^4 x\, \Bigg[\,
s_1 c^a \partial^2 \partial_\mu A_\mu^a + s_2 g \ep^{abc} c^a
\partial^2 A_\mu^b \cdot A_\mu^c + s_3 g \ep^{abc} c^a \partial_\mu
\partial_\nu A_\nu^b \cdot A_\mu^c \\
&&\, + s_4 g^2  c^a \partial_\mu A_\mu^a \cdot A_\nu^b A_\nu^b + s_5
g^2 c^a \partial_\mu A_\nu^a \cdot A_\mu^b A_\nu^b + s_6 g^2 c^a
A_\mu^a \partial_\nu A_\nu^b A_\mu^b \\
&&\, + s_7 g^2 c^a A_\mu^a \partial_\nu
A_\mu^b A_\nu^b + s_8 g^2 c^a A_\mu^a \partial_\mu A_\nu^b A_\nu^b\\
&&\,+ s_{9} g \ep^{abc}
c^a c^b \partial_\mu A_\mu^{c*} + s_{10} g \ep^{abc} \partial_\mu c^a
c^b A_\mu^{c*} + s_{11} g^2 c^a c^b A_\mu^{a*} A_\mu^a \,\Bigg]
\end{eqnarray*}
At first sight we may not be able to make this vanish using only eight
parameters.  But the eleven parameters are not all independent, since
$(\delta_Q \Sigma)_l$ must vanish.  Looking at the asymptotic behavior
of $(\delta_Q \Sigma)_l$, we obtain
\begin{eqnarray*}
(\delta_Q \Sigma)_l &\to& \int d^4 x \, \Bigg[\,
\left( \frac{\delta \s_0}{\delta A_\mu^a} - i \partial_\mu
    \frac{\delta \s_0}{\delta \bar{c}^{a*}} \right) \Ld{A_\mu^{a*}}
\Sigma_l
 + \Ld{A_\mu^{a*}} \s_0 \cdot \left( \frac{\delta}{\delta A_\mu^a}
    - i \partial_\mu \frac{\delta}{\delta \bar{c}^{a*}}\right)
\Sigma_l\\
&&\quad - \s_0 \Rd{c^a} \frac{\delta \Sigma_l}{\delta c^{a*}} -
\frac{\delta \s_0}{\delta c^{a*}} \cdot \Sigma_l \Rd{c^a} \,\Bigg] = 0
\end{eqnarray*}
This constrains the $s$ parameters.  Becchi has shown the possibility
of satisfying the BRST invariance in his Parma lectures.\cite{becchi}
I hope to provide more details in an updated version of the present
notes. ;-)

\newpage
\subsection*{Appendix: Introduction of auxiliary fields $B^a$}

Instead of using the gauge fixing term
\[
- \frac{1}{2 \xi} \int_k A_\mu^a (k) A_\nu^a (-k) \frac{1}{\K{k}}
k_\mu k_\nu
\]
we can gauge fix the theory using auxiliary fields $B^a$ coupled to
$\frac{1}{\xi} k_\mu A_\mu^a$.  $B^a$ do not transform under the BRST,
and we do not introduce their antifields.  The full action can now be
written in the following form:
\begin{eqnarray*}
\s &=& - \frac{1}{2} \int_k A_\mu^a (k) A_\nu^a (-k) \frac{1}{\K{k}}
\left( k^2 \delta_{\mu\nu} - k_\mu k_\nu \right)\\
&& \, + \int_k \frac{1}{\K{k}} B^a (-k) k_\mu A_\mu^a (k) - \frac{\xi}{2}
\int_k \frac{1}{\K{k}} B^a (-k) B^a (k)\\
&&\, + \int_k \left( A_\mu^{a*} (-k) - \frac{1}{\K{k}} k_\mu \bar{c}^a
    (-k) \right) k_\mu c^a (k)
+  \int_k \bar{c}^{a*} (k) B^a (-k) \\
&&\, + \si \Bigg[ A_\mu^a (k) - \frac{1 - \K{k}}{k^2}
    k_\mu \bar{c}^{a*} (k),  A_\mu^{a*} (-k) - \frac{1}{\K{k}}
    k_\mu \bar{c}^a (-k),\\
&&\qquad\qquad c^a (k), c^{a*} (-k) \Bigg]
\end{eqnarray*}
Substituting this into the BRST invariance $\Sigma = 0$ ($\Sigma$ is
defined the same way as before since there is no antifield for $B$),
and setting $\bar{c}=\bar{c}^*=B=0$, we find that $\s$ satisfies the
same BRST invariance as $\s'$ that we have obtained without using $B$.
Thus, there is nothing to gain or lose by introducing the auxiliary
fields.

\newpage
\section{Non-perturbative aspects of the Wilson ERG}

So far our discussions of ERG have been restricted to perturbation
theory.  Actually, Ken Wilson introduced ERG\footnote{He called it
  simply RG.}  to define the continuum limits in quantum field theory
non-perturbatively.  In this final part of the lectures we wish to
discuss the non-perturbative aspects of the ERG.\footnote{We will
  review the materials explained in sects.~11 and 12 of the
  Wilson-Kogut lecture notes.\cite{wk}} We will first introduce a
general framework for ERG differential equations to relate Wilson's
ERG differential equation to Polchinski's.  We then apply ERG to
define the non-trivial continuum limit of the scalar theory in $D=3$.
 
\subsection{Generalized diffusion equation}

We consider a real scalar theory in $D$ dimensional euclidean space.
Given an action $S[\phi]$, we construct a new action $S_t [\phi]$ by
the following integral transformation:
\begin{eqnarray*}
&&\exp \left[ S_t [\phi] \right]\\
&&\equiv \int [d\phi'] \exp\left[ - \frac{1}{2} \int_p \left( A_t (p)
        \phi (p) - B_t (p) \phi' (p) \right) \left( A_t (p) \phi (-p)
        - B_t (p) \phi' (-p) \right) + S [\phi'] \right]\\
&&= \int [d\phi'] \exp \left[ - \frac{1}{2} \int_p A_t (p)^2 \left(
        \phi (p) - \frac{B_t (p)}{A_t (p)} \phi' (p) \right)
\left( \phi (-p) - \frac{B_t (p)}{A_t (p)} \phi' (-p) \right) +
S[\phi'] \right]
\end{eqnarray*}
where we assume both $A_t$ and $B_t$ are functions of $p^2$.
The physical meaning of this transformation is clear: $\phi (p)$ is
diffused around 
\[
\frac{B_t (p)}{A_t (p)} \phi' (p)
\]
with the width $\frac{1}{A_t (p)}$.  The distribution of $\phi'$ is
given in terms of the weight $\e^{S[\phi']}$.  This is analogous to
the solution of the diffusion equation
\[
\partial_t P (t,x) = D \partial_x^2 P (t,x)
\]
for the distribution function $P(t,x)$ in the coordinate space at time
$t$.  The solution is given by the integral formula
\[
P(t,x) = \sqrt{\frac{4 D t}{\pi}} \int_{-\infty}^\infty dy\, \exp \left[
    - \frac{1}{4 D t} (x-y)^2 \right] \, P(0,y)
\]

We now wish to derive the differential equation for $S_t [\phi]$
analogous to the above diffusion equation.  For this, we change
variables to rewrite
\begin{eqnarray*}
&&\exp \left[ S_t [\phi]\right]\\
&& = \int [d\phi'] \exp \left[ - \frac{1}{2} \int_p \phi' (p) \phi'
    (-p) + S\left[ \frac{1}{B_t (p)} \phi' (p) + \frac{A_t (p)}{B_t
          (p)} \phi (p)\right] \right]
\end{eqnarray*}
Differentiating this with respect to $t$, we obtain
\begin{eqnarray*}
&&\partial_t S_t [\phi] \exp [ S_t [\phi] ]\\
&& = \int_p \int [d\phi'] \exp \left[ - \frac{1}{2} \int_p \phi' (p)
    \phi' (-p) \right] 
 \left( \partial_t \frac{1}{B_t (p)}\cdot \phi' (p) + \partial_t
    \frac{A_t (p)}{B_t (p)} \phi (p) \right) \\
&& \qquad\qquad \times \frac{B_t (p)}{A_t (p)}
\frac{\delta}{\delta \phi (p)} \exp \left[ S \left[ \frac{1}{B_t (p)}
        \phi' (p) + \frac{A_t (p)}{B_t (p)} \phi (p)\right] \right]\\
&& = \int_p \frac{B_t (p)}{A_t (p)} \partial_t \frac{1}{B_t (p)} \cdot
\frac{\delta}{\delta \phi (p)} \int [d\phi'] \phi' (p) \exp  \left[ -
    \frac{1}{2} \int_p \phi' (p) \phi' (-p) + S \right]\\
&&\quad + \int_p \partial_t \ln \frac{A_t(p)}{B_t(p)} \cdot \phi (p)
\frac{\delta S_t [\phi]}{\delta \phi (p)} \exp \left[ S_t [\phi] \right]
\end{eqnarray*}
Since
\[
\int [d \phi'] \frac{\delta}{\delta \phi' (-p)} \exp \left[ -
    \frac{1}{2} \int_p \phi' (p) 
    \phi' (-p) + S \left[\frac{1}{B_t (p)}
        \phi' (p) + \frac{A_t (p)}{B_t (p)} \phi (p)\right] \right]= 0
\]
we obtain
\[
 \int [d\phi'] \phi' (p) \exp \left[ - \frac{1}{2} \int_p \phi'
    \phi' + S \right]
= \frac{1}{A_t (p)} \frac{\delta S_t [\phi]}{\delta \phi (-p)}
\exp \left[ S_t [\phi] \right]
\]
Hence, we obtain
\begin{eqnarray*}
\partial_t S_t [\phi] &=& \int_p \Bigg[
\frac{B_t (p)}{A_t (p)^2}\partial_t \frac{1}{B_t (p)} \cdot 
\left\lbrace \frac{\delta S_t [\phi]}{\delta \phi (p)} 
\frac{\delta S_t [\phi]}{\delta \phi (-p)} + \frac{\delta^2 S_t
  [\phi]}{\delta \phi (p) \delta \phi (-p)} \right\rbrace\\
&&\quad  + 
\partial_t \ln \frac{A_t (p)}{B_t (p)} \cdot \phi (p) \frac{\delta S_t
  [\phi]}{\delta \phi (p)} \Bigg]
\end{eqnarray*}
Defining
\[
\left\lbrace
\begin{array}{c@{~\equiv~}l}
F_t (p) & \partial_t \ln \frac{A_t (p)}{B_t (p)}\\
G_t (p) & 2 \frac{B_t (p)}{A_t (p)^2}\partial_t \frac{1}{B_t (p)} =
 - 2 \left\lbrace - \partial_t \ln \frac{A_t (p)}{B_t (p)} +
    \frac{\partial_t A_t (p)}{A_t (p)} \right\rbrace \frac{1}{A_t
  (p)^2} \end{array}\right.
\]
we can write
\begin{eqnarray*}
\partial_t S_t &=& \int_p \Bigg[ \,F_t (p)
    \cdot \phi(p) \frac{\delta S_t}{\delta \phi (p)}\\
&&\quad + G_t (p) 
\cdot \frac{1}{2} \left\lbrace \frac{\delta S_t }{\delta \phi (p)}
    \frac{\delta S_t}{\delta \phi (-p)} + \frac{\delta^2 S_t}{\delta
      \phi (p) \delta \phi (-p)} \right\rbrace \Bigg]
\end{eqnarray*}
This has the familiar form of the ERG differential equation.

We now define the generating functionals of connected correlation functions:
\[
\left\lbrace\begin{array}{c@{~=~}l}
\e^{W[J]} & \int [d\phi] \exp \left[ S[\phi] + i \int_p J(p) \phi (-p)\right]\\
\e^{W_t [J]} & \int [d\phi] \exp \left[ S_t[\phi] + i \int_p J(p)
    \phi (-p) \right]\end{array}\right.
\]
We can compute $W_t [J]$ in terms of $W [J]$ as follows:
\begin{eqnarray*}
\e^{W_t [J]} &=& \int [d\phi] \e^{ i \int_p J(p) \phi (-p) }
\int [d\phi'] \exp \Bigg[ \\
 &&  - \frac{1}{2}
\int_p \left(A_t (p) \phi (p) - B_t (p) \phi' (p)\right)
\left(A_t (p) \phi (-p) - B_t (p) \phi' (-p) \right)
+ S[\phi'] \Bigg]\\
&=& \int [d\phi] \int [d\phi'] \exp \Bigg[ i \int_p 
\frac{J(p)}{A_t (p)}
  \left( A_t (p) \phi (-p) - B_t (p) \phi' (-p) \right)\\
&&\,  - \frac{1}{2} \int_p \left(A_t (p) \phi (p) - B_t (p) \phi' (p)\right)
\left(A_t (p) \phi (-p) - B_t (p) \phi' (-p) \right)\\
&& + S[\phi'] + i \int_p J(p) \frac{B_t (p)}{A_t (p)} \phi' (-p)
\Bigg]\\
&=& \int [d\phi] \exp \left[ - \frac{1}{2} \int_p \phi (p) \phi (-p) +
    i \int_p \frac{J(p)}{A_t (p)} \phi (-p) \right]\\
&& \times \int [d\phi'] \exp \left[ S[\phi'] + i \int_p J(p) \frac{B_t
      (p)}{A_t (p)} \phi' (-p) \right] \\
&=& \exp \left[ - \frac{1}{2} \int_p \frac{1}{A_t (p)^2} J(p) J(-p) +
    W \left[ \frac{B_t (p)}{A_t (p)} J(p) \right]\right]
\end{eqnarray*}
This implies
\[
\left\lbrace\begin{array}{c@{~=~}l}
\vev{\phi (p) \phi (-p)}_{S_t} & \frac{1}{A_t (p)^2} + \frac{B_t
  (p)^2}{A_t (p)^2} \vev{\phi (p) \phi (-p)}_S\\
\vev{\phi (p_1) \cdots \phi (p_n)}^{\mathrm{connected}}_{S_t} & \prod_{i=1}^n
\frac{B_t (p_i)}{A_t (p_i)} \cdot \vev{\phi (p_1) \cdots \phi
  (p_n)}^{conn.}_{S} \quad (n > 1)\end{array}\right.
\]

In all the examples we will discuss below, we choose
\[
\ffbox{F_t (p) \equiv \partial_t \ln \frac{A_t (p)}{B_t (p)}
= \frac{\Delta (p\e^t)}{K(p\e^t)} - \frac{\eta (t)}{2}}
\]
where $\Lambda = \mu \e^{-t}$ is the cutoff scale, and we have chosen
$\mu = 1$ for convenience.  Hence,
\[
\ffbox{\frac{A_t (p)}{B_t (p)} = \frac{K(p)}{K(p\e^t)} \exp 
\left[ - \frac{1}{2} \int_0^t dt'\, \eta (t') \right]}
\]
Note we have chosen
\[
\frac{A_0 (p)}{B_0 (p)} = 1
\]
For
\[
S_0 = S
\]
we must also choose
\[
\frac{1}{A_0 (p)} = 0
\]

With our choice for $F_t (p)$, we obtain
\begin{eqnarray*}
G_t (p) &=& - 2 \left\lbrace - F_t (p) + \partial_t \ln A_t (p)
\right\rbrace \frac{1}{A_t (p)^2}\\
&=& - 2 \left\lbrace - \frac{\Delta (p \e^t)}{K(p \e^t)} +
    \frac{\eta}{2} + \partial_t \ln A_t (p) \right\rbrace \frac{1}{A_t
  (p)^2}\\
&=& - 2 \partial_t \ln \left( A_t (p) \e^{\int^t \frac{\eta}{2}}
    \frac{K(p \e^t)}{K(p)} \right) \frac{1}{A_t (p)^2}
\end{eqnarray*}
Hence, defining
\[
\bar{A}_t (p) \equiv  A_t (p) \e^{\int^t \frac{\eta}{2}}
    \frac{K(p \e^t)}{K(p)} 
\]
we obtain
\[
G_t (p) = - 2 \e^{\int^t \eta} \frac{K(p\e^t)^2}{K(p)^2}
\frac{\partial_t \ln \bar{A}_t (p)}{\bar{A}_t (p)^2} 
\]
Hence,
\[
\partial_t \frac{1}{\bar{A}_t (p)^2} = \e^{- \int^t \eta}
\frac{K(p)^2}{K(p \e^t)^2} G_t (p)
\]

Let us now look at two examples of $G_t (p)$ and calculate the
corresponding $A_t (p)$.

\subsection*{Example 1 (Wilson)}

We choose
\[
G_t (p)  = 2 \e^{2t} \left( \frac{\Delta (p\e^t)}{K(p\e^t)} + 1 -
        \frac{1}{2} \eta (t) \right)
\]
This gives
\begin{eqnarray*}
\partial_t \frac{1}{\bar{A}_t (p)^2} &=& 2 \e^{2 t - \int^t \eta}
\frac{K(p)^2}{K(p\e^t)^2} \left( \frac{\Delta (p\e^t)}{K(p\e^t)} + 1 -
    \frac{\eta}{2} \right)\\
&=& \partial_t \left( \e^{2 t - \int^t \eta} \frac{K(p)^2}{K(p\e^t)^2} \right)
\end{eqnarray*}
Hence, we obtain
\[
\frac{1}{\bar{A}_t (p)^2} = \e^{2 t - \int^t \eta}
\frac{K(p)^2}{K(p\e^t)^2} - 1
\]
using the initial condition $1/A_0 = 0$.  Thus, we obtain
\[
\frac{1}{A_t (p)^2} = \e^{2t} - \frac{K(p\e^t)^2}{K(p)^2}\, \e^{\int_0^t
  \eta}
\]
This implies
\[
\left\lbrace\begin{array}{c@{~=~}l}
\vev{\phi (p)\phi (-p)}_{S_t} - \e^{2t} &
\e^{\int_0^t \eta}\, \frac{K(p\e^t)^2}{K(p)^2} \left(
\vev{\phi (p)\phi (-p)}_S - 1 \right)\\
\vev{\phi (p_1) \cdots \phi (p_{2n})}_{S_t} &
\e^{n \int_0^t \eta}\, \prod_{i=1}^{2n} \frac{K(p_i\e^t)}{K(p_i)}
\cdot \vev{\phi (p_1) \cdots \phi (p_{2n})}_{S}
\end{array}\right.
\]

This is the original choice of Wilson discussed in sect.~11 of
\cite{wk}.  In this choice, the two-point function behaves as
\[
\vev{\phi (p) \phi (-p)}_{S_t} \stackrel{t \to
  \infty}{\longrightarrow} \e^{2 t}
\]

\subsection*{Example 2 (Polchinski)}

We choose
\[
G_t (p) = \frac{1}{p^2 + m^2 (t)} \left\lbrace \Delta (p\e^t) - \left( \eta
        (t) + \frac{b_m (t)}{p^2 + m^2 (t)} \right) K(p\e^t) \left( 1
        - K(p\e^t)\right)\right\rbrace
\]
where
\[
b_m (t) \equiv \frac{d m^2 (t)}{dt}
\]
This gives the Polchinski ERG differential equation if $\eta = 0$ and
$m^2$ is $t$-independent.

Let us find the corresponding $A_t (p)$.  We find
\begin{eqnarray*}
\partial_t \frac{1}{\bar{A}_t (p)^2} &=& \e^{- \int^t \eta}
\frac{K(p)^2}{K(p\e^t)^2} \\
&& \,\times \frac{1}{p^2 + m^2 (t)} \left\lbrace \Delta (p\e^t) - \left( \eta
        (t) + \frac{b_m (t)}{p^2 + m^2 (t)} \right) K(p\e^t) \left( 1
        - K(p\e^t)\right)\right\rbrace\\
&=& K(p)^2 \partial_t \left( \frac{\e^{- \int^t \eta}}{p^2 + m^2 (t)}
\left(\frac{1}{K(p\e^t)}-1 \right)  \right)
\end{eqnarray*}
Hence, 
\[
\frac{1}{\bar{A}_t (p)^2} = K(p)^2 \left( \frac{\e^{- \int^t
        \eta}}{p^2 + m^2 (t)} \left(\frac{1}{K(p\e^t)}-1 \right)
- \frac{1}{p^2 + m^2 (0)} \left( \frac{1}{K(p)}-1 \right) \right)
\]
Thus, we obtain
\[
\frac{1}{A_t (p)^2} = K(p\e^t)^2 \left\lbrace \frac{1}{p^2 + m^2 (t)}
    \left( \frac{1}{K(p\e^t)} - 1 \right) - \frac{\e^{\int_0^t
        \eta}}{p^2 + m^2 (0)} \left( \frac{1}{K(p)} - 1 \right)
\right\rbrace
\]
Hence, we obtain
\begin{eqnarray*}
&&\vev{\phi (p) \phi (-p)}_{S_t} - \frac{1}{p^2 + m^2 (t)} K(p\e^t)
\left( 1 - K(p\e^t)\right)\\
&& = \e^{\int_0^t \eta} \,\frac{K(p\e^t)^2}{K(p)^2} \left(
\vev{\phi (p)\phi (-p)}_S - \frac{1}{p^2 + m^2 (0)} K(p) (1 -
K(p))\right)
\end{eqnarray*}
and
\[
\vev{\phi (p_1) \cdots \phi (p_{2n})}_{S_t}
= \e^{n \int_0^t \eta} \prod_{i=1}^{2n} \frac{K(p_i
  \e^t)}{K(p_i)}\cdot
\vev{\phi (p_1) \cdots \phi (p_{2n})}_{S}
\]
With this choice of $G_t (p)$, we obtain the characteristic asymptotic
behavior
\[
\vev{\phi (p) \phi (-p)}_{S_t} \stackrel{t \to
  \infty}{\longrightarrow} \frac{K(p\e^t)}{p^2 + m^2 (t)}
\]

Before concluding this subsection, let us consider the relation
between two different formalisms of ERG: one given by $A_t, B_t$ and
another given by $C_t, D_t$.  Let
\begin{eqnarray*}
&&\exp [S_t [\phi]]\\
&&\, \equiv \int [d\phi'] \exp \left[ - \frac{1}{2} (A_t (p) \phi (p) -
    B_t \phi' (p))(A_t (p) \phi (-p) - B_t (p) \phi' (-p)) + S[\phi']
\right]\\
&&\exp [S'_t [\phi]]\\
&&\, \equiv \int [d\phi'] \exp \left[ - \frac{1}{2} (C_t (p) \phi (p) -
    D_t \phi' (p))(C_t (p) \phi (-p) - D_t (p) \phi' (-p)) + S[\phi']
\right]
\end{eqnarray*}
We wish to derive the relation between $S_t$ and $S'_t$.  To derive
the relation, we recall the following results on the generating
functional:
\[
\left\lbrace\begin{array}{c@{~=~}l}
\e^{W_t [J]} & \exp \left[ W \left[\frac{B_t (p)}{A_t (p)}
        J(p)\right] - \frac{1}{2} \int_p \frac{1}{A_t (p)^2} J(p)
    J(-p) \right]\\
\e^{W'_t [J]} & \exp \left[ W \left[\frac{D_t (p)}{C_t (p)}
        J(p)\right] - \frac{1}{2} \int_p \frac{1}{C_t (p)^2} J(p)
    J(-p) \right]\end{array}\right.
\]
Hence, we obtain
\[
\e^{W'_t [J]} = \exp \left[ W_t \left[ 
        \frac{A_t D_t}{B_t C_t} J \right] - \frac{1}{2} \int_p J(p)
    J(-p) \frac{1}{C_t (p)^2} \left( 1 - \frac{D_t (p)^2}{B_t (p)^2}
    \right) \right]
\]
This implies
\begin{eqnarray*}
&&\e^{S'_t [\phi]} = \int [d\phi'] \exp \Bigg[ S_t [\phi']\\
&& - \frac{1}{2} \int_p
    \frac{1}{\frac{1}{C_t (p)^2} - \frac{1}{A_t (p)^2} R_t (p)^2}
\left( \phi (p) - R_t (p) \phi' (p) \right)
\left( \phi (-p) - R_t (p) \phi' (-p)\right) \Bigg]
\end{eqnarray*}
where
\[
R_t (p) \equiv \frac{A_t (p) D_t (p)}{B_t (p) C_t (p)}
\]

Let us apply this result to the two examples given above.  The first
example is given by
\[
\left\lbrace\begin{array}{c@{~=~}l}
\frac{1}{A_t (p)^2} & \e^{2t} - \frac{K (p \e^t)^2}{K(p)^2}
\e^{\int_0^t dt' \eta_W (t')}\\
\frac{A_t (p)}{B_t (p)} & \frac{K(p)}{K(p \e^t)} \e^{- \frac{1}{2}
  \int_0^t dt' \eta_W (t')} 
\end{array}\right.
\]
and the second by
\[
\left\lbrace\begin{array}{c@{~=~}l} 
\frac{1}{C_t (p)^2} & K(p\e^t)^2
        \left\lbrace \frac{1}{p^2 + m^2 (t)} \left( \frac{1}{K(p\e^t)}
                - 1 \right) - \frac{\e^{\int_0^t
                \eta_P}}{p^2 + m^2 (0)} \left( \frac{1}{K(p)} - 1
            \right) \right\rbrace\\ 
        \frac{C_t (p)}{D_t (p)} & \frac{K(p)}{K(p \e^t)} \e^{-
          \frac{1}{2} \int_0^t dt' \eta_P (t')}
\end{array}\right.
\]
Thus, we obtain
\[
R_t (p) = \e^{\frac{1}{2} \int_0^t dt'\, \left( \eta_P (t') - \eta_W
      (t') \right)}
\]

\subsection{Modification of the Polchinski ERG equation}

The second example introduced in the previous subsection suggests a
possibility of generalizing the Polchinski ERG differential equation:
\[
\partial_t S_t = \int_p \frac{\Delta (p \e^t)}{p^2 + m^2} \left(
\frac{p^2 + m^2}{K(p \e^t)} \phi (p) \frac{\delta S_t}{\delta \phi
  (p)}
+ \frac{1}{2} \left\lbrace \frac{\delta S_t}{\delta \phi (p)}
    \frac{\delta S_t}{\delta \phi (-p)} + \frac{\delta^2 S_t}{\delta
      \phi (p) \delta \phi (-p)} \right\rbrace \right)
\]
for which $m^2$ is a constant, and $\eta$ vanishes.

Suppose we have a solution $S_t (m^2)$ of the above ERG differential
equation.  We then modify it infinitesimally by
\[
\delta S_t = \frac{\delta z}{2} \N  
- \delta m^2 \int_p
  \frac{K(p \e^t)(1 -
    K(p \e^t))}{(p^2 + m^2)^2} \frac{1}{2} \left\lbrace \frac{\delta
        S_t}{\delta \phi (p)} \frac{\delta S_t}{\delta \phi (-p)} +
      \frac{\delta^2 S_t}{\delta \phi (p) \delta \phi (-p)}
  \right\rbrace
\]
where
\[
\N \equiv - \int_p \left(
  \phi (p) \frac{\delta S_t}{\delta \phi (p)} + 2 \frac{K(p \e^t)(1 -
    K(p \e^t))}{p^2 + m^2} \frac{1}{2} \left\lbrace
\frac{\delta S_t}{\delta \phi (p)} \frac{\delta S_t}{\delta \phi (-p)}
+ \frac{\delta^2 S_t}{\delta \phi (p) \delta \phi (-p)} \right\rbrace
\right)
\]
is the composite operator changing the normalization of the field, and
the operator proportional to $\delta m^2$ is a type 2 operator
changing only the two-point function.

The modified action $S_t + \delta S_t$ is no longer a solution of the
ERG differential equation, but it is a solution of the ERG
differential equation for the modified squared mass $m^2 + \delta
m^2$.  Moreover, the $t$-independent correlation functions are the
same up to normalization of $\phi$.

\no \textbf{HW\#18}: Show that 
\begin{eqnarray*}
&&\frac{1}{K(p\e^t)^2} \vev{\phi (p) \phi (-p)}_{S_t + \delta S_t} +
\frac{1-1/K(p\e^t)}{p^2 + m^2 + \delta m^2} \\
&& \quad = (1 + \delta z)
\left(\frac{1}{K(p\e^t)^2} \vev{\phi (p) \phi (-p)}_{S_t} +
\frac{1-1/K(p\e^t)}{p^2 + m^2} \right)
\\
&&\prod_{i=1}^{2n} \frac{1}{K(p_i \e^t)} \cdot \vev{\phi (p_1) \cdots
  \phi (p_{2n})}_{S_t + \delta S_t}\\
&&\quad= (1 + n \delta z) \prod_{i=1}^{2n} \frac{1}{K(p_i \e^t)} 
\cdot \vev{\phi (p_1) \cdots \phi (p_{2n})}_{S_t}
\end{eqnarray*}
This proves what is stated in the above paragraph.

We now choose an arbitrary functions $m^2 (t')$.  Let $S_t (m^2 (t'))$
be solutions of the Polchinski ERG differential equation for a fixed
squared mass $m^2 (t')$.  As we compare $S_t (m^2 (t'))$ and $S_t (m^2
(t'+\delta t'))$, we can arrange the solutions so that 
\begin{eqnarray*}
&&\partial_{t'} S_t (m^2 (t'))
= \frac{\eta (t')}{2} \N (t') \\
&& \quad - \frac{d m^2 (t')}{dt'} \int_p
\frac{K(p\e^t) (1 - K(p\e^t))}{(p^2 + m^2 (t'))^2}
\frac{1}{2} \left\lbrace \frac{\delta
        S_t}{\delta \phi (p)} \frac{\delta S_t}{\delta \phi (-p)} +
      \frac{\delta^2 S_t}{\delta \phi (p) \delta \phi (-p)}
  \right\rbrace
\end{eqnarray*}
where $\N (t')$ is the $\N$ operator constructed with $S_t (m^2
(t'))$, and $\eta (t')$ is an arbitrary function.

Finally, if we look at the $t$-dependence of $S_t (m^2 (t))$, where
$t$ and $t'$ are set equal, we obtain the differential equation of the
second example in the previous subsection.
\begin{figure}[t]
\begin{center}
\epsfig{file=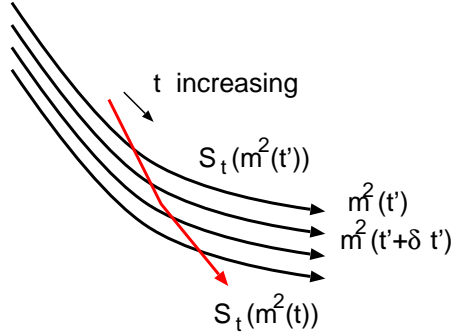}
\caption{Physically equivalent Polchinski ERG flows with different
  values of the squared mass parameter.  $S_t (m^2(t))$ crosses each
  trajectory at a single point.}
\end{center}
\end{figure}

\subsection{Renormalization conditions}

In this subsection we introduce criteria to choose the $t$-dependence
of the squared mass $m^2 (t)$ and the anomalous dimension $\eta (t)$.
The criteria, which amount to renormalization conditions, are by no
means unique.  Let us begin with what is the easiest to understand
physically.

\subsection*{1st version}

We adopt the following two conditions
\[
\left\lbrace\begin{array}{l}
 \V_2 (t;0,0) = 0\\
 \frac{\partial}{\partial p^2} \V_2 (t;p,-p)\Big|_{p^2=0} = 0
\end{array}\right.
\]
The second condition is a condition on the normalization of $\phi$.
Applying the above conditions on the ERG differential equation
\begin{eqnarray*}
\partial_t S_t &=& \int_p \Bigg[ \left(\frac{\Delta (p\e^t)}{K(p \e^t)} -
    \frac{\eta}{2} \right) \phi (p) \frac{\delta S_t}{\delta \phi
  (p)}\\
&& \, + \left( \frac{\Delta (p\e^t)}{p^2 + m^2 (t)} - \left(
        \eta + \frac{b_m (t)}{p^2 + m^2 (t)} \right)
    \frac{K(p\e^t)(1-K(p\e^t))}{p^2 + m^2 (t)} \right)\\
&&\qquad \times \frac{1}{2} \left\lbrace
\frac{\delta S_t}{\delta \phi (p)} \frac{\delta S_t}{\delta \phi (-p)}
+ \frac{\delta^2 S_t}{\delta \phi (p) \delta \phi (-p)} \right\rbrace
\Bigg]
\end{eqnarray*}
we obtain
\begin{eqnarray*}
&& - b_m (t) - \eta (t) m^2 (t)  = \frac{1}{2} \int_q \frac{1}{q^2 +
  m^2 (t)} \Bigg\lbrace \Delta (q \e^t) \\
&&\qquad\qquad - K(q \e^t) (1 - K(q \e^t)) 
\left( \eta (t) + \frac{b_m (t)}{q^2 +
      m^2 (t)} \right) \Bigg\rbrace  \V_4 (t; q,-q,0,0) \\
&& - \eta (t) = \frac{1}{2}  \frac{\partial}{\partial p^2} \int_q
    \frac{1}{q^2+m^2 (t)} \left\lbrace
  \Delta (q \e^t) - K(q \e^t)(1-K(q\e^t))
 \left( \eta (t) + \frac{b_m (t)}{q^2+m^2}\right)\right\rbrace\\
&&\qquad\qquad\qquad\qquad \times
 \V_4 (t; q,-q,p,-p)\Big|_{p^2=0} 
\end{eqnarray*}
Thus, we can determine $b_m (t)$ and $\eta (t)$ in terms of the
four-point vertex.

\subsection*{2nd version}

We introduce a version that is more analogous to the Wilsonian ERG
differential equation which has no squared mass parameter.  We set to
zero the squared mass appearing in the differential equation:
\[
m^2 (t) = 0
\]
Then, the ERG differential equation becomes
\begin{eqnarray*}
&&\partial_t S_t = \int_p \Bigg[ \left( \frac{\Delta
      (p\e^t)}{K(p\e^t)} - \frac{\eta (t)}{2} \right) \phi (p)
\frac{\delta S_t}{\delta \phi (p)}\\
&&\, + \frac{1}{p^2} \left( \Delta (p\e^t) - \eta (t) K(p\e^t) (1 -
    K(p \e^t)) \right) \frac{1}{2} \left\lbrace
\frac{\delta S_t}{\delta \phi (p)} \frac{\delta S_t}{\delta \phi (-p)}
+ \frac{\delta^2 S_t}{\delta \phi (p) \delta \phi (-p)} \right\rbrace
\Bigg]
\end{eqnarray*}
We adopt the normalization condition:
\[
\frac{\partial}{\partial p^2} \V_2 (t; p,-p)\Big|_{p^2 = 0} = 0
\]
This determines $\eta (t)$ as
\[
\eta (t) \equiv \frac{ - \frac{\partial}{\partial p^2} \int_q
  \frac{\Delta (q\e^t)}{q^2} \V_4 (t; q,-q,p,-p)
  \Big|_{p^2=0}}{1 - \frac{\partial}{\partial p^2} \int_q \frac{K(q\e^t) (1 -
    K(q\e^t))}{q^2} \V_4 (t; q,-q,p,-p) \Big|_{p^2=0}}\,,
\]

\subsection*{Wilson ERG}

The Wilson ERG differential equation is given by
\begin{eqnarray*}
\partial_t S_t &=& \int_p \Bigg[
\left\lbrace 
       \left( \frac{\Delta (p\e^t)}{K(p\e^t)} - \frac{\eta(t)}{2} \right) 
\phi (p) \right\rbrace \frac{\delta
  S(t)}{\delta \phi (p)} \\
&& \, + \left( \frac{\Delta (p\e^t)}{K(p\e^t)} + 1 - \frac{\eta (t)}{2} \right)
\left\lbrace
    \frac{\delta S_t}{\delta \phi (p)} \frac{\delta S_t}{\delta \phi
      (-p)} + \frac{\delta^2 S_t}{\delta \phi (p)\delta \phi (-p)}
\right\rbrace \Bigg]
\end{eqnarray*}
We denote the functional derivatives of the action as
\[
u_{2n} (t;p_1,\cdots,p_{2n}) (2\pi)^D \delta^{(D)} (p_1 + \cdots +
p_{2n}) \equiv \frac{\delta^{2n} S_t}{\delta \phi (p_1) \cdots \phi
  (p_{2n})} \Big|_{\phi = 0}
\]
Then, we adopt the normalization condition
\[
\frac{\partial}{\partial p^2} u_2 (t;p,-p) \Big|_{p^2=0} = -1
\]
Applying this to the ERG equation, we obtain
\[
\frac{1}{2} \eta (t) = \frac{2 u_2 (t;0,0)
- \frac{1}{2} \frac{\partial}{\partial p^2} \int_q\,
\left( \frac{\Delta (q\e^t)}{K(q\e^t)} + 1 \right)
  u_4 (t; q,-q,p,-p) \Big|_{p^2=0}}{1 + 2 u_2 (t;0,0) - \frac{1}{2}
  \frac{\partial}{\partial p^2}  \int_q u_4 (t; q,-q,p,-p)
  \Big|_{p^2=0}} 
\]
This depends on both the two- and four-point vertices.

\subsection{Rescaling of space}

Now, in non-perturbative applications of the ERG, it is the fixed
points of the ERG transformation which play the most important roles.
A fixed point is an action $S [\phi]$ that does not transform.
To define a continuum limit, we need a fixed point.  We also need to
know the critical exponents characterizing the fixed point.

The ERG transformations we have formulated so far have no fixed point
for an obvious reason.  The cutoff of the action decreases along the
ERG trajectory.  For ERG to have fixed points, we must introduce
rescaling of space (or momentum) so that the cutoff is fixed under
ERG.  The purpose of this subsection is to understand how rescaling
modifies the ERG differential equations.

The necessary rescaling does not depend on interactions, and it is
enough to study the massless free theory:
\[
S_t = - \frac{1}{2} \int_p \phi (p) \phi (-p) \frac{p^2}{K(p\e^t)}
\]
which has a cutoff $\e^{-t}$.  In order to fix the cutoff at $1$, we
must measure the momentum $p$ in units of the cutoff.  Hence, we must
change $p$ to
\[
p' \equiv \frac{p}{\e^{-t}} = p \e^t
\]
Substituting this into $S_t$, we obtain
\[
S_t = - \frac{1}{2} \int_{p'} \e^{- (D+2) t} \phi (p' \e^{-t}) \phi (-
p' \e^{-t}) \frac{{p'}^2}{K(p')}
\]
We now define
\[
\phi' (p') \equiv \e^{- \frac{D+2}{2} t} \phi (p' \e^{-t})
\]
so that
\[
S_t = - \frac{1}{2} \int_{p'} \phi' (p') \phi' (-p')
\frac{{p'}^2}{K(p')}
\]
becomes independent of $t$.

We now consider a generic theory:
\begin{eqnarray*}
S_t &=& \sum_n \frac{1}{(2n)!} \int_{p_1,\cdots,p_{2n}}
\phi (p_1) \cdots \phi (p_{2n})\, u_{2n} (t; p_1,\cdots,p_{2n}) \\
&&\qquad\qquad \cdot (2\pi)^D \delta^{(D)} (p_1+\cdots+p_{2n})
\end{eqnarray*}
Using $\phi'$ instead of $\phi$, we obtain
\begin{eqnarray*}
S_t &=& \sum_n \frac{1}{(2n)!} \int_{p_1,\cdots,p_{2n}}
\e^{n(D+2)t} \phi' (p_1 \e^t) \cdots \phi' (p_{2n} \e^t)\, u_{2n} (t;
p_1,\cdots,p_{2n}) \\ 
&&\qquad\qquad \cdot (2\pi)^D \delta^{(D)} (p_1+\cdots+p_{2n})
\end{eqnarray*}
Hence, regarding $S_t$ as a functional of $\phi'$, we obtain
\[
\partial_t S_t = \int_{p'} \left\lbrace p'_\mu \frac{\partial \phi'
      (p')}{\partial p'_\mu} + \frac{D+2}{2} \phi' (p') \right\rbrace
\frac{\delta S_t}{\delta \phi' (p')}
\]
This is the modification due to rescaling.

We now note
\[
\frac{\delta}{\delta \phi' (p')} = \e^{- \frac{D-2}{2} t}
\frac{\delta}{\delta \phi (p' \e^{-t})}
\]
\no \textbf{HW\#19}: Derive this result.  (Hint: Apply the above to
$\phi' (q') = \e^{- \frac{D+2}{2} t} \phi (q' \e^{-t})$.)

Hence, the rescaling of space and field modifies the ERG differential
equation to
\begin{eqnarray*}
\partial_t S_t &=&
\int_{p'} \Bigg[ \left\lbrace p'_\mu \frac{\partial \phi'
      (p')}{\partial p'_\mu} 
+ \left( \frac{D+2}{2} + F'_t (p') \right) \phi' (p')\right\rbrace
 \frac{\delta S_t}{\delta \phi' (p')} \\
&&\quad + G'_t (p') \frac{1}{2} \left\lbrace \frac{\delta S_t}{\delta
      \phi' (p')} 
    \frac{\delta S_t}{\delta \phi' (-p')} + \frac{\delta^2 S_t}{\delta
      \phi' (p') \delta \phi' (-p')} \right\rbrace \Bigg]
\end{eqnarray*}
where
\[
F'_t (p') \equiv F_t (p' \e^{-t}) =
\frac{\Delta (p')}{K(p')} - \frac{\eta(t)}{2}
\]
and
\[
G'_t (p') \equiv \e^{- 2t} G_t (p' \e^{-t})
\]

\subsection*{Example 1 (Wilson)}

\[
G'_t (p) = 2 \left( \frac{\Delta (p)}{K(p)} + 1 - \frac{1}{2} \eta (t)
\right)
\]

\subsection*{Example 2 (Polchinski)}

\[
G'_t (p) = \frac{1}{p^2 + \e^{2 t} m^2 (t)} \left\lbrace
\Delta (p) - \left( \eta (t) + \frac{b_m (t)}{p^2 + \e^{2t} m^2 (t)}
\right) K(p) (1 - K(p)) \right\rbrace
\]
where
\[
\frac{d}{dt} \left( \e^{2 t} m^2 (t) \right)
= \left( 2 + b_m (t) \right) \e^{2t} m^2 (t)
\]
From now, we call $\e^{2t} m^2 (t)$ as $m^2 (t)$.  We will also omit
the primes from $\phi'$ and $F_t', G_t'$.

\no \textbf{HW\#20}: Given
\begin{eqnarray*}
S_t [\phi] &=& \sum_{n=1}^\infty \frac{1}{(2n)!}
\int_{p_1,\cdots,p_{2n}} \phi (p_1) \cdots \phi (p_{2n}) \,\V_{2n} (t;
p_1, \cdots, p_{2n})\\
&&\qquad \times (2\pi)^D \delta^{(D)} (p_1 + \cdots + p_{2n})
\end{eqnarray*}
before rescaling, derive
\begin{eqnarray*}
\underbrace{S'_t [\phi']}_{\equiv S_t [\phi]}
 &=& \sum_{n=1}^\infty \frac{1}{(2n)!}
\int_{p_1,\cdots,p_{2n}} \phi' (p_1) \cdots \phi' (p_{2n}) \,\V'_{2n} (t;
p_1, \cdots, p_{2n})\\
&&\qquad \times (2\pi)^D \delta^{(D)} (p_1 + \cdots + p_{2n})
\end{eqnarray*}
where
\[
\V'_{2n} (t; p_1,\cdots, p_{2n}) = \e^{y_{2n} t} \V_{2n} (t; p_1
\e^{-t}, \cdots, p_{2n} \e^{-t} )
\]
and
\[
y_{2n} \equiv D - n (D-2)
\]

\no\textbf{HW\#21}: Show that in the Polchinski formalism $\V_2$ and $\V_4$
satisfy the following ERG differential equations:
\begin{eqnarray*}
&&\partial_t \V_2 (t;p,-p) = \eta (p^2 + m^2) + b_m 
 + \left( 2 - \eta + 2 u (p) - 2 p^2 \frac{\partial}{\partial
      p^2} \right) \V_2 (t;p,-p)\\
&& \qquad+ \V_2 \frac{(\Delta - K u) (p)}{p^2 + m^2} \V_2
 + \frac{1}{2} \int_q \frac{(\Delta - K u)(q)}{q^2+m^2} \V_4 (t;q,-q,p,-p)\\ 
&&\partial_t \V_4 (t;p_1,\cdots,p_4) = \left( (4-D) - 2 \eta 
    + \sum_{i=1}^4
    \left( u(p_i) - p_{i \mu} \frac{\partial}{\partial p_{i\mu}}
    \right)\right) \V_4\\
&& \quad + \V_4 \sum_{i=1}^4 \frac{(\Delta- K u)(p_i)}{p_i^2 + m^2} \V_2
(p_i) + \frac{1}{2} \int_q \frac{(\Delta - K u)(q)}{q^2 + m^2} \V_6
(t;q,-q,p_1,\cdots,p_4) 
\end{eqnarray*}
where
\[
u(t;p) \equiv (1 - K(p)) \left( \eta (t) + \frac{b_m (t)}{p^2 + m^2
      (t)} \right)
\]

\subsection{The Wilson-Fisher fixed point for $D=3$}

For $D=4$, the only fixed point of the ERG is the massless free
theory.  We will construct the fixed point explicitly for any $D$ in
the appendix at the end of this section.  For $D=3$, there is a
non-trivial fixed point, discovered by Wilson and Fisher.

As the initial action at $t=0$, let us consider
\[
S_0 \equiv - \frac{1}{2} \int_p \phi (p) \phi (-p) \frac{p^2 +
  m_0^2}{K(p)} - \frac{\lambda_0}{4!} \int_{p_1,p_2,p_3} \phi (p_1) \phi
(p_2) \phi (p_3) \phi (-p_1-p_2-p_3)
\]
For a fixed $\lambda_0 > 0$, we consider starting the ERG trajectories
with various values of $m_0^2$.  For a generic choice of $m_0^2$, the
theory is massive, and the ERG trajectory leads to an infinitely
massive theory with no correlations since the physical mass increases
exponentially along the ERG trajectory.  However, for a specific value
of $m_0^2 = m_{cr}^2 (\lambda_0)$, the theory is massless, and the ERG
trajectory goes into the Wilson-Fisher fixed point as $t \to \infty$.
\begin{figure}[h]
\begin{center}
\epsfig{file=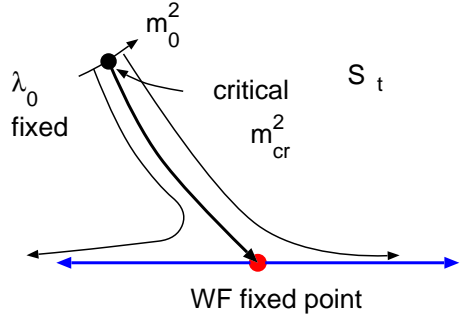}
\caption{At criticality $m_0^2 = m_{cr}^2 (\lambda_0)$, the ERG
  trajectory leads to the Wilson-Fisher fixed point.}
\end{center}
\end{figure}
For $m_0^2 > m_{cr}^2 (\lambda_0)$, the theory has $\mathbf{Z}_2$
invariance, but it is broken spontaneously for $m_0^2 < m_{cr}^2
(\lambda_0)$.

In the modified Polchinski ERG, we have a running squared mass $m^2
(t)$ whose initial value is given by $m^2 (0) = m_0^2$.  We expect
that
\[
m^2 (t) \stackrel{t \to \infty}{\longrightarrow} m^{2*}
\]
for $m_0^2 = m_{cr}^2 (\lambda_0)$.  Here $m^{2*}$ is the squared mass
parameter of the fixed point action.  

\subsection{Perturbative determination of critical exponents}

We finally discuss a perturbative construction of the Wilson-Fisher
fixed point.  Especially we wish to compute the anomalous dimensions
(critical exponents) of the squared mass and $\phi$ using perturbation
theory.

For perturbation theory, the modified Polchinski ERG differential
equation is the most convenient.\footnote{Our scheme is not mass
  independent.  There is a mass independent variation, and perhaps
  that is the simplest for perturbation theory.}
\begin{eqnarray*}
&&\partial_t S_t = \int_p \left\lbrace p_\mu \frac{\partial \phi
      (p)}{\partial p_\mu} + \left( \frac{D+2}{2} - \frac{\eta
      (t)}{2} + \frac{\Delta (p)}{K(p)} \right) \phi (p) \right\rbrace
\frac{\delta S_t}{\delta \phi (p)}\\
&&\quad + \int_p \frac{1}{p^2 + m^2 (t)} \left\lbrace \Delta (p) -
    \left(\eta (t) 
+ \frac{b_m (t)}{p^2 + m^2 (t)}\right) K(p) \left( 1 -
        K(p)\right) \right\rbrace \\
&&\qquad\qquad\qquad\qquad \times \frac{1}{2} \left(
\frac{\delta S_t}{\delta \phi (p)} \frac{\delta S_t}{\delta \phi
  (-p)}
+ \frac{\delta^2 S_t}{\delta \phi (p) \delta \phi (-p)} \right)
\end{eqnarray*}
where $m^2 (t)$ satisfies
\[
\frac{d}{d t} m^2 (t) = 2 m^2 (t) + b_m (t)
\]

The two normalization conditions
\[
\left\lbrace\begin{array}{c@{~=~}l}
 \V_2 (m^2; p,-p)\Big|_{p^2=0} & 0\\
 \frac{\partial}{\partial p^2} \V_2 (m^2; p,-p)\Big|_{p^2=0} & 0
\end{array}\right.
\]
determine $b_m (t)$ and $\eta (t)$ as follows:
\begin{eqnarray*}
&& - b_m (t) - \eta (t) m^2 (t)\\
&&\quad  = \frac{1}{2} \int_q \frac{1}{q^2 + m^2 (t)} \left\lbrace
\Delta (q) - K(q) (1 - K(q)) \left( \eta (t) + \frac{b_m (t)}{q^2 +
      m^2 (t)} \right) \right\rbrace \\
&&\qquad\qquad\qquad\qquad \times \V_4 (t; q,-q,0,0) \\
&& - \eta (t) = \frac{1}{2}  \frac{\partial}{\partial p^2} \int_q
    \frac{1}{q^2+m^2 (t)} \left\lbrace
  \Delta (q) - K(q)(1-K(q))
 \left( \eta (t) + \frac{b_m (t)}{q^2+m^2}\right)\right\rbrace\\
&&\qquad\qquad\qquad\qquad \times
 \V_4 (t;    q,-q,p,-p)\Big|_{p^2=0} 
\end{eqnarray*}

Under the above ERG flow, the correlation functions obtain the
following simple $t$-dependence:
\begin{eqnarray*}
&&\frac{1}{K(p \e^t)^2} \left( \vev{\phi (p \e^t) \phi (- p\e^t)}_{S(t)}
    - \frac{K(p \e^t)}{p^2 \e^{2t} + m^2 (t)} \right) 
+ \frac{1}{p^2 \e^{2t} + m^2 (t)}\\
&& \quad= \exp \left[ - 2 t + \int^t dt'\, \eta (t') \right]
\underbrace{\vev{\phi (p) \phi (-p)}}_{\textrm{$t$-independent}}\\
&& \prod_{i=1}^{n > 2} \frac{1}{K (p_i \e^t)} \cdot \vev{\phi (p_1
  \e^t) \cdots \phi (p_n \e^t)}_{S (t)}\\
&&\quad = \exp \left[ t \left( D - n \frac{D+2}{2} \right) + \frac{n}{2}
    \int^t dt'\, \eta (t') \right] \cdot \underbrace{\vev{\phi (p_1)
    \cdots \phi (p_n)}}_{\textrm{$t$-independent}}
\end{eqnarray*}
This is valid for any $D$, and we take $D=3$ here.

We wish to find the infrared (IR) fixed point using perturbation
theory.  It is convenient (and perhaps essential) to impose the UV
renormalizability of the theory so that the theory can be
characterized by only two parameters. If we solve the ERG differential
equation using an initial condition such as given in the previous
section, we need to keep track of every aspect of the action.  It
would be inconvenient for perturbative calculations.

The scalar theory in $D=3$ is a truly renormalizable theory, and UV
renormalizability amounts to the asymptotic condition
\[
S_t \stackrel{t \to - \infty}{\longrightarrow} S_G^* \equiv -
\frac{1}{2} \int_p \phi (p) \phi (-p) \frac{p^2}{K(p)}
\]
where $S_G^*$ is the gaussian fixed point theory, corresponding to the
free massless theory.

The UV renormalizability familiar to us is formulated in terms of the
action before rescaling of space: the six- and higher-point vertices
should vanish asymptotically as $t \to - \infty$.  Hence, using the
result of HW\#20, we obtain
\[
\e^{(n-3) t} \V_{2n} (t; p_1 \e^t, \cdots, p_{2n} \e^t )
\stackrel{t \to - \infty}{\longrightarrow} 0
\]
for $2n \ge 6$.  

We can now parameterize the theory with $m^2 (t)$ and the running
coupling constant
\[
\lambda (t) \equiv - \V_4 (t; 0,0,0,0)
\]
These two parameters specify the UV asymptotic behavior of the action
unambiguously, and hence the entire action is specified.
Thus, we can write
\[
S_t = S (m^2 (t), \lambda (t))
\]
The ERG differential equation gives
\[
\frac{d \lambda (t)}{dt} = \lambda (t) + \beta (t)
\]
where $\beta (t)$ is defined by
\begin{eqnarray*}
&&- \beta (t) - 2 \lambda (t) \cdot \eta (t) \\
&& = \frac{1}{2} \int_q \frac{1}{q^2 +
  m^2 (t)} \left\lbrace \Delta (q) - K(q)(1-K(q)) \left( \eta (t) +
        \frac{b_m (t)}{q^2 + m^2 (t)} \right) \right\rbrace \\
&&\qquad\qquad\qquad\qquad \times \V_6 (t; q,-q,0,0,0,0)
\end{eqnarray*}

Since the action is completely characterized by the two parameters
$m^2 (t), \lambda (t)$, the ERG differential equation reduces to the
two ordinary differential equations\footnote{In a mass independent
  scheme, we will have $b_m (m^2, \lambda) = m^2 \beta_m (\lambda)$,
  and $\beta$ depends only on $\lambda$.}:
\[
\left\lbrace\begin{array}{c@{~=~}l}
\frac{d m^2}{dt} & 2 m^2 + b_m (m^2, \lambda)\\
\frac{d \lambda}{dt} & \lambda + \beta (m^2, \lambda)
\end{array}\right.
\]
The fixed point $(m^{2*}, \lambda^*)$ is determined by the two
conditions:
\[
\left\lbrace\begin{array}{l@{~=~0}}
 2 m^{2*} + b_m (m^{2*}, \lambda*)\\
\lambda^* + \beta (m^{2*}, \lambda*)
\end{array}\right.
\]
\begin{figure}[h]
\begin{center}
\epsfig{file=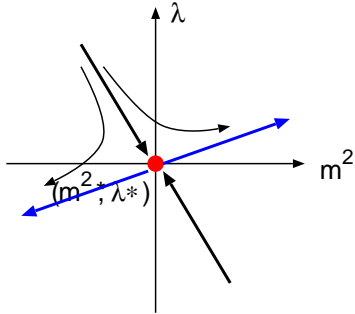}
\caption{ERG flows in the two-dimensional parameter space}
\end{center}
\end{figure}
To get the anomalous dimension of the squared mass, we linearize the
RG equations:
\[
\begin{array}{c@{~=~}l}
\frac{d}{dt} \left( m^2 - m^{2*} \right) & \left( 2 + \frac{\partial
      b_m (m^2,\lambda)}{\partial m^2}\Big|^* \right) (m^2 - m^{2*}) +
+ \frac{\partial b_m (m^2, \lambda)}{\partial \lambda}\Big|^* (\lambda
- \lambda^*)\\
\frac{d}{dt} \left( \lambda - \lambda^* \right) &
\frac{\partial \beta (m^2, \lambda)}{\partial m^2}\Big|^* (m^2 -
m^{2*}) + \frac{\partial \beta (m^2, \lambda)}{\partial
  \lambda}\Big|^* (\lambda - \lambda^*)
\end{array}
\]
This has two eigenvalues, one positive, and the other negative.  The
positive eigenvalue, $2 + \beta_m^*$, gives the scale dimension of the
relevant parameter (conjugate to something like $\phi^2$), and the
negative eigenvalue gives that of the least irrelevant parameter
(conjugate to $\phi^4$).  We call $\beta_m^*$ the \textbf{anomalous
  dimension} of the squared mass parameter.

Let us now calculate $b_m, \eta, \beta$, which are all functions of
$m^2, \lambda$.  At 1-loop, we obtain
\[
\left\lbrace\begin{array}{c@{~=~}l}
b_m (m^2, \lambda) & \frac{\lambda}{2} \int_q \frac{\Delta (q)}{q^2
  + m^2}\\
\beta (m^2, \lambda) & - 3 \lambda^2 \int_q \frac{\Delta (q) (1 -
  K(q))}{(q^2 + m^2)^2}\\
\eta (m^2, \lambda) & 0
\end{array}\right.
\]
Hence, the RG equations of the parameters become
\[
\left\lbrace\begin{array}{c@{~=~}l}
\frac{d m^2}{dt}&
2 m^{2} + \frac{\lambda}{2} \int_q \frac{\Delta (q)}{q^2 + m^{2}}
\\
\frac{d\lambda}{dt}&
\lambda - 3 \lambda^{2} \int_q \frac{\Delta (q) (1-K(q))}{(q^2 +
  m^{2})^2}
\end{array}\right.
\]
This implies
\[
m^{2*} = \mathrm{O} (\lambda)
\]
Hence, near the fixed point we can approximate the RG equations
further as
\[
\left\lbrace\begin{array}{c@{~=~}l}
\frac{d m^2}{dt}&
\left( 2 - \frac{\lambda}{2} \int_q \frac{\Delta (q)}{q^4} \right)
 m^{2} + \frac{\lambda}{2} \int_q \frac{\Delta (q)}{q^2}
\\
\frac{d\lambda}{dt}&
\lambda - 3 \lambda^{2} \int_q \frac{\Delta (q) (1-K(q))}{q^4}
\end{array}\right.
\]
Thus, we obtain
\[
\left\lbrace
\begin{array}{c@{~=~}l}
\lambda^* & \frac{1}{3 \int_q \frac{\Delta (q) (1-K(q))}{q^4}}\\
m^{2*} & - \frac{\lambda^*}{4} \int_q \frac{\Delta (q)}{q^2}
\end{array}\right.
\]
and the anomalous dimension of the squared mass is given by
\[
\beta_m^* = - \frac{\lambda^*}{2} \int_q \frac{\Delta (q)}{q^4}
\]

The fixed point values $\lambda^*, m^{2*}$ have no physical meaning,
and they depend on the choice of the cutoff function $K$.  However,
the anomalous dimension $\beta_m^*$ is independent of the choice, and
we can prove the independence by using an argument analogous to the
one given in subsect.~2.12 for universality.  Nevertheless, at any
finite order, the perturbative result depends on $K$.  To calculate
$\beta_m^*$, we make the simplest choice
\[
K (q) \to \theta (1-q^2)
\]
Then, we obtain
\[
\Delta (q) \equiv - 2 q^2 \frac{d}{dq^2} K(q) = 2 \delta (q^2-1)
\]
Thus,
\[
\int_q \frac{\Delta (q)}{q^4} = \frac{1}{(2\pi)^2} \int_0^\infty q
dq^2 \, \frac{1}{q^4} 2 \delta (q^2-1) = \frac{1}{2 \pi^2}
\]
Since
\[
\Delta (q) K(q)^n = - 2 q^2 \frac{d}{dq^2} \frac{K(q)^{n+1}}{n+1}
\]
we obtain
\[
\int_q \frac{\Delta (q) K(q)^n}{q^4} = \frac{1}{n+1} \cdot \frac{1}{2
  \pi^2}
\]
Thus, we obtain
\[
\beta_m^* = - \lambda^* \int_q \frac{\Delta (q) (1 - K(q))}{q^4} = -
\frac{1}{3}
\]
This is the same as what we get from the $\ep$ expansions at the
first order.

\newpage
\subsection*{Appendix: The gaussian fixed point}

In this appendix we would like to construct the gaussian fixed point
explicitly using three formalisms: Polchinski with $m^2$, Polchinski
with $m^2 = 0$, and Wilson.  In all the formalisms, we look for a
solution of the respective ERG differential equation corresponding to
the free massive theory.  The gaussian fixed point is the massless
limit.

\subsection*{Polchinski with $m^2$}

We look for a solution quadratic in fields:
\[
S(t, m^2) = \frac{1}{2} \int_p \phi (p) \phi (-p) \left( - \frac{p^2 +
      m^2 (t)}{K(p)} + \V_2 (t; p,-p) \right)
\]
In the absence of quartic interaction terms, both $b_m$ and $\eta$
vanish, and we obtain
\[
m^2 (t) = m^2 \e^{2t}
\]
where $m^2$ is a constant.  The vertex $\V_2$ satisfies the two
conditions:
\[
\left\lbrace\begin{array}{c@{~=~0}}
\V_2 (t, m^2;p,-p)\Big|_{p^2=0} \\
\frac{\partial}{\partial p^2} \V_2 (t, m^2;p,-p)\Big|_{p^2=0}
\end{array}\right.
\]
and the ERG differential equation
\[
\frac{\partial}{\partial t} \V_2 (t,m^2;p,-p) = \left(2 - 2 p^2
    \frac{\partial}{\partial p^2} \right) \V_2 (t,m^2;p,-p) + \frac{\Delta
  (p)}{p^2 + m^2 \e^{2t}} \V_2 (t,m^2; p,-p)^2
\]
The general solution is given by
\[
\V_2 (t, m^2;p,-p) = - \frac{\e^{2t}}{\frac{1}{p^4 \e^{-4t} f(p \e^{-t})} +
  \frac{1 - K(p)}{p^2 \e^{-2t} + m^2}}
\]
where $f(p)$ is an arbitrary function regular at $p^2=0$.

For any $m^2 \ge 0$, we obtain
\[
\lim_{t \to \infty} \V_2 (t; p,-p) = 0
\]
so that the action is asymptotically given by the free action
\[
S(t, m^2) \stackrel{t \gg 1}{\longrightarrow}
 - \frac{1}{2} \int_p \phi (p) \phi (-p) \frac{p^2 + m^2
  \e^{2t}}{K(p)} 
\]

The critical point is given by $m^2 = 0$, which gives the fixed point
action:
\[
\lim_{t \to \infty} S(t,0) = S_G^* \equiv - \frac{1}{2} \int_p \phi (p) \phi
(-p) \frac{p^2}{K(p)}
\]
The continuum limit is obtained as
\[
\Sf (\bar{m}^2) = \lim_{t \to \infty} S(t, m^2 = \bar{m}^2 \e^{-2t}) = -
\frac{1}{2} \int_p \phi (p) \phi (-p) \frac{p^2 + \bar{m}^2}{K(p)}
\]

\subsection*{Polchinski with $m^2=0$}

We assume that the action is quadratic in fields:
\[
S_t = \frac{1}{2} \int_p \phi (p) \phi (-p) \left(
- \frac{p^2}{K(p)} + \V_2 (t; p,-p) \right)
\]
where the vertex $\V_2$ satisfies the condition
\[
\frac{\partial}{\partial p^2} \V_2 (t; p,-p)\Big|_{p^2=0} = 0
\]
and the ERG differential equation
\[
\frac{\partial}{\partial t} \V_2 (t;p,-p) =
\left( 2 - 2 p^2 \frac{\partial}{\partial p^2} \right) \V_2 (t;p,-p)
+ \frac{\Delta (p)}{p^2} \V_2 (t;p,-p)^2
\]
The general solution is given by
\[
\V_2 (t; p,-p) = - \frac{1}{\frac{1}{\e^{2t} \left( m^2 + p^4 \e^{-4t}
        f(p \e^{-t}) \right)} + \frac{1 - K(p)}{p^2}}
\]
where $m^2 \ge 0$ is an arbitrary constant, and $f (p)$ is an
arbitrary function of $p^2$ regular at $p^2 = 0$.

$m^2 = 0$ is the critical point:
\[
\lim_{t \to \infty} \V_2 (t; p,-p) \Big|_{m^2=0} = 0
\]
so that the fixed point action is given by the same action as for the
Polchinski with $m^2$:
\[
S_G^* = - \frac{1}{2} \int_p \phi (p) \phi (-p)  \frac{p^2}{K(p)}
\]

The massive free theory is obtained as the continuum limit:
\[
\lim_{t \to \infty} \V_2 (t; p,-p)\Big|_{m^2 = \bar{m}^2 \e^{- 2t}}
= - \frac{p^2 \bar{m}^2}{p^2 + \bar{m}^2 (1 - K(p))}
\]
The action of the theory is given by
\[
S_P (\bar{m}^2) \equiv - \frac{1}{2} \int_p \phi (p) \phi (-p)
\frac{p^2+\bar{m}^2}{K(p)} \frac{p^2}{p^2 + \bar{m}^2 (1-K(p))}
\]
The cutoff-independent correlation function is obtained as
\[
\frac{1}{p^2} + \frac{1}{K(p)^2} \left(
\vev{\phi (p) \phi (-p)}_{S(\bar{m}^2)} - \frac{K(p)}{p^2} \right)
= \frac{1}{p^2 + \bar{m}^2}
\]

\subsection*{Wilson}

Let us assume a quadratic action:
\[
S_t = \frac{1}{2} \int_p \phi (p) \phi (-p)\,u(t;p,-p) 
\]
In the Wilsonian case the anomalous dimension depends on both the
quadratic and quartic terms.  Hence, it may not vanish even for the
quadratic theory.  Hence, the ERG differential equation of Wilson
gives
\begin{eqnarray*}
\partial_t u_2 (t;p,-p) &=& \left(2 - 2 p^2 \frac{\partial}{\partial
      p^2} + 2 \frac{\Delta (p)}{K(p)} - \eta (t) \right) u_2 (t;
p,-p) \\
&& + 2 \left( \frac{\Delta (p)}{K(p)} + 1 - \frac{1}{2} \eta
    (t)\right) u_2 (t;p,-p)^2
\end{eqnarray*}
Given the initial condition
\[
u_2 (0; p,-p) = - \omega (p) = - \left(\omega (0) + p^2 + \cdots\right)
\]
the solution is given by
\[
u_2 (t;p,-p) = - \frac{\e^{2 t} \omega (p \e^{-t})}{\e^{2 t} \omega (p
  \e^{-t}) + \e^{\int_0^t dt'\,\eta(t')} \frac{K(p)^2}{K(p \e^{-t})^2}
 \left(1 - \omega (p \e^{-t})\right)}
\]
where
\[
\e^{\int_0^t dt'\,\eta(t')} = \frac{1}{2 \left(1 - \omega
      (0)\right)^2}
\left\lbrace 1 - 2 \e^{2t} \omega (0)\left(1 - \omega (0)\right) +
\sqrt{1 - 4 \e^{2t} \omega (0) \left( 1 - \omega (0)\right)}
\right\rbrace
\]
The solution is valid for any $t > 0$ if $\omega (0) = 0$.  For
$\omega (0) > 0$, however, the above solution is valid only for $t >
0$ that satisfies
\[
4 \e^{2t} \omega (0)\left(1 - \omega (0)\right) < 1
\]

At the critical point $\omega (0)=0$, we obtain
\[
\lim_{t\to \infty} u_2 (t;p,-p) = - \frac{p^2}{p^2 + K(p)^2}
\]
corresponding to the fixed point action:
\[
S_G^* = - \frac{1}{2} \int_p \phi (p) \phi (-p) \frac{p^2}{p^2 + K(p)^2}
\]
Physically this is the same Gaussian fixed point as in the Polchinski
with or without $m^2$, but the concrete form depends on the formalism.
To obtain the massive continuum limit we choose the $t$-dependent
initial condition
\[
\omega (0) = \bar{m}^2 \e^{-2t}
\]
so that
\[
\lim_{t\to\infty} u_2 (t;p,-p) = - \frac{p^2 + \bar{m}^2}{p^2 +
  \bar{m}^2 + K(p)^2 z (\bar{m}^2)}
\]
where
\[
z (\bar{m}^2) \equiv \frac{1}{2} \left( 1 - 2 \bar{m}^2 + \sqrt{1 - 4
      \bar{m}^2} \right)
\]
Obviously, we must restrict $\bar{m}^2$ to 
\[
4 \bar{m}^2 < 1
\]
With the continuum limit action
\[
S_W (\bar{m}^2) = - \frac{1}{2} \int_p \phi (p) \phi (-p)
 \frac{p^2 + \bar{m}^2}{p^2 + \bar{m}^2 + K(p)^2 z (\bar{m}^2)}
\]
the cutoff independent correlation function is obtained as
\[
\frac{1}{K(p)^2} \left( \vev{\phi (p) \phi (-p)}_{S(\bar{m}^2)} - 1
\right) = \frac{z(\bar{m}^2)}{p^2 + \bar{m}^2}
\]
We get a mass dependent field normalization in this case.

The above continuum limit action $S_W$ can be obtained from the
continuum action $S_P$ of the Polchinski formalism with $m^2=0$ as
follows:
\begin{eqnarray*}
\e^{S_W [\phi]} &=& \int [d\phi'] \exp \Bigg[ S_P [\phi']\\
&&\, - \frac{1}{2} \int_p A
    (p)^2 \left( \phi (p) - \frac{B(p)}{A(p)} \phi' (p) \right)
\left( \phi (-p) - \frac{B(p)}{A(p)} \phi' (-p) \right) \Bigg]
\end{eqnarray*}
where
\[
\left\lbrace\begin{array}{c@{~=~}l}
 \frac{1}{A(p)^2} & 1 - z (\bar{m}^2) \frac{K(p) (1-K(p))}{p^2}\\
 \frac{B(p)^2}{A(p)^2} & z (\bar{m}^2)
\end{array}\right.
\]

\newpage

\noindent{\Large\textbf{Acknowledgment}}

\vspace{0.5cm} I would like to thank Drs.~L.~Akant and K.~\"Ulker for
giving me an opportunity to present the lectures in a stimulating
environment.  I also thank Bekir Can L\"utf\"uo\u{g}lu for much help,
and Dr.~H.~Gies for a list of review articles.

\end{document}